# Foundations of photonic quantum computation

Martin Bombardelli[1], Gérard Fleury[1]
Philippe Lacomme[1] and Bogdan Vulpescu[2]

[1]*Université Clermont Auvergne, UMR 6158 LIMOS, 1 rue de la Chébarde, Aubière, 63178, France*
*Martin.bombardelli@isima.fr, Gerard.fleury@isima.fr, philippe.lacomme@isima.fr*
[2]*Université Clermont Auvergne, UMR 6533 LPCA, 4 Av. Blaise Pascal, 63170 Aubière, France*
*bogdan.vulpescu@clermont.in2p3.fr*

## Abstract

This work aims to introduce the fundamental concepts required to perform computations on photonic quantum computers by presenting the gates specific to this architecture and highlighting the connections between standard Pauli gates and those available in photonic systems. The introduction navigates between physical considerations related to the optical components used, theoretical aspects concerning quantum operators, and a more applied section introducing implementations using the Perceval library developed by Quandela. This paper is intended for engineers and researchers familiar with Pauli gates and standard quantum concepts, looking at a clear and compact introduction to photonic components. A second part aims to introduce the concept of polarization, not from a theoretical perspective, but through its practical applications. To do so, we compare the similarities and differences between the original Grover's algorithm formulation and a version that leverages polarization. Components specific to polarization are introduced and described in the context of the computations involved in Grover's algorithm. The description provided is as mathematical as possible and deliberately avoids physical considerations, in order to allow researchers familiar with "conventional" quantum circuits to more easily grasp the concepts.

## 1. Introduction

### *1.1. Photonic quantum computing*

Simulating quantum physics on a computer is wildly believed to be an algorithmically difficult task. This has been the case as early as 1980, this led Yuri Manin (Manin, 1980) and (Nature Rev Phys, 2022) to mention quickly the idea of using quantum systems to make quantum automatons. The same year, Paul Benioff (Benioff,1980) introduced quantum Turing machines. One of the problems that arise when one wants to simulate quantum physics accurately is the amount of memory needed to fully represent the quantum state of the system (Buluta et al., 2009). This difficulty in simulating quantum systems led Richard Feynman to imagining a "computer" made from and for quantum systems in his foundational article "Simulating Physics with Computers" (Feynman, R.P.,2018). The concept was then extended to what is known as gate model quantum computing (QC) by David Deutsch (Deutsch, 1985). The first advantage of this quantum model of computation was found by leveraging the query model on Deutsch's problem. QC allows us to check if a given function $f:\{0,1\} \to \{0,1\}$ is constant ($f(0) = f(1)$) or if it is balanced ($f(0) \neq f(1)$) in one query whereas its classical counterpart needs two measurements. Then an extended version was made by (Deutsch et al.,1992) with an even greater improvement over the classical counterpart.

Building on these early successes, researchers came up with other quantum algorithms which had a lower algorithmic complexity than their classical counterparts. The canonical and most famous examples are Lov Grover's (Grover, 1996) search algorithm and Shor's factoring algorithm (Shor, 1994).

Photonic QC is one out of many physical realizations of a quantum computer. The first draft came from (Chuang et al., 1995) who introduced the "dual rail encoding" allowing and to encode qubits into Fock states (Fock, 2004). This led Knill, Laflamme and Milburn to come up with an entire protocol to physically realize a gate model quantum computer using optical components known as KLM protocol (Knill et al., 2001). This protocol was enriched the following years by (Knill, 2002) and (Ralph et al.,2001). Through the scope of theses quantum computers, one can realize the two most famous algorithms of the field, namely Grover's search realized by (Samsonov et al.,2020) and Shor's factoring algorithm realized by (Politi et al., 2009). But light encompasses another property which we call polarization which arises from the electromagnetic wave associated with it. For a good lecture, see (Roussel,2023). One of the most interesting things using photonic quantum computing is the numerous numbers of ways in which you can encode qubits. In (Heurtel et al., 2023), Heurtel N. said: *"Photons have many degrees of freedom and offer a rich variety of ways to encode qubits. One of the most common approaches is to use the dual-rail path encoding."*

Staying in the realm of KLM protocol is not leveraging to the fullest the possibilities offered by photonic quantum computing. Coming from here, (Kiwat et al., 2000) proposed an implementation of the Grover algorithm exploiting the polarization properties of light by encoding two qubits on two spatial modes and two polarization modes. One can even consider the realm of continuous variables quantum computing with the GKP encoding (Gottesman., 2001) and later on, with its experimental realization by (Konno et al., 2023).

Hardware development underwent astonishing growth with numerous actors showing their progress (AbuGhanem,2024) (Fyrillas et al., 2025). Gate realisation through optical component also saw progress, with (Crespi, 2011) in 2011 demonstrating a CNOT gate for polarization encoded qubits of success probability $\frac{1}{9}$.

## *1.2. Understanding Modes in the Specific Case of Two Spatial Modes*

To encode qubits in a photonic quantum computer, the first method proposed was path encoding, introduced by (Chuang and Yamamoto, 1995). The locations of the photons - referred to as spatial modes - define the state of the qubit. Spatial modes can, for instance, correspond to optical fibers. Throughout this chapter, we use the term mode to refer specifically to spatial mode, to simplify the notation. We use the notation $|n_0, n_1, \ldots n_{m-1}\rangle_{01\ldots(m-1)}$ where $0,1,\ldots(m-1)$ denote m spatial modes occupied by $n_0, n_1, \ldots n_{m-1}$ photons, respectively. The state described by this notation is called a Fock state.

**Remark**

When possible and in cases where ambiguity may arise, we use the notation $|n_0 n_1 \ldots n_{m-1}\rangle_F$ to indicate explicitly that the set of modes corresponds to a Fock state and omitting the names associated with the spatial modes.

It is common to omit the commas between each component of the state, so that $|n_0, n_1 \ldots, n_{m-1}\rangle_f$ is often written as $|n_0 n_1 \ldots n_{m-1}\rangle_f$

The definition of a qubit in the dual rail encoding relies on a single photon distributed over two spatial modes:

$$|0\rangle = |1,0\rangle_{01} = |1,0\rangle_f$$
$$|1\rangle = |0,1\rangle_{01} = |0,1\rangle_f$$

When the photon is in mode 0, this corresponds to the Fock state $|1,0\rangle_{01}$ sometimes written simply as $|1,0\rangle$ (although this notation should be avoided to prevent confusion between qubit states and Fock states). The Fock state $|1,0\rangle_{01}$ is denoted as $|0\rangle$ and a representation is provided in Figure 1.

We are considering here a specific case where a single photon is distributed over two modes.

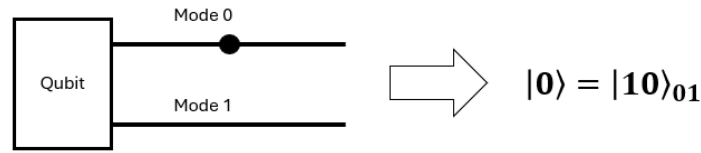

*Figure 1. State $|0\rangle = |10\rangle_{01}$*

Similarly, the Fock state $|01\rangle_{01}$ is defined as $|1\rangle$ and a representation is provided in Figure 2.

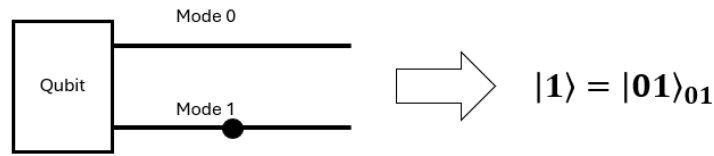

*Figure 2. State $|1\rangle = |01\rangle_{01}$*

## Qubit Encoding with Fock States in Perceval

Using the Perceval library, one can verify the consistency between the theoretical formalism and the implementation provided by Quandela. The test is limited to two modes and takes advantages of:

- One procedure `toFockState(qubitState)` to transform one quantum state into a Fock state;
- One procedure `strState(state)` to print.

To install the right Perceval library, USE pip install perceval-quandela and NOT pip install perceval !

It is thus possible to test $|0\rangle \otimes |0\rangle$, $|1\rangle \otimes |0\rangle$, $|0\rangle \otimes |1\rangle$ and $|1\rangle \otimes |1\rangle$

```python
import perceval as pcvl

def toFockState(qubitState):
       pe = {0:[1,0],  1:[0,1]}
       res =  pe[qubitState[0]] + pe[qubitState[1]]
       return res

def strState(state):
    return str(pcvl.BasicState(state))

qubit_istate = [[0,0], [0,1], [1,0],[1,1]]
for oqstate in output_qubit_states:
       istate = toFockState(qubit_istate)

       print("Input qubit state:", strState(qubit_istate))
       print("Corresponding input Fock state:", strState(istate))
```

The results are shown in Figure 3 and correspond to the expected outcomes, namely:

$$|0\rangle \otimes |0\rangle = |1010\rangle_{0123}$$
$$|0\rangle \otimes |1\rangle = |1001\rangle_{0123}$$
$$|1\rangle \otimes |0\rangle = |0110\rangle_{0123}$$
$$|1\rangle \otimes |1\rangle = |0101\rangle_{0123}$$

The results of the Python program are consistent with the expected outcomes, as shown in Figure 3. Let us note there can exist Fock states that do not represent qubits including for example $|11\rangle_{01}$ that corresponds to two photons (one in each of the modes 0 and 1).

```
Input qubit state: |0,0>
Corresponding input Fock state: |1,0,1,0>
Input qubit state: |0,1>
Corresponding input Fock state: |1,0,0,1>
Input qubit state: |1,0>
Corresponding input Fock state: |0,1,1,0>
Input qubit state: |1,1>
Corresponding input Fock state: |0,1,0,1>
```

*Figure 3. Representation of qubits and their corresponding Fock states using Perceval*

### 1.3. Creation and Annihilation Operators for the Modes

The creation and annihilation operators associated with a mode $m$ are denoted $\hat{a}_m^\dagger$ and $\hat{a}_m$, respectively. Their actions are as follows:

- The creation operator $\hat{a}_m^\dagger$ allows the transition from $|0\rangle_m$ to $|1\rangle_m$ in spatial mode 0. For example, $\hat{a}_0^\dagger$ transforms the vacuum state $|00\rangle_{01}$ into $|10\rangle_{01}$.
- The annihilation operator $\hat{a}_m$ allows the transition from $|1\rangle_m$ to $|0\rangle_m$. For example, $\hat{a}_0$ transforms $|10\rangle_{01}$ into the vacuum state $|00\rangle_{01}$.

These operators correspond to the emission or annihilation of a photon in a given mode. They play a crucial role when describing the effects of complex optical components.

## 2. The Generic Beam Splitter Component

### 2.1. Physical Description of the Action of the Beam Splitter

A beam splitter is an optical component used in photonic circuits. It acts on the fields associated with the photons we send on it's two output modes and results in two output modes. Let $E_a$ and $E_b$ denote the amplitudes of the two input electric fields associated with an arbitrary number of photons in the two input spatial modes $a$ and $b$. Also, let $E_c$ and $E_d$ the amplitudes of the two output fields for the two spatial modes $c$ and $d$, as illustrated in Figure 4. Let $r_{ac}$ and $r_{bd}$ be the reflection coefficients, $t_{ad}$ and $t_{bc}$ the transmission coefficients.

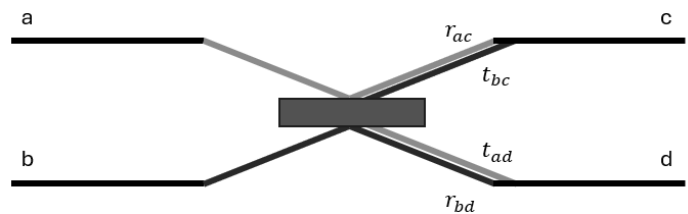

*Figure 4. Representation of a Beam Splitter with Two Input Modes and Two Output Modes*

The output field amplitudes can be expressed in terms of the input amplitudes through the so-called continuity relations:

$$E_c = r_{ac}E_a + t_{bc}E_b \qquad (1)$$

$$E_d = t_{ad}E_a + r_{bd}E_b \qquad (2)$$

where

$r_{ac}$ : is the contribution of $E_a$ found at the output $E_c$

$r_{bd}$ : is the contribution of $E_b$ found at the output $E_d$

$t_{ad}$ : is the contribution of $E_a$ found at the output $E_d$

$t_{bc}$ : is the contribution of $E_b$ found at the output $E_c$

The conservation of energy (proportional to the squared modulus of $\boldsymbol{E_x}$) leads to:

$$|E_a|^2 + |E_b|^2 = |E_c|^2 + |E_d|^2 \qquad (3)$$

Using (1.1) and (1.2) the action of the beam splitter can be described in matrix form based on the field amplitudes of each mode, as shown in *Figure* 5.

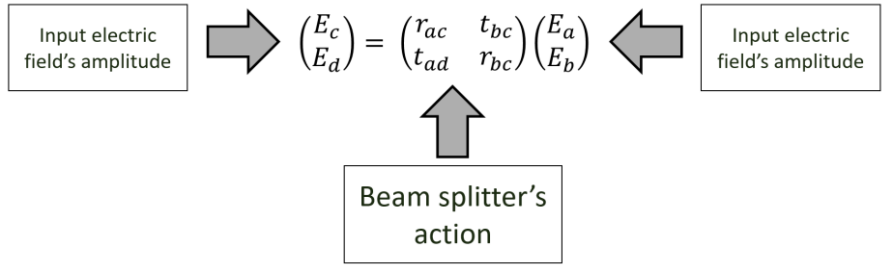

*Figure 5. Action of a Beam Splitter*

Let us note:

$$U = \begin{bmatrix} r_{ac} & t_{bc} \\ t_{ad} & r_{bd} \end{bmatrix}$$

If there is a single input with the entire field distributed over solely the mode $\boldsymbol{a}$ (*i.e.* $|E_a|^2 \neq 0$ and $|E_b|^2 = 0$), Then:

$$\begin{pmatrix} r_{ac} & t_{bc} \\ t_{ad} & r_{bd} \end{pmatrix} \cdot \begin{pmatrix} E_a \\ 0 \end{pmatrix} = \begin{pmatrix} E_c \\ E_d \end{pmatrix}$$

It follows that:

$$r_{ac} \cdot E_a = E_c$$

$$t_{ad} \cdot E_a = E_d$$

Still in the case where the input is on one mode (here on $a$), we have $|E_b|^2 = 0$ in energy conservation:

$$|E_a|^2 = |E_c|^2 + |E_d|^2$$

because $r_{ac} \cdot E_a = E_c$ and $t_{ad} \cdot E_a = E_d$:

$$|E_a|^2 = |r_{ac}E_a|^2 + |t_{ad}E_a|^2$$

So

$$|E_a|^2 = (|r_{ac}|^2 + |t_{ad}|^2) \cdot |E_a|^2$$

$$1 = |r_{ac}|^2 + |t_{ad}|^2$$

By the same reasoning, with $E_a = 0$ and $E_b \neq 0$:
$$\begin{pmatrix} r_{ac} & t_{bc} \\ t_{ad} & r_{bd} \end{pmatrix} \cdot \begin{pmatrix} 0 \\ E_b \end{pmatrix} = \begin{pmatrix} E_c \\ E_d \end{pmatrix}$$

One would obtain:
$$1 = |r_{bd}|^2 + |t_{bc}|^2$$

□

Given (1.1) and (1.2)
$$E_c = r_{ac}E_a + t_{bc}E_b$$
$$E_d = t_{ad}E_a + r_{bd}E_b$$

It is possible to calculate:
$$|E_c|^2 = |r_{ac}E_a + t_{bc}E_b|^2 = |r_{ac}E_a|^2 + |t_{bc}E_b|^2 + r_{ac}E_a \cdot (t_{bc}E_b)^* + t_{bc}E_b \cdot (r_{ac}E_a)^*$$

Similarly
$$|E_d|^2 = |t_{ad}E_a|^2 + |r_{bd}E_b|^2 + t_{ad}E_a \cdot r_{bd}^*E_b^* + r_{bd}E_b \cdot t_{ad}^*E_a^*$$

Adding the two lines together:
$$|E_c|^2 + |E_d|^2 = (|r_{ac}|^2 + |t_{ad}|^2)|E_a|^2 + (|r_{bd}|^2 + |t_{bc}|^2)|E_b|^2$$
$$+ r_{ac}E_a \cdot (t_{bc}E_b)^* + t_{bc}E_b \cdot (r_{ac}E_a)^* + t_{ad}E_a \cdot r_{bd}^*E_b^* + r_{bd}E_b \cdot t_{ad}^*E_a^*$$
$$= \underbrace{(|r_{ac}|^2 + |t_{ad}|^2)}_{=1 \text{ (see the previous remark)}} |E_a|^2 + \underbrace{(|r_{bd}|^2 + |t_{bc}|^2)}_{=1} |E_b|^2$$
$$+ (t_{ad}r_{bd}^* + r_{ac}t_{bc}^*)E_aE_b^* + (r_{bd}t_{ad}^* + t_{bc}r_{ac}^*)E_a^*E_b$$

so
$$|E_c|^2 + |E_d|^2 = |E_a|^2 + |E_b|^2 + (t_{ad}r_{bd}^* + r_{ac}t_{bc}^*)E_aE_b^* + (r_{bd}t_{ad}^* + t_{bc}r_{ac}^*)E_a^*E_b$$

**Reminder**

According to (3), we should have:
$$|E_a|^2 + |E_b|^2 = |E_c|^2 + |E_d|^2$$

By substituting $|E_c|^2 + |E_d|^2$ with its summed expression:
$$|E_a|^2 + |E_b|^2 = |E_a|^2 + |E_b|^2 + (t_{ad}r_{bd}^* + r_{ac}t_{bc}^*)E_aE_b^* + (r_{bd}t_{ad}^* + t_{bc}r_{ac}^*)E_a^*E_b$$

And so
$$(t_{ad}r_{bd}^* + r_{ac}t_{bc}^*)E_aE_b^* + (r_{bd}t_{ad}^* + t_{bc}r_{ac}^*)E_a^*E_b = 0$$

Since this expression must hold for any choice of $E_a$ and $E_b$, it implies that:
$$(t_{ad}r_{bd}^* + r_{ac}t_{bc}^*) = 0 \qquad (5)$$
$$(r_{bd}t_{ad}^* + t_{bc}r_{ac}^*) = 0 \qquad (6)$$

**Let us prove that $U$ is unitary**

Consider the matrix:

$$U^\dagger U = \begin{bmatrix} r_{ac}^* & t_{ad}^* \\ t_{bc}^* & r_{bd}^* \end{bmatrix} \cdot \begin{bmatrix} r_{ac} & t_{bc} \\ t_{ad} & r_{bd} \end{bmatrix}$$

$$U^\dagger U = \begin{bmatrix} r_{ac} r_{ac}^* + t_{ad} t_{ad}^* & r_{ac}^* t_{bc} + t_{ad}^* r_{bd} \\ t_{bc}^* r_{ac} + r_{bd}^* t_{ad} & t_{bc}^* t_{bc} + r_{bd} r_{bd}^* \end{bmatrix}$$

$$U^\dagger U = \begin{bmatrix} |r_{ac}|^2 + |t_{ad}|^2 & r_{ac}^* t_{bc} + t_{ad}^* r_{bd} \\ t_{bc}^* r_{ac} + r_{bd}^* t_{ad} & |t_{bc}|^2 + |r_{bd}|^2 \end{bmatrix}$$

Since it was previously demonstrated that

$$1 = (|r_{ac}|^2 + |t_{ad}|^2)$$
$$1 = (|r_{bd}|^2 + |t_{bc}|^2)$$

So

$$U^\dagger \cdot U = \begin{bmatrix} 1 & r_{ac}^* t_{bc} + t_{ad}^* r_{bd} \\ t_{bc}^* r_{ac} + r_{bd}^* t_{ad} & 1 \end{bmatrix}$$

And since we have also shown that:

$$(t_{ad} r_{bd}^* + r_{ac} t_{bc}^*) = 0$$
$$(r_{bd} t_{ad}^* + t_{bc} r_{ac}^*) = 0$$

We have:

$$U^\dagger \cdot U = \begin{bmatrix} 1 & 0 \\ 0 & 1 \end{bmatrix}$$

Therefore, the beam splitter's operation is unitary.

## 2.2. Parameterization of the Beam Splitter Matrix $U$

The gate $U$ representing the action of a beam splitter can then be denoted as $BS$:

$$BS = \begin{pmatrix} r_{ac} & t_{bc} \\ t_{ad} & r_{bd} \end{pmatrix}$$

There are several conventions for the beam splitter, some of which belong to $U(2)$ but not to $SU(2)$. In all cases, the operator must be unitary since it must belong to $U(2)$ either to $SU(2)$ so that:

$$BS^\dagger BS = Id$$

As the coefficients are complex numbers, each can be represented in exponential form. So

$$t_{bc} = |t_{bc}|.e^{i.\phi_{bc}}$$
$$r_{ac} = |r_{ac}|.e^{i.\phi_{ac}}$$
$$r_{bd} = |r_{bd}|.e^{i.\phi_{bd}}$$
$$t_{ad} = |t_{ad}|.e^{i.\phi_{ad}}$$

For the complex conjugates, we have:

$$t_{bc}^* = |t_{bc}|.e^{-i.\phi_{bc}}$$

$$r_{ac}^* = |r_{ac}|.e^{-i.\phi_{ac}}$$
$$r_{bd}^* = |r_{bd}|.e^{-i.\phi_{bd}}$$
$$t_{ad}^* = |t_{ad}|.e^{-i.\phi_{ad}}$$

As demonstrated in the previous section:

$$t_{bc}^*.r_{ac} + r_{bd}^*.t_{ad} = 0$$

Which can be rewritten as:

$$|t_{bc}|.e^{-i.\phi_{bc}}.|r_{ac}|.e^{i.\phi_{ac}} + |r_{bd}|.e^{-i.\phi_{bd}}.|t_{ad}|.e^{i.\phi_{ad}} = 0$$
$$|t_{bc}|.|r_{ac}|.e^{i.(\phi_{ac}-\phi_{bc})} + |r_{bd}|.|t_{ad}|.e^{i.(\phi_{ad}-\phi_{bd})} = 0$$

So

$$|t_{bc}|.|r_{ac}|.e^{i.(\phi_{ac}-\phi_{bc})} = -|r_{bd}|.|t_{ad}|.e^{i.(\phi_{ad}-\phi_{bd})}$$
$$|t_{bc}|.|r_{ac}| = -|r_{bd}|.|t_{ad}|.e^{i.(\phi_{ad}-\phi_{bd}-\phi_{ac}+\phi_{bc})}$$
$$\frac{|r_{ac}|}{|t_{ad}|} = -\frac{|r_{bd}|}{|t_{bc}|}.e^{i.(\phi_{ad}-\phi_{bd}-\phi_{ac}+\phi_{bc})}$$

It is now possible to analyze:

$$\frac{|r_{ac}|}{|t_{ad}|} = -\frac{|r_{bd}|}{|t_{bc}|}.e^{i.(\phi_{ad}-\phi_{bd}-\phi_{ac}+\phi_{bc})}$$

All magnitudes are positive real numbers, so $e^{i.(\phi_{ad}-\phi_{bd}-\phi_{ac}+\phi_{bc})}$ can only take the value $-1$.

So

$$\phi_{ad} - \phi_{bd} - \phi_{ac} + \phi_{bc} = \pi$$

And

$$\frac{|r_{ac}|}{|t_{ad}|} = \frac{|r_{bd}|}{|t_{bc}|}$$

Then

$$\frac{|r_{ac}|^2}{|t_{ad}|^2} = \frac{|r_{bd}|^2}{|t_{bc}|^2}$$

It was previously shown that:

$$|r_{ac}|^2 + |t_{ad}|^2 = 1$$
$$|t_{bc}|^2 + |r_{bd}|^2 = 1$$

So

$$\frac{1-|t_{ad}|^2}{|t_{ad}|^2} = \frac{1-|t_{bc}|^2}{|t_{bc}|^2}$$

i.e.

$$\frac{1}{|t_{ad}|^2} - 1 = \frac{1}{|t_{bc}|^2} - 1$$

It follows that

$|t_{ad}| = |t_{bc}|$ and we note $|t_{ad}| = |t_{bc}| = t$

The equation $\frac{|r_{ac}|}{|t_{ad}|} = \frac{|r_{bd}|}{|t_{bc}|}$ can be rewritten $\frac{|r_{ac}|}{t} = \frac{|r_{bd}|}{t}$ which implies that:

$$|r_{ac}| = |r_{bd}| \text{ and we note } |r_{ac}| = |r_{bd}| = r$$

So we get

$$r^2 + t^2 = 1$$

Thus, the BS operator can be parameterized as:
$$BS = \begin{pmatrix} r_{ac} & t_{bc} \\ t_{ad} & r_{bd} \end{pmatrix}$$

so

$$t_{bc} = |t_{bc}|.e^{i.\phi_{bc}}$$
$$r_{ac} = |r_{ac}|.e^{i.\phi_{ac}}$$
$$r_{bd} = |r_{bd}|.e^{i.\phi_{bd}}$$
$$t_{ad} = |t_{ad}|.e^{i.\phi_{ad}}$$

And then

$$BS = \begin{pmatrix} |r_{ac}|.e^{i.\phi_{ac}} & |t_{bc}|.e^{i.\phi_{bc}} \\ |t_{ad}|.e^{i.\phi_{ad}} & |r_{bd}|.e^{i.\phi_{bd}} \end{pmatrix}$$

Moreover

$$|t_{ad}| = |t_{bc}| \text{ and we note } |t_{ad}| = |t_{bc}| = t$$
$$|r_{ac}| = |r_{bd}| \text{ and we note } |r_{ac}| = |r_{bd}| = r$$

And by consequence:

$$BS = \begin{pmatrix} r.e^{i.\phi_{ac}} & t.e^{i.\phi_{bc}} \\ t.e^{i.\phi_{ad}} & r.e^{i.\phi_{bd}} \end{pmatrix}$$

**Remark on the Physical Meaning of the Coefficients $r$ and $t$**

The coefficients $r^2 = R$ and $t^2 = T$ correspond to:

The reflectance $R = \frac{P_{Reflected}}{P_{Incident}}$

The transmittance $T = \frac{P_{Transmitted}}{P_{Incident}}$

We can also define:

The absorbance $A = \frac{P_{Absorbed}}{P_{Incident}}$ where the $P$ values represent powers.

Energy conservation requires $R + T + A = 1$ and throughout this document, we consider lossless beam splitters (i.e. $A = 0$)

The BS operator is completely defined by the six parameters:

$$r, t, \phi_{ac}, \phi_{bc}, \phi_{ad}, \phi_{bd}$$

And by the constraints:
$$r^2 + t^2 = 1$$
$$\phi_{ad} - \phi_{bd} - \phi_{ac} + \phi_{bc} = \pi$$

Because:
$$\begin{cases} r^2 + t^2 = 1 \\ r = |r_{ac}| \geq 0 \; et \; t \geq 0 \end{cases} \Rightarrow \exists \theta \in [0, +\frac{\pi}{2}[, r = \sin(\theta) \; and \; t = \cos(\theta)$$
$$with \; \theta = \arctan\left(\frac{r}{t}\right)$$

One can write:
$$BS = \begin{pmatrix} r.e^{i.\phi_{ac}} & t.e^{i.\phi_{bc}} \\ t.e^{i.\phi_{ad}} & r.e^{i.\phi_{bd}} \end{pmatrix} = \begin{pmatrix} \sin(\theta).e^{i.\phi_{ac}} & \cos(\theta).e^{i.\phi_{bc}} \\ \cos(\theta).e^{i.\phi_{ad}} & \sin(\theta).e^{i.\phi_{bd}} \end{pmatrix}$$

Let us note that (Campos et al., 1989) suggest a change of variables:
$$\phi_t = \frac{1}{2}(\phi_{ad} - \phi_{bc})$$
$$\phi_r = \frac{1}{2}(\phi_{ac} - \phi_{bd} - \pi)$$
$$\phi_0 = \frac{1}{2}(\phi_{ad} + \phi_{bc})$$

To obtain
$$\phi_{ac} = \phi_0 + \phi_r$$
$$\phi_{ad} = \phi_0 + \phi_t$$
$$\phi_{bc} = \phi_0 - \phi_t$$
$$\phi_{bd} = \phi_0 - \phi_r - \pi$$

**Verification**

The verifications for $\phi_{ad}$ and $\phi_{bc}$ are straightforward; let us expand $\phi_{ac}$ and $\phi_{bd}$

For $\phi_{ac}$ :
$$\phi_0 + \phi_r = \frac{1}{2}(\phi_{ad} + \phi_{bc}) + \frac{1}{2}(\phi_{ac} - \phi_{bd} - \pi)$$
$$= \frac{1}{2}(\phi_{ad} + \phi_{bc} + \phi_{ac} - \phi_{bd} - \pi)$$
$$\phi_0 + \phi_r = \frac{1}{2}\left(\underbrace{\phi_{ad} - \phi_{ac} + \phi_{bc} - \phi_{bd}}_{\pi} - \pi + \phi_{ac} + \phi_{ac}\right)$$
$$\phi_0 + \phi_r = \phi_{ac}$$

And for $\phi_{bd}$ :
$$\phi_0 - \phi_r - \pi = \frac{1}{2}(\phi_{ad} + \phi_{bc}) - \frac{1}{2}(\phi_{ac} - \phi_{bd} - \pi) - \pi$$

$$\phi_0 - \phi_r - \pi = \frac{1}{2}\left(\underbrace{\phi_{ad} - \phi_{ac} + \phi_{bc} - \phi_{bd}}_{\pi} + \phi_{bd} + \phi_{bd} + \pi\right) - \pi = \frac{(2\phi_{bd} + 2\pi)}{2} - \pi$$

$$\phi_0 - \phi_r - \pi = \phi_{bd}$$

We can therefore rewrite by replacing each of the expressions:

$$BS = \begin{pmatrix} \sin(\theta).e^{i.\phi_{ac}} & \cos(\theta).e^{i.\phi_{bc}} \\ \cos(\theta).e^{i.\phi_{ad}} & \sin(\theta).e^{i.\phi_{bd}} \end{pmatrix} = \begin{pmatrix} \sin(\theta).e^{i.(\phi_0+\phi_r)} & \cos(\theta).e^{i.(\phi_0-\phi_t)} \\ \cos(\theta).e^{i.(\phi_0+\phi_t)} & \sin(\theta).e^{i.(\phi_0-\phi_r-\pi)} \end{pmatrix}$$

Thus, a common expression for the noiseless beam splitter becomes:

$$BS_1 \equiv BS = e^{i.\phi_0}\begin{pmatrix} \sin(\theta).e^{i.\phi_r} & \cos(\theta).e^{-i.\phi_t} \\ \cos(\theta).e^{i.\phi_t} & -\sin(\theta).e^{-i.\phi_r} \end{pmatrix}$$

### 2.3. Important Remark on Different Conventions

The definitions of $\phi_r$, $\phi_t$ and $\phi_0$ are not universal; for example, by changing the definition of $\phi_r$:

$$\phi_r = \frac{1}{2}(\phi_{ac} - \phi_{bd} + \pi)$$

We find

$$BS_2(\theta, \phi_r, \phi_t, \phi_0) = e^{i.\phi_0}\begin{pmatrix} -\sin\left(\frac{\theta}{2}\right).e^{i.\phi_r} & \cos\left(\frac{\theta}{2}\right).e^{-i.\phi_t} \\ \cos\left(\frac{\theta}{2}\right).e^{i.\phi_t} & \sin\left(\frac{\theta}{2}\right).e^{-i.\phi_r} \end{pmatrix}$$

Or alternatively, by defining:

$$r^2 + t^2 = 1 \Rightarrow \exists \theta \in [0, \pi[, r = \cos\left(\frac{\theta}{2}\right) \text{ and } t = \sin\left(\frac{\theta}{2}\right) \quad \text{with } \theta = \arctan\left(\frac{t}{r}\right)$$

One could also have obtained:

$$BS_3(\theta, \phi_r, \phi_t, \phi_0) = e^{i.\phi_0}\begin{pmatrix} \cos\left(\frac{\theta}{2}\right).e^{i.\phi_r} & \sin\left(\frac{\theta}{2}\right).e^{-i.\phi_t} \\ \sin\left(\frac{\theta}{2}\right).e^{i.\phi_t} & -\cos\left(\frac{\theta}{2}\right).e^{-i.\phi_r} \end{pmatrix}$$

By considering $\phi_r = \phi_t = \phi_0 = 0$ and $\theta = \frac{\pi}{2}$ we recover the expression of the Hadamard gate.

$$BS_3\left(\frac{\pi}{2}, 0, 0, 0\right) = \begin{pmatrix} \cos\left(\frac{\pi}{4}\right) & \sin\left(\frac{\pi}{4}\right) \\ \sin\left(\frac{\pi}{4}\right) & -\cos\left(\frac{\pi}{4}\right) \end{pmatrix} = \begin{pmatrix} \frac{\sqrt{2}}{2} & \frac{\sqrt{2}}{2} \\ \frac{\sqrt{2}}{2} & -\frac{\sqrt{2}}{2} \end{pmatrix} = H$$

### 2.4. Quandela Conventions

Note that the conventions presented above are not exhaustive. Perceval, the library proposed by (Heurtel et al., 2022), offers by default three conventions.

Let us now consider, in each case, all phases set to zero.

**Quandela Convention $H$ :**

$$BS_H(\theta \in [0;\pi], 0,0,0) = \begin{pmatrix} \cos\left(\frac{\theta}{2}\right) & \sin\left(\frac{\theta}{2}\right) \\ \sin\left(\frac{\theta}{2}\right) & -\cos\left(\frac{\theta}{2}\right) \end{pmatrix}$$

In particular, $BS_H\left(\frac{\pi}{2}, 0,0,0\right) = H$ defines the Hadamard gate. This convention can be obtained with the same definitions of $\phi_r, \phi_t$ and $\phi_0$ but by choosing:

$$\theta \in [0,\pi[\,, r = \cos\left(\frac{\theta}{2}\right) et\ t = \sin\left(\frac{\theta}{2}\right)$$

**Quandela Convention $R_x$ :**

$$BS_{Rx}(\theta \in [0;\pi], 0,0,0) = \begin{pmatrix} \cos\left(\frac{\theta}{2}\right) & i\sin\left(\frac{\theta}{2}\right) \\ i\sin\left(\frac{\theta}{2}\right) & \cos\left(\frac{\theta}{2}\right) \end{pmatrix}$$

The operator $BS_{Rx}$ is obtained by again choosing $\theta \in [0,\pi[\,, r = \cos\left(\frac{\theta}{2}\right)$ and $t = \sin\left(\frac{\theta}{2}\right)$, in the previous calculations, along with the selections of $\phi_{t,r,0}$ given in Table 1.

**New parameters for $\phi_{t,r,0}$ to define $BS_{Rx}$**

$$\phi_t = \frac{1}{2}(\phi_{ad} - \phi_{bc}),$$

$$\phi_r = \frac{1}{2}(\phi_{ac} - \phi_{bd}),$$

$$\phi_0 = \frac{1}{2}(\phi_{ad} + \phi_{bc} - \pi)$$

which can be rewritten

$$\phi_{ac} = \phi_0 + \phi_r$$
$$\phi_{bd} = \phi_0 - \phi_r$$
$$\phi_{ad} = \phi_0 + \phi_t + \frac{\pi}{2}$$
$$\phi_{bc} = \phi_0 - \phi_t + \frac{\pi}{2}$$

The Table 1 provides the correspondence between the old definitions of $\phi_{t,r,0}$ and the new definitions.

**Table 1: Choice of the new $\phi_{t,r,0}$**

| New parameters $\phi_{t,r,0}$ | Previous parameters $\phi_{t,r,0}$ introduced previously |
|---|---|
| $\phi_t = \frac{1}{2}(\phi_{ad} - \phi_{bc})$ | $\phi_t = \frac{1}{2}(\phi_{ad} - \phi_{bc})$ |
| $\phi_r = \frac{1}{2}(\phi_{ac} - \phi_{bd})$ | $\phi_r = \frac{1}{2}(\phi_{ac} - \phi_{bd} - \pi)$ |
| $\phi_0 = \frac{1}{2}(\phi_{ad} + \phi_{bc} - \pi)$ | $\phi_0 = \frac{1}{2}(\phi_{ad} + \phi_{bc})$ |

**Quandela Convention $R_y$:**

$$BS_{Ry}(\theta \in [0;\pi], 0,0,0) = \begin{pmatrix} \cos\left(\frac{\theta}{2}\right) & -\sin\left(\frac{\theta}{2}\right) \\ \sin\left(\frac{\theta}{2}\right) & \cos\left(\frac{\theta}{2}\right) \end{pmatrix}$$

This corresponds exactly to the $RY$ gate according to IBM's notation. This expression arises through similar transformations.

## 2.5. Symmetric Beam Splitter

In the first convention,

$$BS_1(\theta, \phi_r, \phi_t, \phi_0) = e^{i.\phi_0} \begin{pmatrix} \sin(\theta).e^{i.\phi_r} & \cos(\theta).e^{-i.\phi_t} \\ \cos(\theta).e^{i.\phi_t} & -\sin(\theta).e^{-i.\phi_r} \end{pmatrix}$$

A beam splitter is said to be symmetric if it is assigned $r = \frac{1}{\sqrt{2}}$ and $t = \frac{1}{\sqrt{2}}$ that is, in the convention, $\theta = \frac{\pi}{4}$ which corresponds to the photon having a "one in two chance of being reflected" and a "one in two chance of being transmitted".

From a mathematical point of view, we choose $\phi_r = \phi_t = \phi_0 = 0$ and we obtain, for example:

$$BS_1\left(\frac{\pi}{4}, 0,0,0\right) = 1 \begin{pmatrix} \frac{1}{\sqrt{2}}.1 & \frac{1}{\sqrt{2}}.1 \\ \frac{1}{\sqrt{2}}.1 & -\frac{1}{\sqrt{2}}.1 \end{pmatrix} = \frac{1}{\sqrt{2}} \begin{pmatrix} 1 & 1 \\ 1 & -1 \end{pmatrix} = H$$

In this same convention, by again taking $\theta = \frac{\pi}{4}$, but with $\phi_t = 0, \phi_r = -\frac{\pi}{2}$ and $\phi_0 = \frac{\pi}{2}$ we obtain, for example:

$$BS_1\left(\frac{\pi}{4}, -\frac{\pi}{2}, 0, \frac{\pi}{2}\right) = i \begin{pmatrix} \frac{1}{\sqrt{2}}.-i & \frac{1}{\sqrt{2}}.1 \\ \frac{1}{\sqrt{2}}.1 & -\frac{1}{\sqrt{2}}.i \end{pmatrix} = \frac{1}{\sqrt{2}} \begin{pmatrix} 1 & i \\ i & 1 \end{pmatrix}$$

The values of the symmetric beam splitter in the Quandela convention (for $\theta = \frac{\pi}{2}$ and $\phi_x = 0 \ \forall x$) are:

$$BS_H\left(\frac{\pi}{2}, 0,0,0,0\right) = \frac{1}{\sqrt{2}} \begin{pmatrix} 1 & 1 \\ 1 & -1 \end{pmatrix} \text{ which represents the Hadamard gate}$$

$$BS_{Rx}\left(\frac{\pi}{2}, 0,0,0,0\right) = \frac{1}{\sqrt{2}} \begin{pmatrix} 1 & i \\ i & 1 \end{pmatrix}$$

$$BS_{Ry}\left(\frac{\pi}{2}, 0,0,0,0\right) = \frac{1}{\sqrt{2}} \begin{pmatrix} 1 & -1 \\ 1 & 1 \end{pmatrix}$$

All these conventions are equivalent in the sense that one can be transformed into another. Let us clarify a subtlety that was previously overlooked. The Perceval implementation is as follows:

$$BS_{Rx}(\theta, \Phi_{tl}, \Phi_{bl}, \Phi_{tr}, \Phi_{br}) = \begin{pmatrix} e^{i.(\Phi_{tl}+\Phi_{tr})} \cos\left(\frac{\theta}{2}\right) & i.e^{i.(\Phi_{bl}+\Phi_{tr})} \sin\left(\frac{\theta}{2}\right) \\ i.e^{i.(\Phi_{br}+\Phi_{tl})} \sin\left(\frac{\theta}{2}\right) & e^{i.(\Phi_{bl}+\Phi_{br})} \cos\left(\frac{\theta}{2}\right) \end{pmatrix}$$

$$BS_{H}(\theta, \Phi_{tl}, \Phi_{bl}, \Phi_{tr}, \Phi_{br}) = \begin{pmatrix} e^{i.(\Phi_{tl}+\Phi_{tr})} \cos\left(\frac{\theta}{2}\right) & e^{i.(\Phi_{bl}+\Phi_{tr})} \sin\left(\frac{\theta}{2}\right) \\ e^{i.(\Phi_{br}+\Phi_{tl})} \sin\left(\frac{\theta}{2}\right) & -e^{i.(\Phi_{bl}+\Phi_{br})} \cos\left(\frac{\theta}{2}\right) \end{pmatrix}$$

$$BS_{Ry}(\theta, \Phi_{tl}, \Phi_{bl}, \Phi_{tr}, \Phi_{br}) = \begin{pmatrix} e^{i.(\Phi_{ti}+\Phi_{tr})} \cos\left(\frac{\theta}{2}\right) & -e^{i.(\Phi_{bl}+\Phi_{tr})} \sin\left(\frac{\theta}{2}\right) \\ e^{i.(\Phi_{br}+\Phi_{tl})} \sin\left(\frac{\theta}{2}\right) & e^{i.(\Phi_{bl}+\Phi_{br})} \cos\left(\frac{\theta}{2}\right) \end{pmatrix}$$

With the special cases:

$$BS_H(\theta, 0,0,0,0) = BS_H(\theta) = \frac{1}{\sqrt{2}} \begin{pmatrix} \cos\left(\frac{\theta}{2}\right) & \sin\left(\frac{\theta}{2}\right) \\ \sin\left(\frac{\theta}{2}\right) & -\cos\left(\frac{\theta}{2}\right) \end{pmatrix}$$

$$BS_{Rx}(\theta, 0,0,0,0) = BS_{Rx}(\theta) = \frac{1}{\sqrt{2}} \begin{pmatrix} \cos\left(\frac{\theta}{2}\right) & i.\sin\left(\frac{\theta}{2}\right) \\ i.\sin\left(\frac{\theta}{2}\right) & \cos\left(\frac{\theta}{2}\right) \end{pmatrix}$$

$$BS_{Ry}(\theta, 0,0,0,0) = BS_{Ry}(\theta) = \frac{1}{\sqrt{2}} \begin{pmatrix} \cos\left(\frac{\theta}{2}\right) & -\sin\left(\frac{\theta}{2}\right) \\ \sin\left(\frac{\theta}{2}\right) & \cos\left(\frac{\theta}{2}\right) \end{pmatrix}$$

It should be noted that the library allows passing 4 phases as parameters, which differs from the 3 phases described previously. Figure 6 shows on the left the placement of these four phases, to be compared with the parameters of the generic beam splitter on the right.

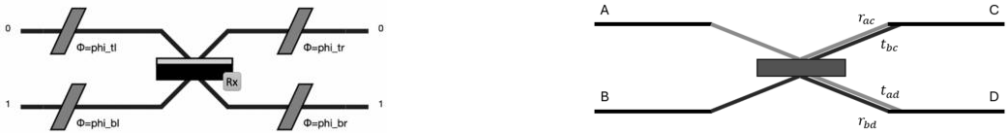

*Figure 6. Differences between the phase parameters of the beam splitter provided by the Perceval library and those considered previously*

The phases considered are those associated with each complex coefficient describing the action of the beam splitter, whereas Perceval adds a phase to each mode (inputs and outputs) that can be controlled using phase shifters (described in the next section). This allows them to obtain a more tunable beam splitter.

By default, in the library, the parameters are $\theta = \frac{\pi}{2}$ and $\phi = 0$ for all $\phi$, so that:

$$BS_H(\theta) = \frac{1}{\sqrt{2}} \begin{pmatrix} \cos\left(\frac{\theta}{2}\right) & \sin\left(\frac{\theta}{2}\right) \\ \sin\left(\frac{\theta}{2}\right) & -\cos\left(\frac{\theta}{2}\right) \end{pmatrix}$$

$$BS_H() = BS_H\left(\frac{\pi}{2}, 0,0,0,0\right) = \frac{1}{\sqrt{2}}\begin{pmatrix} 1 & 1 \\ 1 & -1 \end{pmatrix}$$

The notation $BS_H()$ corresponds to calling the procedure $BS_H()$ without specifying the parameters.

## 3. Phase Shifter.

Apart from the beam splitter, there is a second essential optical component that enables transformations on qubits. It is also possible to use phase shifters: a transparent plate acting on a single input mode and allowing the phase of the output mode to be changed if a photon is here. Physically, a phase shifter is a device that "slows down" the photon, creating a phase shift relative to the rest of the system. There are several ways to implement it. One way is simply to place a transparent plate in the photon's path, as stressed in the figure below (Figure 7).

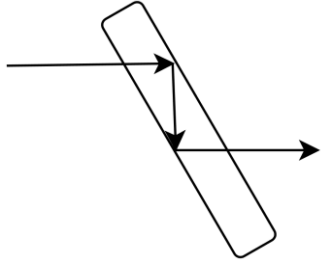

*Figure 7. Phase Shifter representation*

The phase shifter acts on a single input mode and produces a single output mode. In other words, its action can be described in terms of Fock states. Recall that the basis states of qubits are encoded on two modes by

$$|0\rangle = |10\rangle_{01}$$

and

$$|1\rangle = |01\rangle_{01}$$

Let $PS_p(\phi)$ denote the operator representing the application of a phase shifter with phase $\phi$ to a mode $p$. Its action on Fock states can be described as follows:

$$PS_p(\phi)|0\rangle_p \mapsto |0\rangle_p \; et \; PS_p(\phi)|1\rangle_p \mapsto e^{i\phi}|1\rangle_p$$

Its action on Fock states can be described as follows:

$$PS_1(\phi)|0\rangle = PS_1(|10\rangle_{01}) = |10\rangle_{01} = |0\rangle$$

This time, applying it to the qubit $|1\rangle$:

$$PS_1(\phi)|1\rangle = PS_1(\phi)|01\rangle_{01} = e^{i\phi}|01\rangle_{01} = e^{i\phi}|1\rangle$$

We thus have the action on qubits :

$$PS_0(\phi) = \begin{pmatrix} e^{i.\phi} & 0 \\ 0 & 1 \end{pmatrix} \text{ and } PS_1(\phi) = \begin{pmatrix} 1 & 0 \\ 0 & e^{i\phi} \end{pmatrix}$$

**Reminders**

According to the definition of $R_z$:

$$R_z(\phi) = e^{-\frac{iZ\theta}{2}} = \begin{pmatrix} e^{-\frac{i\phi}{2}} & 0 \\ 0 & e^{\frac{i\phi}{2}} \end{pmatrix}$$

This corresponds (up to a global phase) to the $P$ and $R_1$ gates in Qiskit:

$$P(\phi) = R_1(\phi) = \begin{pmatrix} 1 & 0 \\ 0 & e^{i\phi} \end{pmatrix}$$

The action of $PS_1(\phi)$ is exactly that of the gate $P(\phi) = e^{\frac{i\phi}{2}} R_z(\phi)$ that is, in matrix form:

$$\begin{pmatrix} 1 & 0 \\ 0 & e^{i\phi} \end{pmatrix} = e^{\frac{i\phi}{2}} \begin{pmatrix} e^{-\frac{i\phi}{2}} & 0 \\ 0 & e^{\frac{i\phi}{2}} \end{pmatrix}$$

We also immediately obtain:

$$PS_1(\phi) = R_1(\phi)$$

From a matrix perspective, we have the following:

$$PS_1(\phi)|0\rangle = \begin{pmatrix} 1 & 0 \\ 0 & e^{i.\phi} \end{pmatrix} \begin{pmatrix} 1 \\ 0 \end{pmatrix} = \begin{pmatrix} 1 \\ 0 \end{pmatrix} = |0\rangle$$

$$PS_1(\phi)|1\rangle = \begin{pmatrix} 0 \\ e^{i.\phi} \end{pmatrix} = e^{i.\phi}|1\rangle$$

and

$$PS_0(\phi)|0\rangle = e^{i.\phi}|0\rangle$$
$$PS_0(\phi)|1\rangle = |1\rangle$$

**Property**

Applying a phase shift to each mode is equivalent to applying a global phase shift to the qubit.
Let $|\psi\rangle = \begin{pmatrix} \alpha \\ \beta \end{pmatrix}$ be a qubit in an arbitrary state, then:

$$e^{i.\phi}.|\psi\rangle = PS_0(\phi) \ PS_1(\phi).\begin{pmatrix} \alpha \\ \beta \end{pmatrix}$$

**Proof.**

$$|\psi_0\rangle = PS_0(\phi)PS_1(\phi).\begin{pmatrix} \alpha \\ \beta \end{pmatrix}$$

$$|\psi_2\rangle = PS_0(\phi).\begin{pmatrix} \alpha \\ e^{i.\phi}\beta \end{pmatrix}$$

$$|\psi_3\rangle = \begin{pmatrix} e^{i.\phi}\alpha \\ e^{i.\phi}\beta \end{pmatrix} = e^{i.\phi}.\begin{pmatrix} \alpha \\ \beta \end{pmatrix}$$

☐

**Important Remark**

In particular, using $PS_1(\pi)$, the Z gate can be easily defined:

$$PS_1(\pi)|0\rangle = |0\rangle = Z.|0\rangle$$
$$PS(\pi)|1\rangle = e^{i.\pi}|1\rangle = -|1\rangle = Z.|1\rangle$$

## 4. Modeling of Single-Qubit Gates

### 4.1. Graphical Representations of the Beam Splitter and the Phase Shifter

Photonics systems can be described using various graphical representations of their optical components. To recognize them, here are a few examples. In the following diagrams, each line represents an optical mode. Let us begin with the beam splitter, shown in Figure 8 and let us continue with the phase shifter in Figure 9.

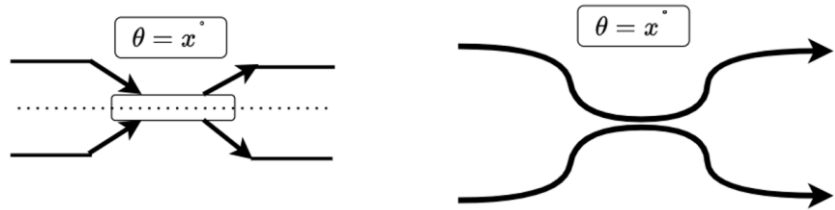

Figure 8. Two representations of the beam splitter

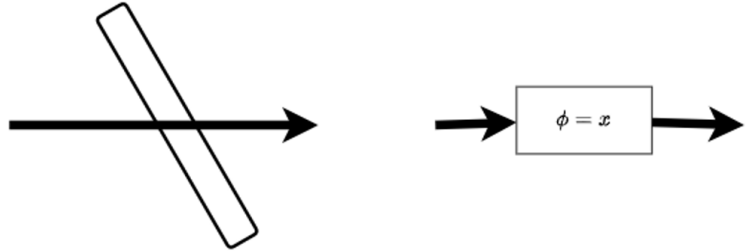

Figure 9. Two representations of the phase shifter

### 4.2. Definition of the Permutation Component

The permutation component is an optical element whose action is very simple: it exchanges the number of photons between the modes given as parameters.

For example, for $PERM_{10}$:

$$|n_0 n_1\rangle_{01} \mapsto |n_1 n_0\rangle_{01}$$

Another example involving more modes:
$$PERM_{2301}(|n_0 n_1 n_2 n_3\rangle_{0123}) = |n_2 n_3 n_0 n_1\rangle_{0123}$$
2301 corresponds to the permutation
$$\begin{matrix} 0 & 1 & 2 & 3 \\ \downarrow & \downarrow & \downarrow & \downarrow \\ 2 & 3 & 0 & 1 \end{matrix}$$
One can observe that $PERM_{01} = Id$ as it corresponds to the permutation:
$$\begin{matrix} 0 & 1 \\ \downarrow & \downarrow \\ 0 & 1 \end{matrix}$$

**Important Remark**

In the remainder of this document, the objective is to reconstruct standard quantum gates using optical components. To test the implementation of quantum gates using these optical elements, we use a circuit that simulates a Fock state with 2 qubits, i.e., 4 modes. In many instances, the circuits include additional modes that are not strictly necessary. This choice was made to help the reader better understand how to embed gates into a larger circuit. By default, all the code examples below refer to this standard circuit.

### 4.3. Implementation of the X Gate

Our goal is to construct an $X$-gate using the **permutation** component (Figure 10). This representation must be clearly differentiated from that of a Beam Splitter. By swapping the modes and maintaining the convention that "mode 0 is the top mode in the diagram" we apply the transformation $\text{PERM}_{10}: |n_0 n_1\rangle_{01} \mapsto |n_1 n_0\rangle_{01}$

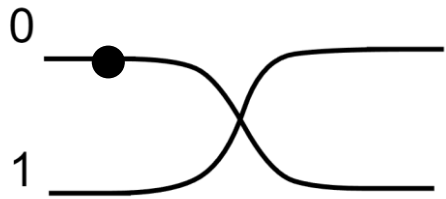

*Figure 10. Permutation of the two modes of a Fock state realizing the* X-*gate*

For an arbitrary qubit state $(\alpha|0\rangle + \beta|1\rangle)$ this corresponds to the action of the $X$-gate:

$$PERM_{10}(\alpha|0\rangle + \beta|1\rangle) = PERM_{10}(\alpha|10\rangle_{01} + \beta |01\rangle_{01})$$
$$= \alpha\, PERM_{10}(|10\rangle_{01}) + \beta\, PERM_{10}(|01\rangle_{01})$$
$$= \alpha|01\rangle_{01} + \beta |10\rangle_{01}$$
$$= \alpha |1\rangle + \beta |0\rangle$$

We indeed recover the effect of the Pauli $X$-gate (Figure 11).

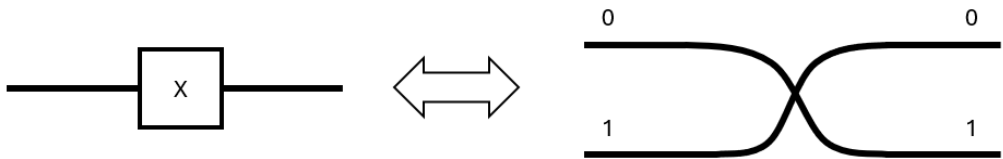

*Figure 11. Equivalent of the* X-*gate on qubits using PERM on the Fock state*

Example with 4 modes, numbered 0, 1, 2, 3:
- modes 0 and 1 represent qubit 0;
- modes 2 and 3 represent qubit 1.

To define an $X$ gate applied to **qubit 0**, i.e. the operation $X_0 = (X \otimes \text{Id})$, we must apply the following permutation:

$$\begin{matrix} 0 & 1 & 2 & 3 \\ \downarrow & \downarrow & \downarrow & \downarrow \\ 1 & 0 & 2 & 3 \end{matrix}$$

(Sometimes the shorthand notation $PERM_{1023} = PERM_{10}$ is used)

A corresponding **Perceval implementation** is described below.

**Step 1. Circuit Initialization**
This can be done by initializing qubits and then converting them into a **Fock state**. For example, starting from the state $|10\rangle$ the corresponding code would be:
```
qubit_istate = [1,0]
istate = toFockState(qubit_istate)
```

**Step 2. Define the circuit composed of 2 qubits, i.e., 4 Fock modes**
```
circuit = pcvl.Circuit(4)
```

**Step 3. Add an *X* gate to the circuit, applied to the first qubit**
```
circuit.add(0,PERM([1,0, 2,3]))
#or circuit.add(0,PERM([1,0]))
```

**Step 4. Display the circuit**
```
pcvl.pdisplay(circuit.compute_unitary())
print("")
pcvl.pdisplay(circuit)
```

**Step 5. Set the computation to do**
```
backend = pcvl.BackendFactory().get_backend("Naive")
backend.set_circuit(circuit)
backend.set_input_state(pcvl.BasicState(istate))
```

**Etape 6. Define the possible output states**
```
output_qubit_states = [
    [x1,x2]
    for x1 in [0,1] for x2 in [0,1]
]
```

**Etape 7. Display the results**
```
for oqstate in output_qubit_states:
    ostate = toFockState(oqstate)
    a = backend.prob_amplitude(pcvl.BasicState(ostate))
    print(strState(ostate)," -> ",strState(oqstate), a)

print('end')
```

Figure 12 shows the result of the code execution, showing that $X_0.|10\rangle = |00\rangle$ which is the expected outcome on the qubit states.

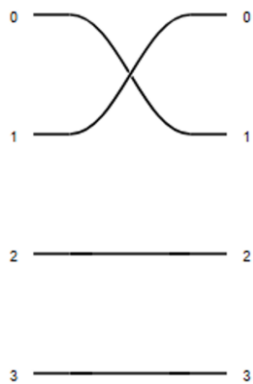

*Figure 12. Circuit obtained from the Perceval implementation of* $X_0 = (X \otimes Id)$

It is possible to modify the initial state and define $|10\rangle$, as follow:
```
qubit_istate = [1,1]
```
and then verify that $X_0.|11\rangle = |01\rangle$ (Figure 13).

```
Output state amplitudes: (post-selected on qubit states, not renormalized)
|x1,x2>
|1,0,1,0>  ->  |0,0> 0j
|1,0,0,1>  ->  |0,1> (1+0j)
|0,1,1,0>  ->  |1,0> 0j
|0,1,0,1>  ->  |1,1> 0j
```

*Figure 13. Result of the Perceval implementation of $X_0.|11\rangle = |01\rangle$*

### 4.4. Implementation of the SWAP Gate

We now focus on the standard SWAP gate applied to qubits. Compared to the $X$ gate described earlier, we must redefine the permutation over all four Fock modes, rather than just two. Since the circuit simulates Fock states, we have 4 modes for 2 qubits. To define a gate that swaps qubit 1 with qubit 2, i.e., performs the following operation:

$$SWAP|10\rangle = |01\rangle$$

i.e. the following permutations:

$$\begin{matrix} 0 & 1 & 2 & 3 \\ \downarrow & \downarrow & \downarrow & \downarrow \\ 2 & 3 & 0 & 1 \end{matrix}$$

and a visualization is given in Figure 14.

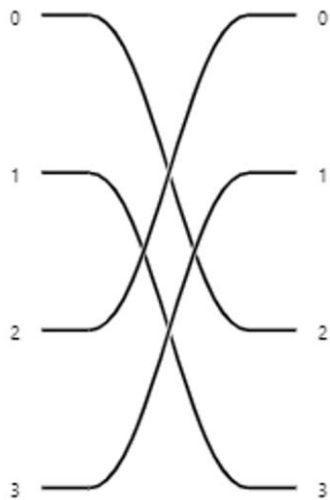

*Figure 14. Permutations to define a SWAP gate*

To test this gate, one can simply replace the lines:

```
# 3. Adding a X gate
circuit.add(0,PERM([1,0]))
```

by :

```
# 3. Adding a Swap gate
circuit.add(0,PERM([2,3, 0,1]))
```

As shown in Figure 15 we have $Swap|10\rangle = |01\rangle$.

```
Output state amplitudes: (post-selected on qubit states, not renormalized)
|x1,x2>
|1,0,1,0>  ->  |0,0> 0j
|1,0,0,1>  ->  |0,1> 0j
|0,1,1,0>  ->  |1,0> 0j
|0,1,0,1>  ->  |1,1> (1+0j)
```

*Figure 15. Result of the SWAP gate implemented with Perceval and applied to $|10\rangle$*

### 4.5. Implementation of the H Gate

Let us consider the expression obtained for the Beam Splitter introduced in section 2 :

$$BS_H = \frac{1}{\sqrt{2}}\begin{pmatrix} \cos\left(\frac{\theta}{2}\right) & \sin\left(\frac{\theta}{2}\right) \\ \sin\left(\frac{\theta}{2}\right) & -\cos\left(\frac{\theta}{2}\right) \end{pmatrix}$$

If we choose $\theta = \frac{\pi}{2}$ and consider $\phi = 0$ for all the phases involved, then

$$BS_H\left(\theta = \frac{\pi}{2}\right) = H = \frac{1}{\sqrt{2}}\begin{pmatrix} 1 & 1 \\ 1 & -1 \end{pmatrix}$$

and a representation is introduced in Figure 16.

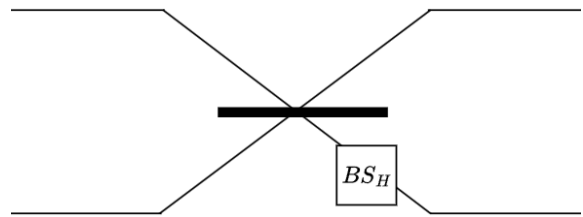

*Figure 16. Realizing the H gate using $BS_H$*

**Validation with Perceval**

We use the same code as for the $Swap$ and $X$ with the goal of defining the $H$ gate on qubit 0. So

$$|\psi\rangle = (H \otimes Id)|00\rangle$$

$$|\psi\rangle = \frac{1}{\sqrt{2}}(|0\rangle + |1\rangle) \otimes |0\rangle$$

$$|\psi\rangle = \frac{1}{\sqrt{2}}(|00\rangle + |10\rangle)$$

Only **step 3** of the code changes and includes a call to the $H-gate$ of the $BS$ class for Beam Splitter.

```
# 3.definition of H-gate
circ_bs_H=BS.H()
circuit = pcvl.Circuit(4)
circuit.add(0,circ_bs_H)
```

The execution of the circuit with Perceval produces the result shown in Figure 17 which matches the expected outcome $|\psi\rangle = \frac{1}{\sqrt{2}}(|00\rangle + |10\rangle)$.

```
Output state amplitudes: (post-selected on qubit states, not renormalized)
|x1,x2>
|1,0,1,0>  ->  |0,0> (0.7071067811865476+0j)
|1,0,0,1>  ->  |0,1> 0j
|0,1,1,0>  ->  |1,0> (0.7071067811865475+0j)
|0,1,0,1>  ->  |1,1> 0j
```

*Figure 17. Result of the Perceval implementation of $(H \otimes Id)|00\rangle$*

In the Perceval library, the graphical symbol of the Beam Splitter $BS_H$ on modes 0 and 1, which implements the $H - gate$ is shown in Figure 18.

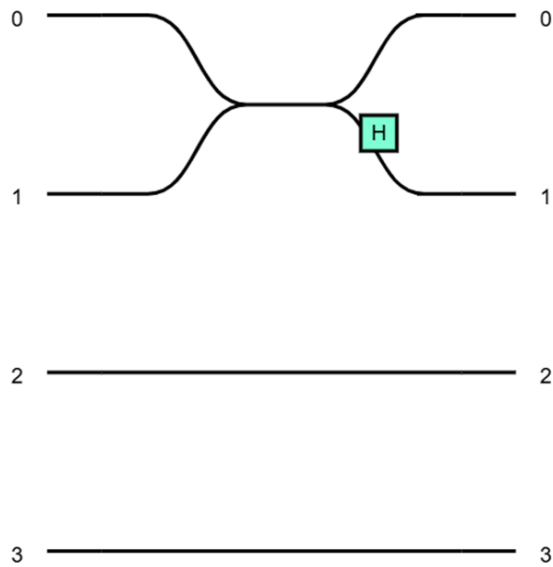

*Figure 18. Representation of the $H \otimes Id$*

### 4.6. Representation of the Pauli Z Gate

To implement $Z_0$ ($Z$ on the first qubit), one must apply a phase shift of $\pi$ on mode 1:

$$PS_1(\pi)|0\rangle = PS_1(\pi)(|01\rangle_{01}) = |01\rangle_{01} = |0\rangle$$
$$PS_1(\pi)|1\rangle = PS_1(\pi)(|01\rangle_{01}) = -|01\rangle_{01} = -|1\rangle$$

Again, in the code, only step 3 changes, which now contains a call to the method $PS$ for the phase shifter.

```
# 3. Definition of the gate Z_1
circuit = pcvl.Circuit(4)
circuit_Z = PS(np.pi)
circuit.add(1,circuit_Z)
```

Note that the phase modification is applied to mode 1 of qubit 1 (`circuit.add(1,circuit_Z)`). The experiments are performed in the context of 2 qubits, i.e., 4 modes as in the previous examples and Figure 19 shows the corresponding circuit.

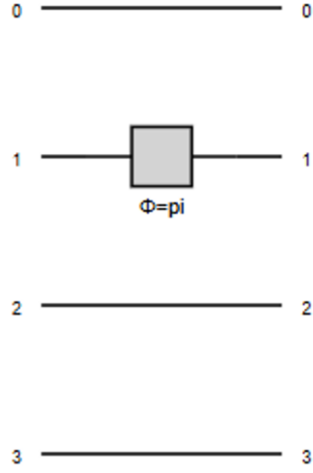

*Figure 19.* Application of a *PS component representing* $Z_0 = (Z \otimes \text{Id})$

*using a Phase Shifter on mode 1 of qubit 0*

We test the circuit in Figure 20 using $Z$ applied to $|11\rangle$ and we get the expected result $(-|1\rangle \otimes |1\rangle)$

```
Output state amplitudes: (post-selected on qubit states, not renormalized)
|x1,x2>
|1,0,1,0>  ->  |0,0> 0j
|1,0,0,1>  ->  |0,1> 0j
|0,1,1,0>  ->  |1,0> 0j
|0,1,0,1>  ->  |1,1> (-1+1.2246467991473532e-16j)
```

*Figure 20. Results for* $|11\rangle = |0101\rangle_{0123}$ *as input*

### 4.7. Implementation of the Y Gate

**Reminder**

$$Y|0\rangle = Y\begin{pmatrix}1\\0\end{pmatrix} = \begin{pmatrix}0 & -i\\i & 0\end{pmatrix}\begin{pmatrix}1\\0\end{pmatrix} = \begin{pmatrix}0\\i\end{pmatrix} = i.|1\rangle$$

$$Y|1\rangle = Y\begin{pmatrix}0\\1\end{pmatrix} = \begin{pmatrix}0 & -i\\i & 0\end{pmatrix}\begin{pmatrix}0\\1\end{pmatrix} = \begin{pmatrix}-i\\0\end{pmatrix} = -i.|0\rangle$$

**Property**

$$Y = PS_1\left(\frac{\pi}{2}\right) PS_0\left(-\frac{\pi}{2}\right) Perm_{10}$$

**Proof.**

Let us consider $|\psi\rangle = \begin{pmatrix}\alpha\\\beta\end{pmatrix}$

We have:

$$PS_1\left(\frac{\pi}{2}\right).PS_0\left(-\frac{\pi}{2}\right).Perm_{10}(\alpha|0\rangle + \beta|1\rangle)$$

$$= PS_1\left(\frac{\pi}{2}\right).PS_0\left(-\frac{\pi}{2}\right).Perm_{10}(\alpha|10\rangle_{01} + \beta|01\rangle_{01})$$

$$= PS_1\left(\frac{\pi}{2}\right).PS_0\left(-\frac{\pi}{2}\right).(\alpha|01\rangle_{01} + \beta|10\rangle_{01})$$

$$= PS_1\left(\frac{\pi}{2}\right).PS_0\left(-\frac{\pi}{2}\right).(\alpha|1\rangle + \beta|0\rangle)$$

$$= PS_1\left(\frac{\pi}{2}\right).PS_0\left(-\frac{\pi}{2}\right)\binom{\beta}{\alpha}$$

$$= PS_1\left(\frac{\pi}{2}\right).\binom{\beta\, e^{-i.t.\frac{\pi}{2}}}{\alpha}$$

$$= \binom{-i\beta}{i\alpha}$$

And by consequence

$$PS_1\left(\frac{\pi}{2}\right).PS_0\left(-\frac{\pi}{2}\right).Perm_{10}\binom{\alpha}{\beta} = \binom{-i\beta}{i\alpha}$$

And

$$Y\binom{\alpha}{\beta} = \begin{pmatrix} 0 & -i \\ i & 0 \end{pmatrix}\binom{\alpha}{\beta} = \binom{-i\beta}{i\alpha}$$

So we conclude that

$$PS_1\left(\frac{\pi}{2}\right).PS_0\left(-\frac{\pi}{2}\right).Perm_{10} = Y$$

□

To implement the gate $Y_0 = (Y \otimes Id)$ modifying the code at step 3 is enough:

$$Y_0 = PS_1\left(\frac{\pi}{2}\right).PS_0\left(-\frac{\pi}{2}\right).Perm_{10}$$

this corresponds to the following code:
```
# 3.definition of Y on qubit 0
circuit = pcvl.Circuit(4)
circuit.add(0,PERM([1,0, 2,3]))
circuit.add(0,PS(-np.pi/2))
circuit.add(1,PS(np.pi/2))
```

This code corresponds to the photonic circuit shown in Figure 21.

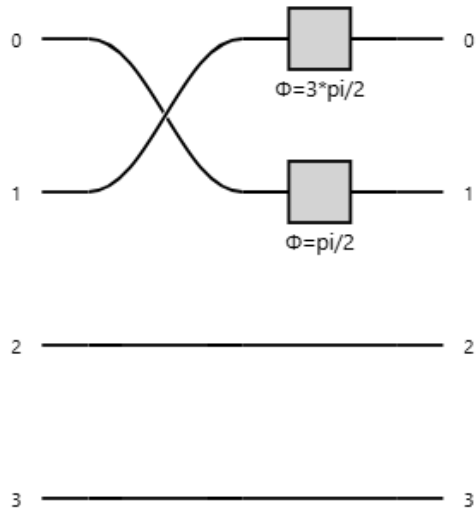

Figure 21. *Photonic circuit for the Y gate on the first qubit.*

You can test the code starting from the state: `qubit_istate = [1,0]`

So

$$Y_0(|1\rangle \otimes |0\rangle) = (Y|1\rangle \otimes |0\rangle) = -i.|0\rangle \otimes |0\rangle$$

We test the circuit in Figure 22 using Y applied to $|10\rangle$ and we get the expected result $(-i|0\rangle \otimes |0\rangle)$

```
Input qubit state: |1,0>
Output state amplitudes: (post-selected on qubit states, not renormalized)
|x1,x2>
|1,0,1,0> -> |0,0> (-1.8369701987210297e-16-1j)
|1,0,0,1> -> |0,1> 0j
|0,1,1,0> -> |1,0> 0j
|0,1,0,1> -> |1,1> 0j
```

*Figure 22.* Application of the components representing $Y$ on qubit 0 and the result for the input state $|10\rangle$

### 4.8. Implementation of the $RX(\theta)$ Gate

Recall that the definition of $RX(\theta)$ is:

$$RX(\theta) = e^{-\frac{iX\theta}{2}} = \begin{pmatrix} \cos\frac{\theta}{2} & -i\sin\frac{\theta}{2} \\ -i\sin\frac{\theta}{2} & \cos\frac{\theta}{2} \end{pmatrix}$$

The Perceval computing library natively provides a gate named $BS_{Rx}$ as the default convention for the Beam Splitter:
`BS.Rx(theta=pcvl.P("theta"))`

However, the definition of the gate provided by the library is

$$BS_{Rx} = \begin{pmatrix} \cos\left(\frac{\theta}{2}\right) & i\sin\left(\frac{\theta}{2}\right) \\ i\sin\left(\frac{\theta}{2}\right) & \cos\left(\frac{\theta}{2}\right) \end{pmatrix}$$

and does not correspond directly to the usual $RX - gate$.

**Property**

$$RX(\theta) = PS_0(\pi).BS_{Rx}(\theta).PS_0(\pi)$$

**Proof.**

We have:

$$RX(\theta)\begin{pmatrix}\alpha\\\beta\end{pmatrix} = \begin{pmatrix} \cos\frac{\theta}{2} & -i\sin\frac{\theta}{2} \\ -i\sin\frac{\theta}{2} & \cos\frac{\theta}{2} \end{pmatrix}\begin{pmatrix}\alpha\\\beta\end{pmatrix}$$

$$RX(\theta)\begin{pmatrix}\alpha\\\beta\end{pmatrix} = \begin{pmatrix} \alpha\cos\frac{\theta}{2} - i\beta\sin\frac{\theta}{2} \\ -i\alpha\sin\frac{\theta}{2} + \beta\cos\frac{\theta}{2} \end{pmatrix}$$

And we have:

$$PS_0(\pi)BS_{Rx}(\theta)PS_0(\pi)\begin{pmatrix}\alpha\\\beta\end{pmatrix} = PS_0(\pi)BS_{Rx}(\theta)\begin{pmatrix}-\alpha\\\beta\end{pmatrix}$$

Because

$$BS_{Rx}(\theta) = \begin{pmatrix} \cos\frac{\theta}{2} & i\sin\frac{\theta}{2} \\ i\sin\frac{\theta}{2} & \cos\frac{\theta}{2} \end{pmatrix}$$

We have:

$$PS_0(\pi).BS_{Rx}(\theta).PS_0(\pi)\begin{pmatrix}\alpha\\\beta\end{pmatrix} = PS_0(\pi).\begin{pmatrix} -\alpha\cos\frac{\theta}{2} + i\beta\sin\frac{\theta}{2} \\ -i\alpha\sin\frac{\theta}{2} + \beta\cos\frac{\theta}{2} \end{pmatrix}$$

$$PS_0(\pi)BS_{Rx}(\theta)PS_0(\pi)\begin{pmatrix}\alpha\\\beta\end{pmatrix} = \begin{pmatrix} \alpha\cos\frac{\theta}{2} - i\beta\sin\frac{\theta}{2} \\ -i\alpha\sin\frac{\theta}{2} + \beta\cos\frac{\theta}{2} \end{pmatrix}$$

□

A photonic circuit representation is shown in Figure 23.

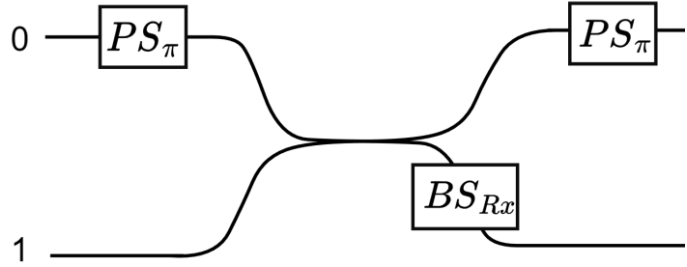

*Figure 23. Quantum circuit for the* RX(θ)

Since the goal is to test the circuit with Perceval, let us choose, for example, $\theta = \frac{\pi}{2}$. So we have:

$$PS_0(\pi)BS_{Rx}\left(\frac{\pi}{2}\right)PS_0(\pi)\begin{pmatrix}\alpha\\\beta\end{pmatrix} = \begin{pmatrix} \alpha\frac{\sqrt{2}}{2} - i\beta\frac{\sqrt{2}}{2} \\ -i\alpha\frac{\sqrt{2}}{2} + \beta\frac{\sqrt{2}}{2} \end{pmatrix}$$

in particular,

$$PS_0(\pi)BS_{Rx}\left(\frac{\pi}{2}\right)PS_0(\pi)\begin{pmatrix}1\\0\end{pmatrix} = \begin{pmatrix} \frac{\sqrt{2}}{2} \\ -i\frac{\sqrt{2}}{2} \end{pmatrix}$$

$$PS_0(\pi)BS_{Rx}\left(\frac{\pi}{2}\right)PS_0(\pi)\begin{pmatrix}0\\1\end{pmatrix} = \begin{pmatrix}-i\frac{\sqrt{2}}{2}\\ \frac{\sqrt{2}}{2}\end{pmatrix}$$

The code to apply the $RX\left(\frac{\pi}{2}\right)$ is given below and once again, only step 3 changes in the code. And for testing purposes, an angle of $\theta = \frac{\pi}{2}$ has been chosen.

```
# 2. Definition of RX(pi/2)
circuit = pcvl.Circuit(4)
circuit.add(0,PS(np.pi))
circuit.add(0,BS.Rx(np.pi/2))
circuit.add(0,PS(np.pi))
```

Note that the parameter of the method `BS.Rx()` is indeed $\frac{\pi}{2}$ and the resulting circuit thus corresponds to $RX\left(\frac{\pi}{2}\right)$. The code can be tested starting from the state: `qubit_istate = [1,0]`

**Expected results** $RX_0\left(\frac{\pi}{2}\right).|10\rangle = -i.\frac{1}{\sqrt{2}}|00\rangle + \frac{1}{\sqrt{2}}|10\rangle$

By definition

$$RX\left(\frac{\pi}{2}\right) = \begin{pmatrix} \cos\frac{\pi}{4} & -i.\sin\frac{\pi}{4} \\ -i.\sin\frac{\pi}{4} & \cos\frac{\pi}{4} \end{pmatrix} = \begin{pmatrix} \frac{1}{\sqrt{2}} & -i\frac{1}{\sqrt{2}} \\ -i\frac{1}{\sqrt{2}} & \frac{1}{\sqrt{2}} \end{pmatrix}$$

So

$$RX_0\left(\frac{\pi}{2}\right)|10\rangle = RX\left(\frac{\pi}{2}\right)|1\rangle \otimes |0\rangle$$

$$RX_0\left(\frac{\pi}{2}\right)|10\rangle = RX\left(\frac{\pi}{2}\right)\begin{pmatrix}0\\1\end{pmatrix} \otimes |0\rangle$$

$$RX_0\left(\frac{\pi}{2}\right)|10\rangle = \begin{pmatrix} \frac{1}{\sqrt{2}} & -i.\frac{1}{\sqrt{2}} \\ -i.\frac{1}{\sqrt{2}} & \frac{1}{\sqrt{2}} \end{pmatrix}\begin{pmatrix}0\\1\end{pmatrix} \otimes |0\rangle$$

$$RX_0\left(\frac{\pi}{2}\right)|10\rangle = \begin{pmatrix}-i\frac{1}{\sqrt{2}}\\ \frac{1}{\sqrt{2}}\end{pmatrix} \otimes \begin{pmatrix}1\\0\end{pmatrix}$$

$$RX_0\left(\frac{\pi}{2}\right)|10\rangle = \begin{pmatrix}-i\frac{1}{\sqrt{2}}\\0\\ \frac{1}{\sqrt{2}}\\0\end{pmatrix}$$

$$RX_0\left(\frac{\pi}{2}\right)|10\rangle = -i\frac{1}{\sqrt{2}}|00\rangle + \frac{1}{\sqrt{2}}|10\rangle$$

The execution result is shown in Figure 24 and it matches the expected outcome, namely $-i\frac{1}{\sqrt{2}}|00\rangle + \frac{1}{\sqrt{2}}|10\rangle$.

```
Input qubit state: |1,0>
Output state amplitudes: (post-selected on qubit states, not renormalized)
|x1,x2>
|1,0,1,0>  ->  |0,0> (-8.659560562354932e-17-0.7071067811865475j)
|1,0,0,1>  ->  |0,1> 0j
|0,1,1,0>  ->  |1,0> (0.7071067811865476+0j)
|0,1,0,1>  ->  |1,1> 0j
```

*Figure 24. Test of the gate $RX_0(\theta)$ with $\theta = \frac{\pi}{2}$*

### 4.9. Representation of the $RY(\theta)$ Gate

Let us note that:

$$RY(\theta) = \begin{pmatrix} \cos\frac{\theta}{2} & -\sin\frac{\theta}{2} \\ \sin\frac{\theta}{2} & \cos\frac{\theta}{2} \end{pmatrix}$$

The Perceval computation library natively provides a $BS_{Ry}$ gate as a method of the BS class:
`BS.Ry(theta=pcvl.P("theta"))`

The definition of the gate provided by the library is exactly that of the standard $RY$ gate. The photonic representation of the gate is shown in Figure 25.

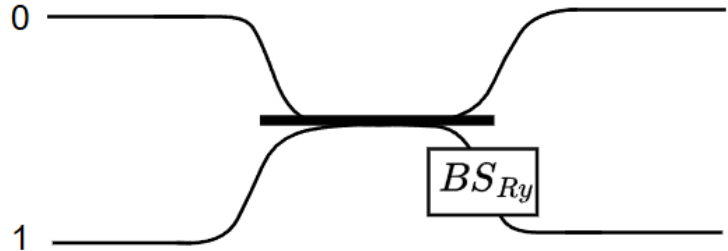

*Figure 25. Representation of the photonic circuit associated with RY*

**Expected result for** $RY_0\left(\frac{\pi}{2}\right)|10\rangle$

By definition

$$RY\left(\frac{\pi}{2}\right) = \begin{pmatrix} \cos\frac{\pi}{4} & -\sin\frac{\pi}{4} \\ \sin\frac{\pi}{4} & \cos\frac{\pi}{4} \end{pmatrix} = \begin{pmatrix} \frac{1}{\sqrt{2}} & -\frac{1}{\sqrt{2}} \\ \frac{1}{\sqrt{2}} & \frac{1}{\sqrt{2}} \end{pmatrix}$$

so

$$RY_0\left(\frac{\pi}{2}\right)|10\rangle = RY\left(\frac{\pi}{2}\right).|1\rangle \otimes |0\rangle$$

$$RY_0\left(\frac{\pi}{2}\right)|10\rangle = \begin{pmatrix} \frac{1}{\sqrt{2}} & -\frac{1}{\sqrt{2}} \\ \frac{1}{\sqrt{2}} & \frac{1}{\sqrt{2}} \end{pmatrix} \begin{pmatrix} 0 \\ 1 \end{pmatrix} \otimes |0\rangle$$

$$RY_0\left(\frac{\pi}{2}\right)|10\rangle = \begin{pmatrix} -\frac{1}{\sqrt{2}} \\ \frac{1}{\sqrt{2}} \end{pmatrix} \otimes \begin{pmatrix} 1 \\ 0 \end{pmatrix}$$

$$RY_0\left(\frac{\pi}{2}\right)|10\rangle = \begin{pmatrix} -\frac{1}{\sqrt{2}} \\ 0 \\ \frac{1}{\sqrt{2}} \\ 0 \end{pmatrix} = -\frac{1}{\sqrt{2}}|00\rangle + \frac{1}{\sqrt{2}}|10\rangle$$

In the code, only step 3 changes again, and for the purposes of the test, an angle of $\frac{\pi}{2}$ has been chosen.

```
# 3. RY(pi/2) gate definition
circuit = pcvl.Circuit(4)
circuit.add(0,BS.Ry(np.pi/2))
```

We can test the code starting from the state: `qubit_istate = [1,0]` and as shown in Figure 26 he results are consistent with the expected ones, namely $-\frac{1}{\sqrt{2}}|00\rangle + \frac{1}{\sqrt{2}}|10\rangle$.

```
Input qubit state: |1,0>
Output state amplitudes: (post-selected on qubit states, not renormalized)
|x1,x2>
|1,0,1,0>  ->  |0,0> (-0.7071067811865475+0j)
|1,0,0,1>  ->  |0,1> 0j
|0,1,1,0>  ->  |1,0> (0.7071067811865476+0j)
|0,1,0,1>  ->  |1,1> 0j
```

*Figure 26. Test of the gate $RY(\theta)$ with $\theta = \frac{\pi}{2}$*

### 4.10. Implementation of the $RZ(\theta)$ gate

Recall that

$$RZ(\theta) = \begin{pmatrix} e^{-i\frac{\theta}{2}} & 0 \\ 0 & e^{i\frac{\theta}{2}} \end{pmatrix}$$

**Property**

$$RZ(\theta) = BS_H.PS_0(\pi).BS_{Rx}(\theta).PS_0(\pi).BS_H$$

**Proof.**

Let us consider $|\psi\rangle = \begin{pmatrix} \alpha \\ \beta \end{pmatrix}$. We have:

$$RZ(\theta)\begin{pmatrix}\alpha\\\beta\end{pmatrix}=\begin{pmatrix}e^{-i\frac{\theta}{2}}&0\\0&e^{i\frac{\theta}{2}}\end{pmatrix}\begin{pmatrix}\alpha\\\beta\end{pmatrix}=\begin{pmatrix}\alpha e^{-i\frac{\theta}{2}}\\\beta e^{i\frac{\theta}{2}}\end{pmatrix}$$

and

$$BS_H.PS_0(\pi).BS_{Rx}(\theta).PS_0(\pi).BS_H.\begin{pmatrix}\alpha\\\beta\end{pmatrix}=BS_H.PS_0(\pi).BS_{Rx}(\theta).PS_0(\pi)\frac{1}{\sqrt{2}}\begin{pmatrix}1&1\\1&-1\end{pmatrix}\begin{pmatrix}\alpha\\\beta\end{pmatrix}$$

$$=BS_H.PS_0(\pi).BS_{Rx}(\theta).PS_0(\pi)\frac{1}{\sqrt{2}}\begin{pmatrix}\alpha+\beta\\\alpha-\beta\end{pmatrix}$$

Because

$$PS_0(\pi)|0\rangle=PS_0(\pi)|10\rangle_{01}=\ e^{i\pi}|10\rangle_{01}=-|0\rangle$$
$$PS_0(\pi)|1\rangle=PS_0(\pi)|01\rangle_{01}=|01\rangle_{01}=|1\rangle$$

We have

$$BS_H.PS_0(\pi).BS_{Rx}(\theta).PS_0(\pi).BS_H.\begin{pmatrix}\alpha\\\beta\end{pmatrix}=BS_H.PS_0(\pi).BS_{Rx}(\theta).\frac{1}{\sqrt{2}}\begin{pmatrix}-(\alpha+\beta)\\\alpha-\beta\end{pmatrix}$$

Because

$$BS_{Rx}(\theta)=\begin{pmatrix}\cos\frac{\theta}{2}&i.\sin\frac{\theta}{2}\\i\sin\frac{\theta}{2}&\cos\frac{\theta}{2}\end{pmatrix}$$

We have

$$BS_H.PS_0(\pi).BS_{Rx}(\theta).PS_0(\pi).BS_H.\begin{pmatrix}\alpha\\\beta\end{pmatrix}=BS_H.PS_0(\pi).\begin{pmatrix}\cos\frac{\theta}{2}&i\sin\frac{\theta}{2}\\i.\sin\frac{\theta}{2}&\cos\frac{\theta}{2}\end{pmatrix}.\frac{1}{\sqrt{2}}\begin{pmatrix}-(\alpha+\beta)\\\alpha-\beta\end{pmatrix}$$

$$BS_H.PS_0(\pi).BS_{Rx}(\theta).PS_0(\pi).BS_H.\begin{pmatrix}\alpha\\\beta\end{pmatrix}=BS_H.PS_0(\pi).\frac{1}{\sqrt{2}}\begin{pmatrix}-(\alpha+\beta)\cos\frac{\theta}{2}+i(\alpha-\beta)\sin\frac{\theta}{2}\\-i.(\alpha+\beta)\sin\frac{\theta}{2}+(\alpha-\beta)\cos\frac{\theta}{2}\end{pmatrix}$$

and

$$BS_H.PS_0(\pi).BS_{Rx}(\theta).PS_0(\pi).BS_H.\begin{pmatrix}\alpha\\\beta\end{pmatrix}=BS_H.\frac{1}{\sqrt{2}}\begin{pmatrix}(\alpha+\beta)\cos\frac{\theta}{2}-i(\alpha-\beta)\sin\frac{\theta}{2}\\-i.(\alpha+\beta)\sin\frac{\theta}{2}+(\alpha-\beta)\cos\frac{\theta}{2}\end{pmatrix}$$

and

$$BS_H.PS_0(\pi).BS_{Rx}(\theta).PS_0(\pi).BS_H.\begin{pmatrix}\alpha\\\beta\end{pmatrix}=\frac{1}{\sqrt{2}}\begin{pmatrix}1&1\\1&-1\end{pmatrix}.\frac{1}{\sqrt{2}}\begin{pmatrix}(\alpha+\beta)\cos\frac{\theta}{2}-i(\alpha-\beta)\sin\frac{\theta}{2}\\-i.(\alpha+\beta)\sin\frac{\theta}{2}+(\alpha-\beta)\cos\frac{\theta}{2}\end{pmatrix}$$

$$=\frac{1}{2}\begin{pmatrix}(\alpha+\beta)\cos\frac{\theta}{2}-i(\alpha-\beta)\sin\frac{\theta}{2}-i.(\alpha+\beta)\sin\frac{\theta}{2}+(\alpha-\beta)\cos\frac{\theta}{2}\\(\alpha+\beta)\cos\frac{\theta}{2}-i(\alpha-\beta)\sin\frac{\theta}{2}+i.(\alpha+\beta)\sin\frac{\theta}{2}-(\alpha-\beta)\cos\frac{\theta}{2}\end{pmatrix}$$

$$= \frac{1}{2} \begin{pmatrix} 2.\alpha \cos\frac{\theta}{2} - 2.i.\alpha \sin\frac{\theta}{2} \\ 2.\beta \cos\frac{\theta}{2} + 2.\beta.i.\sin\frac{\theta}{2} \end{pmatrix}$$

$$= \begin{pmatrix} \alpha \cos\frac{\theta}{2} - i.\alpha \sin\frac{\theta}{2} \\ \beta \cos\frac{\theta}{2} + \beta.i.\sin\frac{\theta}{2} \end{pmatrix}$$

So

$$BS_H.PS_0(\pi).BS_{Rx}(\theta).PS_0(\pi).BS_H. = \begin{pmatrix} e^{-i.\frac{\theta}{2}}\alpha \\ e^{i.\frac{\theta}{2}}\beta \end{pmatrix} = RZ(\theta)\begin{pmatrix} \alpha \\ \beta \end{pmatrix} = \begin{pmatrix} \alpha e^{-i.\frac{\theta}{2}} \\ \beta e^{i.\frac{\theta}{2}} \end{pmatrix}$$

☐

The photonic representation of the gate is shown in Figure 27.

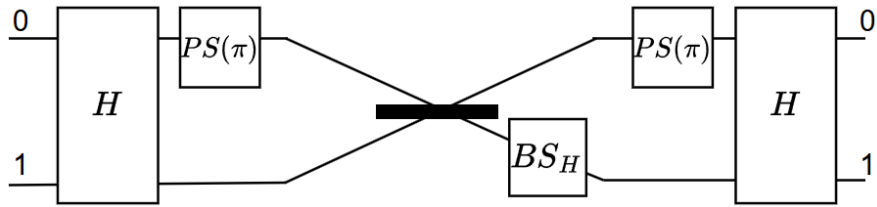

*Figure 27. Representation of the circuit associated with Rz*

In the code, only step 3 changes, and for the purposes of the test, an angle of $\frac{\pi}{2}$ has been choosen.

```
# 3. Definition of RZ(pi/2)
circuit = pcvl.Circuit(4)
circuit.add(0,BS.H())
circuit.add(0,PS(np.pi))
circuit.add(0,BS.Rx(np.pi/2))
circuit.add(0,PS(np.pi))
circuit.add(0,BS.H())
```

**Expected results for $RZ_1\left(\frac{\pi}{2}\right).|10\rangle$**

By definition

$$RZ\left(\frac{\pi}{2}\right) = \begin{pmatrix} e^{-i.\frac{\pi}{4}} & 0 \\ 0 & e^{i.\frac{\pi}{4}} \end{pmatrix} = \begin{pmatrix} \cos\frac{\pi}{4} - i.\sin\frac{\pi}{4} & 0 \\ 0 & \cos\frac{\pi}{4} + i.\sin\frac{\pi}{4} \end{pmatrix}$$

$$= \begin{pmatrix} \frac{1}{\sqrt{2}} - i.\frac{1}{\sqrt{2}} & 0 \\ 0 & \frac{1}{\sqrt{2}} + i.\frac{1}{\sqrt{2}} \end{pmatrix}$$

so

$$RZ_0\left(\frac{\pi}{2}\right).|10\rangle = \begin{pmatrix} \frac{1}{\sqrt{2}} - i.\frac{1}{\sqrt{2}} & 0 \\ 0 & \frac{1}{\sqrt{2}} + i.\frac{1}{\sqrt{2}} \end{pmatrix} \begin{pmatrix} 0 \\ 1 \end{pmatrix} \otimes |0\rangle$$

$$= \begin{pmatrix} 0 \\ \frac{1}{\sqrt{2}} + i.\frac{1}{\sqrt{2}} \end{pmatrix} \otimes |0\rangle$$

$$RZ_0\left(\frac{\pi}{2}\right).|10\rangle = \left(\frac{1}{\sqrt{2}} + i.\frac{1}{\sqrt{2}}\right).|10\rangle$$

We can test the code starting from the state: `qubit_istate = [1,0]`. The results obtained are consistent with the calculations performed (Figure 28).

```
Input qubit state: |1,0>
Output state amplitudes: (post-selected on qubit states, not renormalized)
|x1,x2>
|1,0,1,0>  ->  |0,0> (-2.299347170293092e-17-1.3401577416544657e-16j)
|1,0,0,1>  ->  |0,1> 0j
|0,1,1,0>  ->  |1,0> (0.7071067811865476-0.7071067811865475j)
|0,1,0,1>  ->  |1,1> 0j
```

*Figure 28. Test of the gate $RZ(\theta)$ with $\theta = \frac{\pi}{2}$*

## 5. Controlled gates modelization

### 5.1. Implementation of CX according to (Ralph, 2001)

Consider the definition of CX as defined by (Heurtel et al., 2023) and originally proposed by (Ralph, 2002), with a description provided in Figure 29.

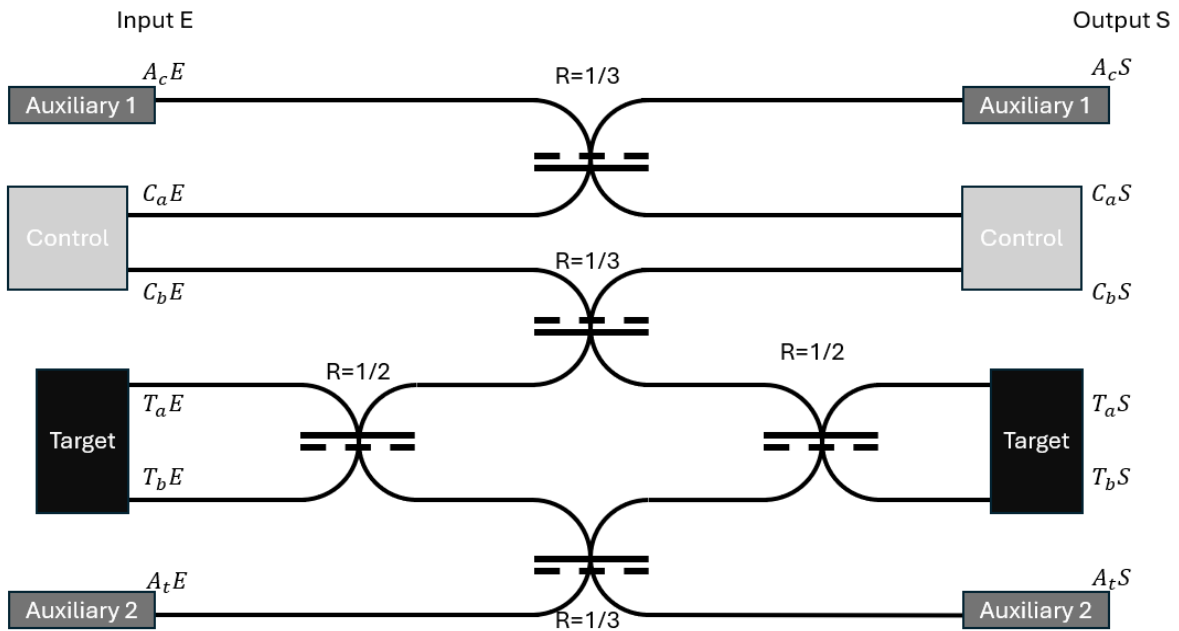

*Figure 29. Circuit for the CNOT from (Ralph 2001)*

The figure presents a method for implementing a CNOT gate. Here, we consider 6 modes representing two qubits and two auxiliary modes. The modes $C_a$ and $C_b$ correspond to the control qubit in the CNOT we wish to realize. Thus $|01\rangle_{C_a C_b}$ represents the Fock state associated with the control qubit in the state $|1\rangle = |01\rangle_{C_a C_b}$. Similarly, we define $T_a$ and $T_b$ as the modes for the "Target" qubit. Additionally, we have two auxiliary modes $A_c$ and $A_t$ which act on the control and target qubits, respectively. Both auxiliary modes are initialized in the states $|0\rangle_{A_c}$ and $|0\rangle_{A_t}$.

To explain the CNOT operation, we need to distinguish between the input and output parts of the CNOT for a given mode. For instance, we distinguish $C_a E$, the first input mode for the control qubit, from $C_a S$, the first output mode for the control qubit. Making this distinction allows us to express the output modes as functions of the input modes.

Before diving into the explanation of the CNOT operation, let us highlight two important elements:

- Each Beam Splitter has a dashed side, which is the side that will receive a phase shift (the side that will receive the minus sign in the chosen convention).
- We consider two types of Beam Splitters: some with a reflectivity of $R = \frac{1}{2}$ (symmetric Beam Splitters) and others with a reflectivity of $R = \frac{1}{3}$ (asymmetric Beam Splitters)

**Notes**

It is important to recall that for a Beam Splitter with reflectivity $\alpha$, and whose graphical representation corresponds to the dashed side facing upward, two output states are defined:
- The state $C$ undergoes a phase shift of value -1 since it is on the dashed side, and the fraction of $A$ that is reflected by the Beam Splitter is $-\sqrt{\alpha}$;
- The fraction of $B$ that "passes through" the Beam Splitter is $\sqrt{1-\alpha}$.

This leads to the output $C = -\sqrt{\alpha}.A + \sqrt{1-\alpha}.B$ with the phase of $B$ unchanged since $B$ corresponds to the side of the Beam Splitter without the dashed lines (Figure 30).

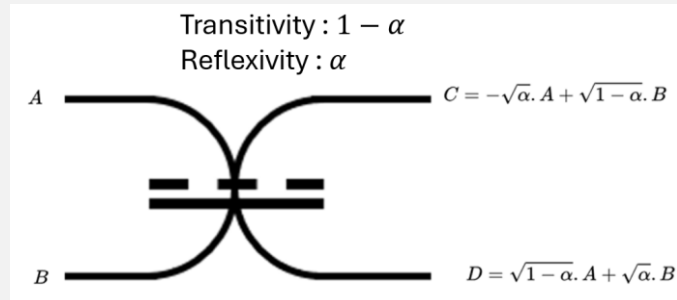

*Figure 30. Output Amplitudes of a Beam Splitter with Reflectivity α*

By observing Figure 29 and using the rules as in Figure 30 we can establish the following relations:

$$A_c S = \frac{1}{\sqrt{3}}(-A_c E + \sqrt{2} C_a E)$$

$$C_a S = \frac{1}{\sqrt{3}}(C_a E + \sqrt{2} A_c E)$$

$$C_b S = -\frac{1}{\sqrt{3}} C_b E + \frac{1}{\sqrt{2}} \times \left( \frac{\sqrt{2}}{\sqrt{3}} T_a E + \frac{\sqrt{2}}{\sqrt{3}} T_b E \right) = \frac{1}{\sqrt{3}} (T_a E + T_b E - C_b E)$$

$$T_a S = \frac{\sqrt{2}}{\sqrt{3}} \times \frac{1}{\sqrt{2}} C_b E + \left( \underbrace{\frac{1}{\sqrt{2}} \times \frac{1}{\sqrt{3}} \times \frac{1}{\sqrt{2}}}_{first\ path} + \underbrace{\frac{1}{\sqrt{2}} \times \frac{1}{\sqrt{3}} \times \frac{1}{\sqrt{2}}}_{second\ path} \right) T_a E$$

$$+ \left( \underbrace{\frac{1}{\sqrt{2}} \times \frac{1}{\sqrt{3}} \times \frac{1}{\sqrt{2}}}_{first\ path} + \underbrace{\left(-\frac{1}{\sqrt{2}}\right) \times \frac{1}{\sqrt{3}} \times \frac{1}{\sqrt{2}}}_{second\ path} \right) T_b E + \frac{\sqrt{2}}{\sqrt{3}} \times \frac{1}{\sqrt{2}} A_t E$$

$$T_a S = \frac{1}{\sqrt{3}} (C_b E + T_a E + A_t E)$$

$$T_b S = \frac{1}{\sqrt{3}} (C_b E + T_b E - A_t E)$$

$$A_t S = \frac{1}{\sqrt{3}} (T_a E + T_b E - A_t E)$$

To each of these modes, we can associate a creation operator. It corresponds to the emission of a photon.

**Remark**

By choosing a different reflectivity $R$, we would obtain different values for each relation.

Example with $R = \frac{1}{4}$ :

$$T_a S = \frac{1}{2\sqrt{2}} C_b E + \frac{\sqrt{3}}{2} T_a E + \frac{1}{2\sqrt{2}} A_t E$$

**Reminder:**

The creation and annihilation operators associated with a mode $m$ are denoted $\hat{a}_m^\dagger$ and $\hat{a}_m$, respectively. Their actions are as follows:

- The creation operator $\hat{a}_m^\dagger$ allows the transition from $|0\rangle_m$ to $|1\rangle_m$ in spatial mode 0. For example, $\hat{a}_0^\dagger$ transforms the vacuum state $|00\rangle_{01}$ into $|10\rangle_{01}$.
- The annihilation operator $\hat{a}_m$ allows the transition from $|1\rangle_m$ to $|0\rangle_m$. For example, $\hat{a}_0$ transforms $|10\rangle_{01}$ into the vacuum state $|00\rangle_{01}$.

**Note 2**

Strictly Speaking, the creation operator on an output mode $m$ should be denoted as $a_m^\dagger$

For convenience and simplicity, we will use the shorthand notation $(i)^\dagger$ for the creation operator on the output mode $i$.

For example:

$$(A_c S)^\dagger \equiv a_{A_c S}^\dagger$$

The creation operator on the input modes can be deduced from the previous relations. For example, we have:

$$\underbrace{(A_c S)^\dagger}_{\substack{Creation\ operator \\ for\ the\ output\ A_c S}} = \frac{1}{\sqrt{3}}(-(A_c E)^\dagger + \sqrt{2}(C_a E)^\dagger)$$

The final state of a two-qubit system (4 modes), to which we add the two auxiliary modes that we reorganize for clarity, can be written as:

$$|\psi_{out}\rangle = \underbrace{\alpha|00\rangle + \beta|01\rangle + \gamma|10\rangle + \delta|11\rangle}_{qubits} \underbrace{|00\rangle_{05}}_{Fock\ state\ of\ auxiliary\ modes}$$

This state can be described in terms of Fock states and creation operators by considering only the 4 input qubits. So:

$$|\psi_{out}\rangle = (\alpha|1010\rangle_{1234} + \beta|1001\rangle_{1234} + \gamma|0110\rangle_{1234} + \delta|0101\rangle_{1234}).|00\rangle_{05}$$

is rewritten:

$$|\psi_{out}\rangle = \Big(\alpha((C_a S)^\dagger (T_a S)^\dagger|0000\rangle_{1234}) + \beta((C_a S)^\dagger (T_b S)^\dagger|0000\rangle_{1234})$$
$$+ \gamma((C_b S)^\dagger (T_a S)^\dagger|0000\rangle_{1234}) + \delta((C_b S)^\dagger (T_b S)^\dagger|0000\rangle_{1234})\Big)|00\rangle_{05}$$

Now, let us consider each term separately.

**For the first term**

$$\alpha((C_a S)^\dagger (T_a S)^\dagger|0000\rangle_{1234})$$

$$= \alpha\left(\left(\frac{1}{\sqrt{3}}((C_a E)^\dagger + \sqrt{2}(A_c E)^\dagger)\right) \times \left(\frac{1}{\sqrt{3}}((C_b E)^\dagger + (T_a E)^\dagger + (A_t E)^\dagger)\right)|0000\rangle_{1234}\right)$$

$$= \frac{\alpha}{3}\Big(((C_a E)^\dagger(C_b E)^\dagger + (C_a E)^\dagger(T_a E)^\dagger + (C_a E)^\dagger (A_t E)^\dagger + \sqrt{2}(A_c E)^\dagger(C_b E)^\dagger$$
$$+ \sqrt{2}(A_c E)^\dagger(T_a E)^\dagger + \sqrt{2}(A_c E)^\dagger(A_t E)^\dagger)|0000\rangle_{1234}\Big)$$

**For the second term**

$$\beta((C_a S)^\dagger (T_b S)^\dagger|0000\rangle_{1234})$$

$$= \beta\left(\left(\frac{1}{\sqrt{3}}((C_a E)^\dagger + \sqrt{2}(A_c E)^\dagger)\frac{1}{\sqrt{3}}((C_b E)^\dagger + (T_b E)^\dagger - (A_t E)^\dagger)\right)|0000\rangle_{1234}\right)$$

$$= \frac{\beta}{3}\Big(((C_a E)^\dagger(C_b E)^\dagger + (C_a E)^\dagger(T_b E)^\dagger - (C_a E)^\dagger(A_t E)^\dagger + \sqrt{2}(A_c E)^\dagger(C_b E)^\dagger$$
$$+ \sqrt{2}(A_c E)^\dagger(T_b E)^\dagger - \sqrt{2}(A_c E)^\dagger(A_t E)^\dagger)|0000\rangle_{1234}\Big)$$

**For the third term**

$$\gamma((C_b S)^\dagger (T_a S)^\dagger|0000\rangle_{1234})$$

$$= \gamma \left( \frac{1}{\sqrt{3}} (T_a E + T_b E - C_b E) \times \frac{1}{\sqrt{3}} (C_b E + T_a E + A_t E) |0000\rangle_F \right)$$

$$= \frac{\gamma}{3} ( ((T_a E)^\dagger (C_b E)^\dagger + (T_a E)^\dagger (T_a E)^\dagger + (T_a E)^\dagger (A_t E)^\dagger + (T_b E)^\dagger (C_b E)^\dagger + (T_b E)^\dagger (T_a E)^\dagger$$
$$+ (T_b E)^\dagger (A_t E)^\dagger - (C_b E)^\dagger (C_b E)^\dagger - (C_b E)^\dagger (T_a E)^\dagger$$
$$- (C_b E)^\dagger (A_t E)^\dagger ) |0000\rangle_{1234} )$$

**For the last term**

$$\delta ((C_b S)^\dagger (T_b S)^\dagger |0000\rangle_{1234})$$

$$= \delta \left( \frac{1}{\sqrt{3}} (T_a E + T_b E - C_b E) \frac{1}{\sqrt{3}} (C_b E + T_b E - A_t E) |0000\rangle_{1234} \right)$$

$$= \frac{\delta}{3} ( ((T_a E)^\dagger (C_b E)^\dagger + (T_a E)^\dagger (T_b E)^\dagger - (T_a E)^\dagger (A_t E)^\dagger + (T_b E)^\dagger (C_b E)^\dagger + (T_b E)^\dagger (T_b E)^\dagger$$
$$- (T_b E)^\dagger (A_t E)^\dagger - (C_b E)^\dagger (C_b E)^\dagger - (C_b E)^\dagger (T_b E)^\dagger$$
$$+ (C_b E)^\dagger (A_t E)^\dagger ) |0000\rangle_{1234})$$

It is necessary to apply a post-selection on the resulting states to retain only the states having:

- Exactly one photon on the control side $(C_a E)^\dagger$ ou $(C_b E)^\dagger$ but not $(C_a E)^\dagger (C_b E)^\dagger$
- Exactly one photon on the target side $(T_a E)^\dagger$ ou $(T_b E)^\dagger$ but not $(T_a E)^\dagger (T_b E)^\dagger$
- No photons on the auxiliary modes: remove all terms involving $(A_c E)^\dagger$ or $(A_t E)^\dagger$

Thus, by grouping the four terms and removing the terms corresponding to the post selection conditions mentioned previously:

$$|\psi_{out}\rangle = [\frac{\alpha}{3} (( \cancel{(C_a E)^\dagger (C_b E)^\dagger} + (C_a E)^\dagger (T_a E)^\dagger + \cancel{(C_a E)^\dagger (A_t E)^\dagger} + \cancel{\sqrt{2}(A_c E)^\dagger (C_b E)^\dagger}$$
$$+ \cancel{\sqrt{2}(A_c E)^\dagger (T_a E)^\dagger} + \cancel{\sqrt{2}(A_c E)^\dagger (A_t E)^\dagger}) |0000\rangle_{1234} )$$

$$+ \frac{\beta}{3} (( \cancel{(C_a E)^\dagger (C_b E)^\dagger} + (C_a E)^\dagger (T_b E)^\dagger - \cancel{(C_a E)^\dagger (A_t E)^\dagger} + \cancel{\sqrt{2}(A_c E)^\dagger (C_b E)^\dagger}$$
$$+ \cancel{\sqrt{2}(A_c E)^\dagger (T_b E)^\dagger} - \cancel{\sqrt{2}(A_c E)^\dagger (A_t E)^\dagger}) |0000\rangle_{1234} )$$

$$+ \frac{\gamma}{3} ( ((T_a E)^\dagger (C_b E)^\dagger + \cancel{(T_a E)^\dagger (T_a E)^\dagger} + \cancel{(T_a E)^\dagger (A_t E)^\dagger} + (T_b E)^\dagger (C_b E)^\dagger + \cancel{(T_b E)^\dagger (T_a E)^\dagger}$$
$$+ \cancel{(T_b E)^\dagger (A_t E)^\dagger} - \cancel{(C_b E)^\dagger (C_b E)^\dagger} - (C_b E)^\dagger (T_a E)^\dagger$$
$$- \cancel{(C_b E)^\dagger (A_t E)^\dagger} ) |0000\rangle_{1234} )$$

$$+ \frac{\delta}{3} ( ((T_a E)^\dagger (C_b E)^\dagger + \cancel{(T_a E)^\dagger (T_b E)^\dagger} - \cancel{(T_a E)^\dagger (A_t E)^\dagger} + (T_b E)^\dagger (C_b E)^\dagger + \cancel{(T_b E)^\dagger (T_b E)^\dagger}$$
$$- \cancel{(T_b E)^\dagger (A_t E)^\dagger} - \cancel{(C_b E)^\dagger (C_b E)^\dagger} - (C_b E)^\dagger (T_b E)^\dagger$$
$$+ \cancel{(C_b E)^\dagger (A_t E)^\dagger} ) |0000\rangle_{1234} ))] |00\rangle_{05}$$

After removing the terms, we obtain:

$$|\psi_{out}\rangle = [\frac{\alpha}{3} ((C_a E)^\dagger (T_a E)^\dagger |0000\rangle_f ) + \frac{\beta}{3} ((C_a E)^\dagger (T_b E)^\dagger |0000\rangle_f )$$
$$+ \frac{\gamma}{3} ( (T_b E)^\dagger (C_b E)^\dagger |0000\rangle_f ) + \frac{\delta}{3} ((T_a E)^\dagger (C_b E)^\dagger |0000\rangle_f )] |00\rangle_{05}$$

$$|\psi_{out}\rangle = \left[\frac{\alpha}{3}(|1010\rangle_f) + \frac{\beta}{3}(|1001\rangle_f) + \frac{\gamma}{3}(|0101\rangle_f) + \frac{\delta}{3}(|0110\rangle_f)\right]|00\rangle_{05}$$

$$|\psi_{out}\rangle = \frac{1}{3}(\alpha|00\rangle + \beta|01\rangle + \gamma|11\rangle + \delta|01\rangle)|00\rangle_{05}$$

We almost recover the expected action of the CNOT gate: $(\alpha|00\rangle + \beta|01\rangle + \gamma|11\rangle + \delta|01\rangle)$ up to a factor of $\frac{1}{3}$.

**Example**

If we started from the state $|\psi\rangle = |10\rangle$ meaning $\gamma = 1$ we obtain:

$$|\psi_{out}\rangle = \frac{1}{3}|11\rangle$$

We do not always find ourselves in this situation whenever we use our CNOT, and the value $\frac{1}{3}$ is the probability amplitude associated with the success of the CNOT.

**Remark**

Ralph's CNOT gate is a "simple" way to implement the CNOT gate on photonic computers using dual-rail encoding. However, the application of this gate succeeds only 1 time out of 9. This limitation is not an error rate that can be corrected with hardware improvements. It is a real fundamental limit intrinsic to the functioning of Ralph's CNOT gate

Figure 31 shows the Perceval implementation as provided by (Heurtel et al., 2023), which corresponds to the one available on the Perceval website.

```
theta_13 = BS.r_to_theta(r=1/3)

cnot = (pcvl.Circuit(6, name = "Ralph CNOT")
        .add((0, 1), BS.H(theta_13, phi_bl = np.pi, phi_tr = np.pi/2, phi_tl = -np.pi/2))
        .add((3, 4), BS.H())
        .add((2, 3), BS.H(theta_13, phi_bl = np.pi, phi_tr = np.pi/2 , phi_tl = -np.pi/2))
        .add((4, 5), BS.H(theta_13))
        .add((3, 4), BS.H()))
```

*Figure 31. Circuit of the CX gate implemented with Perceval (Heurtel et al., 2023)*

Note that the calculation begins by defining $\theta = 2 \times \arccos\left(\sqrt{\frac{1}{3}}\right) \simeq 1.91$ where $1/3$ is the reflectivity (the proportion of photons reflected by the Beam Splitter). Consider the circuit in Figure 29, where we start with two qubits (a CNOT gate acts on two qubits), meaning 4 modes, and we need two auxiliary modes for this definition of the CNOT.

For two qubits in the state $|00\rangle$ as input, from the perspective of Fock states, we have the state:

$$|\psi_0\rangle = |0\rangle_0 \underbrace{(|1\rangle_1|0\rangle_2)}_{control\ qubit\ |0\rangle} \underbrace{(|1\rangle_3|0\rangle_4)}_{target\ qubit\ |0\rangle} |0\rangle_5$$

In this implementation (originally proposed by (Ralph et al., 2001)), the two auxiliary modes are:

- Mode 0 which appears in the first position in Figure 29 ;
- Mode 5 which appears in the last position in Figure 29.

**The code is primarily composed of 4 parts.**

Part 1. CNOT definition
```
cnot = (pcvl.Circuit(6, name = "Ralph CNOT")
        .add((0, 1), BS.H(theta_13, phi_bl = np.pi,
             phi_tr = np.pi/2, phi_tl = -np.pi/2))
        .add((3, 4), BS.H())
        .add((2, 3), BS.H(theta_13, phi_bl = np.pi,
             phi_tr = np.pi/2 , phi_tl = -np.pi/2))
        .add((4, 5), BS.H(theta_13))
        .add((3, 4), BS.H()))
```

Part 2. Circuit definition

The circuit consists of 6 modes, with the first and the last modes corresponding to auxiliary modes.
```
circuit = pcvl.Circuit(6)
```

For readability purposes, the CNOT gate is defined as acting on the first two modes, which implies that the two auxiliary modes must be modes 4 and 5. Therefore, the CNOT call is preceded by permutations to swap the roles of modes 0 and 4.

```
circuit.add(0,PERM([0,1,2,4,3,5]))
circuit.add(0,PERM([0,1,3,2,4,5]))
circuit.add(0,PERM([0,2,1,3,4,5]))
circuit.add(0,PERM([1,0,2,3,4,5]))
# or : circuit.add(0, PERM([4,0,1,2,3,5]))]

circuit.add(0,cnot)

circuit.add(0,PERM([1,0,2,3,4,5]))
circuit.add(0,PERM([0,2,1,3,4,5]))
circuit.add(0,PERM([0,1,3,2,4,5]))
circuit.add(0,PERM([0,1,2,4,3,5]))
# or : circuit.add(0, PERM([1,2,3,4,0,5]))]
```

Part 3. Defining a filter to retain only the states where the CNOT has succeeded

This means two things:

- One have to require that the first four modes correctly represent a qubit, i.e., either mode 0 or mode 1 has a value of 1, and either mode 2 or mode 3 has a value of 1;
- The auxiliary modes, modes 4 and 5, must both have a value of 0.

We thus define a "processor" and apply post-selection to the qubit to verify that the CNOT gate has operated successfully.
```
p = pcvl.Processor("SLOS", circuit)
p.set_postselection(pcvl.PostSelect("[0,1]==1 & [2,3]==1 & [4]==0 & [5]==0"))
```

Part 4. Circuit simulation

It is sufficient to define a `Sampler()` parameterized by the processor $p$, specify the initial state, and then run the simulation..

```
nsample = 50000
from perceval.algorithm import Sampler , Analyzer
sampler = Sampler ( p )
p.with_input ( pcvl.BasicState ([0, 1, 0, 1, 0, 0]))
output = sampler.sample_count (1000)
pcvl.pdisplay ( output['results'], output_format = pcvl.Format.TEXT )
```

**Circuit testing**

The full Python program is provided in Appendix. The first test consists of applying a CX gate to the input state $|11\rangle$, which yields the result shown in Figure 32 where the output state $|10\rangle$ is obtained with 100% probability.

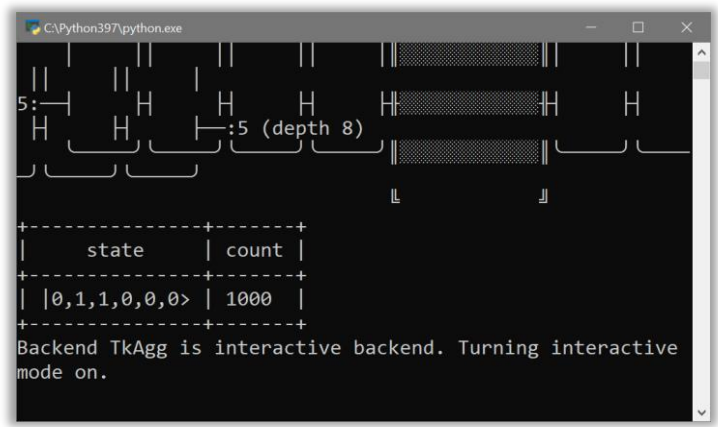

Figure 32. Test of $CX|11\rangle = |10\rangle$

The second test involves applying the CX gate to the input state $|10\rangle$ resulting in the output state $|11\rangle$ as shown in Figure 33.

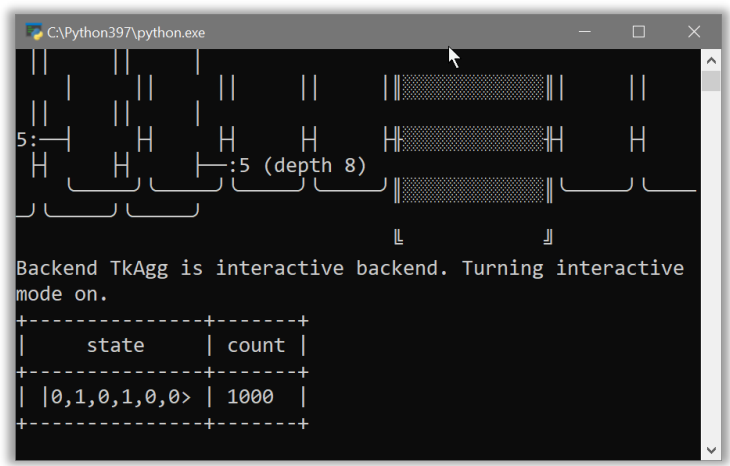

Figure 33. Test of $CX|10\rangle = |11\rangle$

**Conclusion**

The implementation of the $CX$ gate proposed by (Ralph, 2001) is limited to circuits that use only a single CNOT. When a circuit contains two CNOT gates, as shown in Figure 34 and the success conditions depend on modes that are shared with the second CNOT. While the auxiliary modes do not create a problem, the modes shared between the two CNOTs cannot be post-selected in the middle of the circuit. The CNOT gate from (Ralph, 2001) has a success probability of $\frac{1}{9}$.

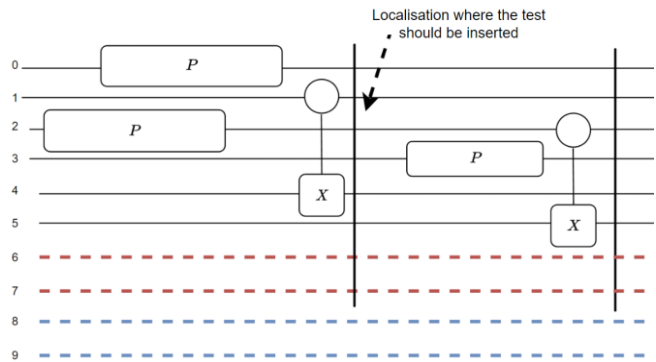

*Figure 34. Incorporating two CNOT gates in a single circuit: infeasibility of implementation using the CNOT scheme by (Ralph, 2001)*

### 5.2. Implementation of a Heralded CX according to (Knill, 2001)

The heralded CNOT gate was proposed by Knill in 2001, and its main advantage lies in the fact that it is sufficient to measure the auxiliary qubits to determine whether the CNOT operation was successful. However, its success probability is only $\frac{2}{27}$. The Perceval Library directly provides the CNOT gate from (Knill, 2001), available in the "catalog" under the Name "heralded cnot".

```
cnot = catalog["heralded cnot"].build_processor()
```

### 5.3. Representation of $CX_{ij}$ using the library (Knill, 2001)

Two control modes are required, which in this case correspond to modes 4 and 5. Therefore, the success conditions for the CNOT operation are that modes 4 and 5 are equal to 1 at the output. Note that unlike the CNOT gate by Ralph (2001), the heralded version requires both auxiliary qubits to be in state 1 at the input.

Thus, the conditions for identifying a successful CNOT operation are:

- Modes 4 and 5 must be equal to 1 **at the output**.

```
p.set_postselection(pcvl.PostSelect("[4]==1 & [5]==1"))
```

The full code of the Herralded Cnot is provided in Appendix and the results introduced in Figure 35 meet the theoretical requirements.

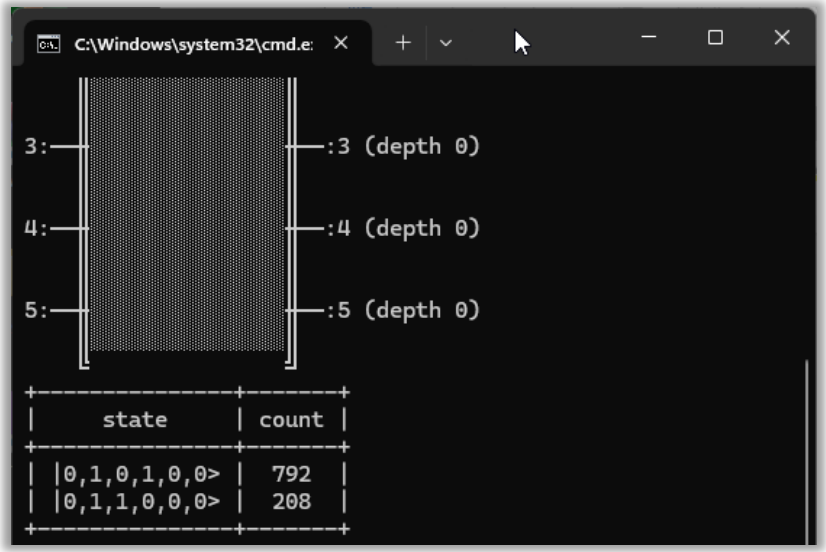

*Figure 35. Test of* $CX|11\rangle = |10\rangle$

## 5.4. Representation of $CX_{ij}$ using the CX of Ralph

The Perceval Library directly provides the CNOT gate from Ralph, available in the "catalog" under the Name " `postprocessed cnot`".
```
cnot = catalog['postprocessed cnot'].build_circuit()
```

The only difference from the previous CNOT lies in the two control modes, which are now located at positions 4 and 5. Therefore, the conditions for identifying a successful CNOT operation are:

- modes 0 and 1 define a qubit;
- modes 2 and 3 define a qubit;
- modes 4 and 5 are equal to 0.

This yields:
```
p.set_postselection(pcvl.PostSelect("[0,1]==1 & [2,3]==1 & [4]==0 & [5]==0"))
```

**Code**

The main modifications compared to the previous case are highlighted (or enclosed).

```python
import perceval as pcvl
import numpy as np
from perceval.components import PS, BS, PERM

import perceval as pcvl
import numpy as np
from perceval.components import BS
from perceval.components import catalog

cnot = catalog['postprocessed cnot'].build_circuit()

# 1. initialization
istate=[0,1,0,1,0,0]
# 2. test of cnot
circuit = pcvl.Circuit(6)
circuit.add(0,cnot)
# 3. Print of the circuit
```

```python
pcvl.pdisplay(circuit.compute_unitary())
print("")
pcvl.pdisplay(circuit)
# 4 Computation
p = pcvl.Processor("SLOS", circuit)
p.set_postselection(pcvl.PostSelect("[0,1]==1 & [2,3]==1 & [4]==0 & [5]==0"))
nsample = 50000
from perceval.algorithm import Sampler , Analyzer

sampler = Sampler ( p )
p.with_input ( pcvl.BasicState (istate)) #Corresponds to logical qubit state

output = sampler.sample_count ( 1000 )

pcvl.pdisplay ( output['results'], output_format = pcvl.Format.TEXT )

print('fin')
```

The Figure 36 gives an example of the CNOT.

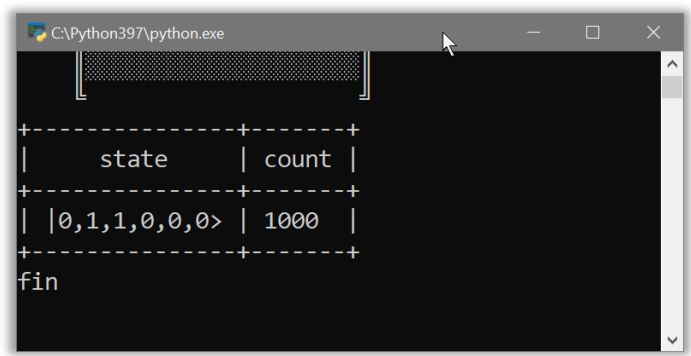

*Figure 36. Test of* $CX|11\rangle = |10\rangle$

## 5.5. Representation of $CCX_{ij}$ : using the Perceval library

The CCNOT gate is also available in the Perceval catalog under the name "`'toffoli'`".
```
ccnot = catalog['toffoli'].build_circuit()
```

It requires 6 control modes, resulting in a test circuit composed of 12 modes. For example, $CX_{12}(|111\rangle)$ is defined by the initial state:
```
istate=[0, 1, 0, 1, 1, 0, 0, 0, 0, 0, 0, 0]
```
where the last 6 zeros correspond to the auxiliary modes. To verify that the CCNOT has succeeded, it is sufficient to test modes 0 and 1 on one hand, modes 2 and 3 on the other, and finally the last 6 modes.
This yields, for example:
```
p.set_postselection(pcvl.PostSelect("[0,1]==1 & [2,3]==1 & [6]==0 & [7]==0 & [8]==0 & [9]==0 & [10]==0 & [11]==0"))
```

**Code**
```python
import perceval as pcvl
import numpy as np
from perceval.components import PS, BS, PERM
from perceval.components import catalog
ccnot = catalog['toffoli'].build_circuit()

# 1. initialisation
istate=[0, 1, 0, 1, 1, 0, 0, 0, 0, 0, 0, 0]
# 2. Cnot
circuit = pcvl.Circuit(12)
circuit.add(0,ccnot)
# 3. print of the circuit
pcvl.pdisplay(circuit.compute_unitary())
```

```
print("")
pcvl.pdisplay(circuit)
# 4 Computation
p = pcvl.Processor("SLOS", circuit)
p.set_postselection(pcvl.PostSelect("[0,1]==1 & [2,3]==1 & [4,5] & [6]==0 &
[7]==0 & [8]==0 & [9]==0 & [10]==0 & [11]==0"))

nsample = 1000
from perceval.algorithm import Sampler , Analyzer
sampler = Sampler ( p )
p.with_input ( pcvl.BasicState (istate)) #Corresponds to logical qubit state
output = sampler.sample_count ( nsample )
print(output['results'])
pcvl.pdisplay ( output['results'], output_format = pcvl.Format.TEXT )

print('fin')
```

Figure 37 shows the execution result for $CCX|110\rangle$.

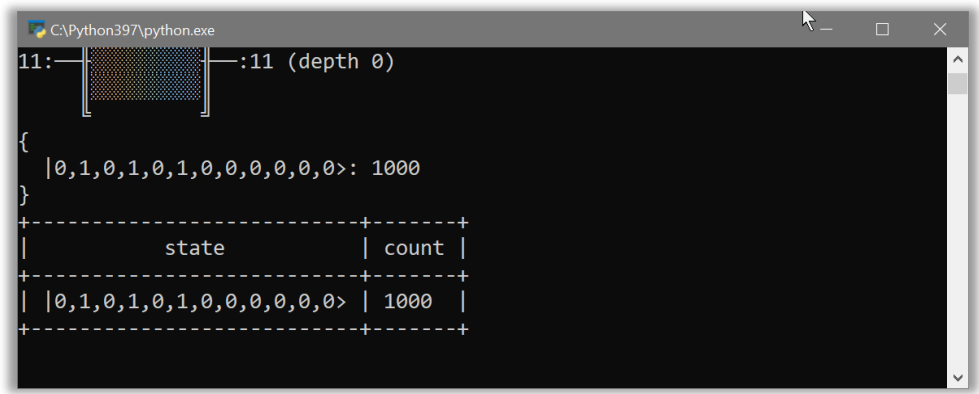

*Figure 37. Test of* $CCX|110\rangle = |111\rangle$

## 5.6. Representation of $CZ_{ij}$ and $CY_{ij}$ gates

The implementation can be done simply since $CZ_{ij} = H_j CX_{ij} H_j$ and $CY_{ij} = S_j CX_{ij} S_j^\dagger = R_{1_j}\left(\frac{\pi}{2}\right) CX_{ij} R_{1_j}\left(-\frac{\pi}{2}\right)$

## *5.7. Summary*

Table 2 provides the correspondence between quantum operators on qubits and the photonic components, along with the Perceval code associated with their implementation.

Table 2. Correspondence between classical gates and their realization using optical components

| Qubit operators | Perceval implementation | Photonic circuit |
|---|---|---|
| $X = \begin{pmatrix} 0 & 1 \\ 1 & 0 \end{pmatrix}$ | `circuit.add(0,PERM([1,0]))` | |
| $Swap_{12} = \begin{pmatrix} 1 & 0 & 0 & 0 \\ 0 & 0 & 1 & 0 \\ 0 & 1 & 0 & 0 \\ 0 & 0 & 0 & 1 \end{pmatrix}$ | `circuit.add(0,PERM([2,3,0,1]))` | |
| $H = \frac{1}{\sqrt{2}}\begin{pmatrix} 1 & 1 \\ 1 & -1 \end{pmatrix}$ | `circuit.add(0, BS.H())` | |
| $Z = \begin{pmatrix} 1 & 0 \\ 0 & -1 \end{pmatrix}$ | `circuit.add(1,PS(np.pi))` | |
| $Y = \begin{pmatrix} 0 & -i \\ i & 0 \end{pmatrix}$ | `circuit.add(0,PERM([1,0]))`<br>`circuit.add(0,PS(-np.pi/2))`<br>`circuit.add(1,PS(np.pi/2))` | |
| $RX(\theta) = \begin{pmatrix} \cos\frac{\theta}{2} & -i.\sin\frac{\theta}{2} \\ -i.\sin\frac{\theta}{2} & \cos\frac{\theta}{2} \end{pmatrix}$ | `circuit.add(0,PS(np.pi/2))`<br>`circuit.add(0,BS.Rx(θ))`<br>`circuit.add(0,PS(np.pi/2))` | |
| $RY(\theta) = \begin{pmatrix} \cos\frac{\theta}{2} & -\sin\frac{\theta}{2} \\ \sin\frac{\theta}{2} & \cos\frac{\theta}{2} \end{pmatrix}$ | `circuit.add(0,BS.Ry(θ))` | |
| $RZ(\theta) = \begin{pmatrix} e^{-i\frac{\theta}{2}} & 0 \\ 0 & e^{i\frac{\theta}{2}} \end{pmatrix}$ | `circuit.add(0,BS.H())`<br>`circuit.add(0,PS(np.pi))`<br>`circuit.add(0,BS.Rx(θ))`<br>`circuit.add(0,PS(np.pi))`<br>`circuit.add(0,BS.H())` | |
| $CX_{ij}\begin{pmatrix} 1 & 0 & 0 & 0 \\ 0 & 1 & 0 & 0 \\ 0 & 0 & 0 & 1 \\ 0 & 0 & 1 & 0 \end{pmatrix}$ | `cnot = catalog['postprocessed cnot'].build_circuit()`<br>`# Ou`<br>`cnot = catalog['heralded cnot'].build_circuit()` | |

# 6. Polarization: basic components and operators

*6.1. Polarization: definition*

When referring to a photon, we are speaking of a wave packet containing exactly one quantum of energy. A wave packet is an electromagnetic wave localized in both time and space. This wave packet has a propagation direction, a magnetic field, and an electric field. A representation is shown in Figure 38, taken from the code provided on the website fempto-physique.fr (Roussel, 2023), illustrating a plane wave linearly polarized along the $O_y$ axis. This website discusses electromagnetic waves in vacuums and provides far more detailed explanations than those presented here.

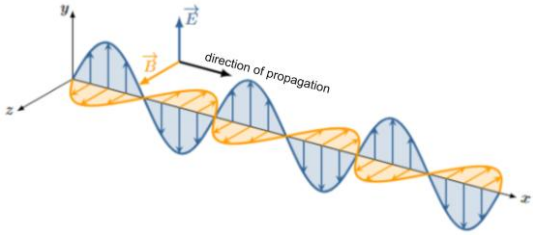

*Figure 38. Structure of a linearly polarized plane wave along $\vec{O_y}$ (Roussel, 2023)*

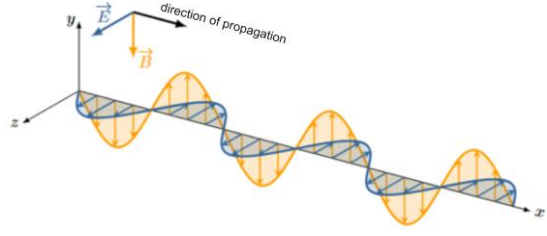

*Figure 39. Structure of a linearly polarized plane wave along $\vec{O_z}$ (Roussel 2023)*

The motion described by the tip of the electric field vector $E$ over time (with $t \sim x$) is referred to as the state of polarization. In the case of Figure 38, the depicted polarization evolves solely along the $y$-axis, and is therefore referred to as linear polarization. One can then consider the $y$-axis as the vertical axis and the $z$-axis as the horizontal axis, as stressed in Figure 39.

Thus, it is possible to associate a linear polarization with each photon. For simplicity, one can restrict the analysis to the $y$- and $z$-axes, corresponding to the horizontal ($H$) and vertical ($V$) polarization states. Other types of polarization (non-linear), as well as polarization along other axes (for example, along the direction $\vec{O_z}+\vec{O_y}$), can also be realized.

*6.2. Fock states encoding*

Polarization is an additional degree of freedom of photons that can be exploited to encode more information within a given mode. To represent the four basis states of a two-qubit system, it is sufficient to use only two modes, each with an associated polarization. As demonstrated by (Heurtel et al., 2023) and (Kwiat et al., 2000), the pair {photon, polarization} can be used to construct the required quantum states. These states are referred to as Fock states, and the polarization can be explicitly specified within each Fock state when a photon is present. For each mode, we have:

$$|0\rangle_f \; or \; \underbrace{|1:H\rangle_f}_{One\ photon\ with\ polarization\ H} \; or \; \underbrace{|1:V\rangle_f}_{One\ photon\ with\ polarization\ V}$$

A correspondence can thus be established between states modeled by qubits and those modeled by Fock states. Table 3 shows a possible correspondence for a two-qubit system.

Table 3. Qubit state and Fock state with 2 qubits

| Qubit state | Fock state |
|---|---|
| $\|00\rangle$ | $\|0, 1:H\rangle_f$ |
| $\|01\rangle$ | $\|0, 1:V\rangle_f$ |
| $\|10\rangle$ | $\|1:H, 0\rangle_f$ |
| $\|11\rangle$ | $\|1:V, 0\rangle_f$ |

### 6.3. Jones' vector definition

The polarization vector (referred to as the *Jones vector*) is a two-dimensional column vector representing, for a given mode *a*, the lack of a photon or the existence of a photon with a given polarization:

$$|0\rangle_a = \begin{pmatrix} 0 \\ 0 \end{pmatrix} \; (vacuum), |1:H\rangle_a = \begin{pmatrix} 1 \\ 0 \end{pmatrix}, |1:V\rangle_a = \begin{pmatrix} 0 \\ 1 \end{pmatrix}$$

In general, when using the Jones vector formalism (Shieh et al., 2010) (Kliger et al., 1990), a state is typically represented by a vector, for example, as $|1:H\rangle_a = \begin{pmatrix} 1 \\ 0 \end{pmatrix}$. It is important not to confuse the notation $\begin{pmatrix} 1 \\ 0 \end{pmatrix}$ which indicates the presence of a single horizontally polarized photon, with the standard Dirac notation.

In the general case, a Jones vector is written as $\begin{pmatrix} \alpha \\ \beta \end{pmatrix}$ and represents:

$$\begin{pmatrix} \alpha \\ \beta \end{pmatrix} = \alpha . |1:H\rangle_a + \beta . |1:V\rangle_a$$

In the particular case where the probability amplitude is concentrated on the horizontal polarization *H*, one can write:

$$|1:H\rangle_a = \begin{pmatrix} 1 \\ 0 \end{pmatrix} = 1. |1:H\rangle_a + 0. |1:V\rangle_a$$

The state $|1:H\rangle_a$ corresponds to a single photon in mode *a*, horizontally polarized. The principle just described can be generalized to an arbitrary number of modes. Thus, with two modes, one obtains the following generalization:

$$|\psi\rangle_{ab} = \alpha|1:H, 0\rangle_{ab} + \beta|1:V, 0\rangle_{ab} + \gamma|0,1:H\rangle_{ab} + \delta|0,1:V\rangle_{ab}$$

### 6.4. New optical components

In this section is given a description of a subset of operators presented in (Shieh et al., 2010) and (Kliger et al., 1990), which provide a more physically oriented treatment. Here, we provide only the matrix-based description and how to apply it to a state.

**Wave Plate** definition

$$WP(\delta, \xi) = \begin{pmatrix} i \sin(\delta) \cos(2\xi) + \cos(\delta) & i \sin(\delta) \sin(2\xi) \\ i \sin(\delta) \sin(2\xi) & -i \sin(\delta) \cos(2\xi) + \cos(\delta) \end{pmatrix}$$

We note that this operator can be rewritten using the classical Pauli gate considering:

$$WP(\delta, \xi) = \cos(\delta) . Id + i \sin(\delta) . \begin{pmatrix} \cos(2\xi) & \sin(2\xi) \\ \sin(2\xi) & -\cos(2\xi) \end{pmatrix}$$

$$WP(\delta, \xi) = \cos(\delta) . Id + i \sin(\delta) . \cos(2\xi) . Z + i \sin(\delta) . \sin(2\xi) . X$$

Its action on a Jones vector corresponding to an arbitrary superposition of polarization states:

$$WP(\delta,\xi).\begin{pmatrix}\alpha\\\beta\end{pmatrix} = \begin{pmatrix}i\sin(\delta)\cos(2\xi)+\cos(\delta) & i\sin(\delta)\sin(2\xi)\\ i\sin(\delta)\sin(2\xi) & -i\sin(\delta)\cos(2\xi)+\cos(\delta)\end{pmatrix}\begin{pmatrix}\alpha\\\beta\end{pmatrix}$$
$$= \begin{pmatrix}\alpha(i\sin(\delta)\cos(2\xi)+\cos(\delta)) + \beta(i\sin(\delta)\sin(2\xi))\\ \alpha(i\sin(\delta)\sin(2\xi)) + \beta(-i\sin(\delta)\cos(2\xi)+\cos(\delta))\end{pmatrix}$$

**Half Wave Plate (HWP)** definition

Let us note that $HWP(\xi) = WP\left(\frac{\pi}{2},\xi\right)$ and we have:

$$HWP(\xi) = \begin{pmatrix}i\cos(2\xi) & i\sin(2\xi)\\ i\sin(2\xi) & -i\cos(2\xi)\end{pmatrix} = i.\begin{pmatrix}\cos(2\xi) & \sin(2\xi)\\ \sin(2\xi) & -\cos(2\xi)\end{pmatrix}$$

$$HWP(\xi) = i\,\cos(2\xi).Z + i\,\sin(2\xi).X$$

**Polarisation rotator** definition

$$PR(\boldsymbol{\theta}) = \begin{pmatrix}\cos\boldsymbol{\theta} & \sin\boldsymbol{\theta}\\ -\sin\boldsymbol{\theta} & \cos\boldsymbol{\theta}\end{pmatrix}$$

And we note that

$$PR\left(\frac{\pi}{2}\right) = \begin{pmatrix}0 & 1\\ -1 & 0\end{pmatrix} = i.Y$$

**Polarized Beam Splitter ($PBS$)** definition

The $PBS$ operator acts on two spatial modes and is represented by the following matrix form:

$$PBS_{ij} = \begin{pmatrix}0 & 0 & 1 & 0\\ 0 & 1 & 0 & 0\\ 1 & 0 & 0 & 0\\ 0 & 0 & 0 & 1\end{pmatrix}$$

where, $i$ and $j$ denote two spatial modes, which may, without loss of generality, be assumed to be consecutive modes, since any other configuration can always be reduced to this case by an appropriate permutation of modes. This matrix can be analyzed straightforwardly by considering a state expressed in terms of creation operators. In the very general case $PBS_{ij}$, the action of the operator leads to:

$$PBS_{ij}\begin{pmatrix}a_{iH}^\dagger\\ a_{iV}^\dagger\\ a_{jH}^\dagger\\ a_{jV}^\dagger\end{pmatrix} = \begin{pmatrix}0 & 0 & 1 & 0\\ 0 & 1 & 0 & 0\\ 1 & 0 & 0 & 0\\ 0 & 0 & 0 & 1\end{pmatrix}\begin{pmatrix}a_{iH}^\dagger\\ a_{iV}^\dagger\\ a_{jH}^\dagger\\ a_{jV}^\dagger\end{pmatrix} = \begin{pmatrix}a_{jH}^\dagger\\ a_{iV}^\dagger\\ a_{iH}^\dagger\\ a_{jV}^\dagger\end{pmatrix}$$

Leading to

$$a_{iH}^\dagger \to a_{jH}^\dagger$$
$$a_{iV}^\dagger \to a_{iV}^\dagger$$
$$a_{jH}^\dagger \to a_{iH}^\dagger$$
$$a_{jV}^\dagger \to a_{jV}^\dagger$$

where $a_{ix}^\dagger \to a_{jy}^\dagger$ which can be interpreted as the presence of a photon in polarization state $x$ in input mode $i$ being mapped to a photon in polarization state $y$ in output mode $j$.

So we have for example, considering mode 0 and 1, we have:

$$PBS_{01} \begin{pmatrix} a_{0H}^\dagger \\ a_{0V}^\dagger \\ a_{1H}^\dagger \\ a_{1V}^\dagger \end{pmatrix} = \begin{pmatrix} 0 & 0 & 1 & 0 \\ 0 & 1 & 0 & 0 \\ 1 & 0 & 0 & 0 \\ 0 & 0 & 0 & 1 \end{pmatrix} \begin{pmatrix} a_{0H}^\dagger \\ a_{0V}^\dagger \\ a_{1H}^\dagger \\ a_{1V}^\dagger \end{pmatrix} = \begin{pmatrix} a_{1H}^\dagger \\ a_{0V}^\dagger \\ a_{0H}^\dagger \\ a_{1V}^\dagger \end{pmatrix}$$

leading to:

$$a_{0H}^\dagger \to a_{1H}^\dagger$$
$$a_{0V}^\dagger \to a_{0V}^\dagger$$
$$a_{1H}^\dagger \to a_{0H}^\dagger$$
$$a_{1V}^\dagger \to a_{1V}^\dagger$$

Thus:

$$PBS_{01}.|1:H,0\rangle_f = PBS_{01}.a_{0H}^\dagger|0,0\rangle_f = a_{1H}^\dagger.|0,0\rangle_f = |0,1:H\rangle_f$$
$$PBS_{01}.|1:V,0\rangle_f = |1:V,0\rangle_f$$
$$PBS_{01}.|0,1:H\rangle_f = |1:H,0\rangle_f$$
$$PBS_{01}.|0,1:V\rangle_f = |0,1:V\rangle_f$$

For each of these operators, the relevant modes are indicated by subscripts. For instance, $PR_0(\theta)$ denotes the application of the polarization rotation operator on mode 0, while $PBS_{0,2}$ denotes the application of a Polarizing Beam Splitter between modes 0 and 2. If no ambiguity arises, many authors omit the specification of the modes to which the operator is applied.

### 6.5. Generalized Beam Splitter formalism for polarized photons

In general, for a circuit whose action is represented by a unitary matrix $U$ acting on the creation operators, one has:

$$\vec{\iota^\dagger} = U^T \vec{o^\dagger}$$

with

- $\vec{\iota^\dagger} = \begin{pmatrix} a_0^\dagger \\ \ldots \\ a_{m-1}^\dagger \end{pmatrix}$ the vector containing the creation operators associated with the input modes.

- $\vec{o^\dagger} = \begin{pmatrix} s_0^\dagger \\ \ldots \\ s_{m-1}^\dagger \end{pmatrix}$ the vector containing the creation operators associated with the output modes.

The unitary matrix associated with the symmetric beam splitter with $\theta = \frac{\pi}{2}$ in the $BS_{Ry}$ convention is:

$$BS_{Ry}(\theta \in [0; \pi], 0,0,0) = \begin{pmatrix} \cos\left(\frac{\theta}{2}\right) & -\sin\left(\frac{\theta}{2}\right) \\ \sin\left(\frac{\theta}{2}\right) & \cos\left(\frac{\theta}{2}\right) \end{pmatrix}$$

$$BS_{Ry}\left(\frac{\pi}{2}, 0,0,0\right) = \begin{pmatrix} \frac{1}{\sqrt{2}} & -\frac{1}{\sqrt{2}} \\ \frac{1}{\sqrt{2}} & \frac{1}{\sqrt{2}} \end{pmatrix}$$

One can consider **the following interpretation of the Beam Splitter action** considering the presence of a single photon in input mode 0 polarized along $H$.

In the relation:

$$\vec{\iota^\dagger} = U^T \vec{o^\dagger}$$

And using $U = BS_{Ry}\left(\frac{\pi}{2}, 0,0,0\right)$ gives:

$$\begin{pmatrix} a_0^\dagger \\ a_1^\dagger \end{pmatrix}_{input} = \begin{pmatrix} \frac{1}{\sqrt{2}} & \frac{1}{\sqrt{2}} \\ -\frac{1}{\sqrt{2}} & \frac{1}{\sqrt{2}} \end{pmatrix} \begin{pmatrix} a_0^\dagger \\ a_1^\dagger \end{pmatrix}_{output}$$

This gives the relationships:

$$\underbrace{a_0^\dagger}_{input} = \underbrace{\frac{1}{\sqrt{2}}(a_0^\dagger + a_1^\dagger)}_{output} \qquad (1)$$

$$\underbrace{a_1^\dagger}_{input} = \underbrace{\frac{1}{\sqrt{2}}(-a_0^\dagger + a_1^\dagger)}_{output} \qquad (2)$$

Using (1) and (2), we get the relations on polarization

$$\underbrace{a_{0H}^\dagger}_{input} = \underbrace{\frac{1}{\sqrt{2}}(a_{0H}^\dagger + a_{1H}^\dagger)}_{output}$$

$$\underbrace{a_{0V}^\dagger}_{input} = \underbrace{\frac{1}{\sqrt{2}}(a_{0V}^\dagger + a_{1V}^\dagger)}_{output}$$

$$\underbrace{a_{1H}^\dagger}_{input} = \underbrace{\frac{1}{\sqrt{2}}(-a_{0H}^\dagger + a_{1H}^\dagger)}_{output}$$

$$\underbrace{a_{1V}^\dagger}_{input} = \underbrace{\frac{1}{\sqrt{2}}(-a_{0V}^\dagger + a_{1V}^\dagger)}_{output}$$

**To conclude, we have:**

$$BS_{Ry}\left(\frac{\pi}{2},0,0,0\right).a_{0H}^{\dagger} = \frac{1}{\sqrt{2}}\left(a_{0H}^{\dagger} + a_{1H}^{\dagger}\right)$$

$$BS_{Ry}\left(\frac{\pi}{2},0,0,0\right).a_{0V}^{\dagger} = \frac{1}{\sqrt{2}}\left(a_{0V}^{\dagger} + a_{1V}^{\dagger}\right)$$

$$BS_{Ry}\left(\frac{\pi}{2},0,0,0\right).a_{1H}^{\dagger} = \frac{1}{\sqrt{2}}\left(-a_{0H}^{\dagger} + a_{1H}^{\dagger}\right)$$

$$BS_{Ry}\left(\frac{\pi}{2},0,0,0\right).a_{1V}^{\dagger} = \frac{1}{\sqrt{2}}\left(-a_{0V}^{\dagger} + a_{1V}^{\dagger}\right)$$

Which is equivalent to

$$BS_{Ry}\left(\frac{\pi}{2},0,0,0\right).|1:H,0\rangle = \frac{1}{\sqrt{2}}(|1:H,0\rangle + |0,1:H\rangle)$$

$$BS_{Ry}\left(\frac{\pi}{2},0,0,0\right).|1:V,0\rangle = \frac{1}{\sqrt{2}}(|1:V,0\rangle + |0,1:V\rangle)$$

$$BS_{Ry}\left(\frac{\pi}{2},0,0,0\right).|0,1:H\rangle = \frac{1}{\sqrt{2}}(-|1:H,0\rangle + |0,1:H\rangle)$$

$$BS_{Ry}\left(\frac{\pi}{2},0,0,0\right).|0,1:V\rangle = \frac{1}{\sqrt{2}}(-|1:V,0\rangle + |0,1:V\rangle)$$

# 7. Grover Implementation

## 7.1. Classical Grover's algorithm

The Grover algorithm permits to make amplification of quantum states previously tagged by an oracle to search in an unstructured database of $n$ entries, a marked element with a quadratic speedup. It was introduced by (Grover, 1996).

The problem consists in finding at least one $x \in E_1$ where $E_1$ is a set of element marked by the Oracle i.e. $E_1 = span(x \in B/f(x) = 1)$ with $f$ a function from $B$ to {0,1} such that $x \rightarrow f(x)$.

The Grover's algorithm is composed of 4 steps:

- the step 0 which is the initialization;
- the step 1 to define a similar amplitude to all states in the computational base $B(|0\rangle;|1\rangle)$;
- the step 2 which is the Oracle definition
- the step 3 which is the Amplification process.

The principle can be illustrated by revisiting the example from (Fleury and Lacomme, 2022), where the definition of the Oracle is not the focus. It is assumed that the Oracle marks the state $|01\rangle$ in a circuit involving only 3 qubits, one of which is the ancillary qubit. This results in a very short circuit for which the associated tensor calculations can be easily carried out (Figure 40).

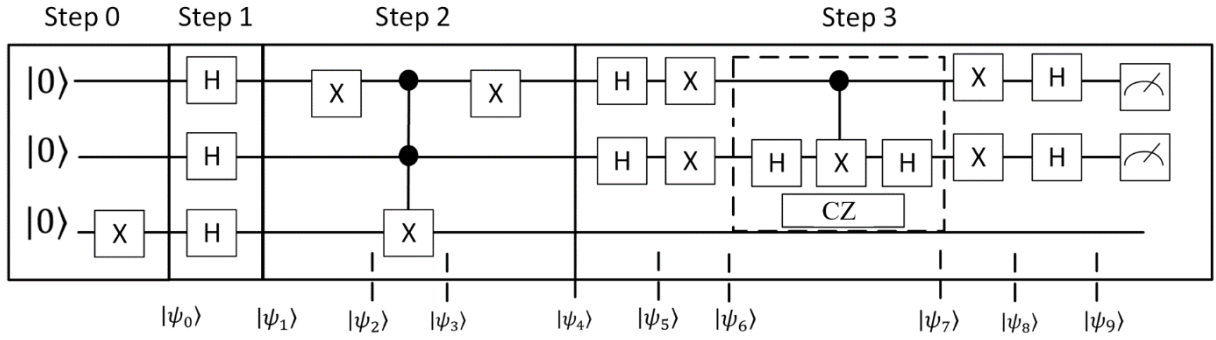

*Figure 40. Example of Grover circuit with 3 qubits*

The tensor calculations in Table 4 show that, at the time of measurement in the basis $B = (|00\rangle,\ldots,|11\rangle)$, the amplitude of the state $|01\rangle$ is $-1$. This implies that a measurement performed in the basis $B = (|0\rangle, |1\rangle)$ will yield the state $|01\rangle$ with 100% probability.

**Table 4. Steps in Grover computation**

|  | Quantum state |
|---|---|
| Step 0 | $|\psi_0\rangle = |001\rangle$ |
| Step 1. Application of $H^{\otimes 3}$ | $|\psi_1\rangle = |pp\rangle \otimes |m\rangle$ |
| **Oracle** | |
| Step 2.1. Application of $X \otimes Id \otimes Id$ | $|\psi_2\rangle = |ppm\rangle$ |
| Step 2.2. Application of $CCX(q_1, q_2; q_3)$ | $|\psi_3\rangle = |ppm\rangle - |11m\rangle$ |
| Step 2.2. Application of $X \otimes Id \otimes Id$ | $|\psi_4\rangle = |ppm\rangle - |01m\rangle = (|pp\rangle - |01\rangle) \otimes |m\rangle$ |
| **Amplification** | |
| Step 3.1. Application of $H^{\otimes 2} \otimes Id$ | $|\psi_5\rangle = (|00\rangle - |pm\rangle) \otimes |m\rangle$ |
| Step 3.2. Application of $X^{\otimes 2} \otimes Id$ | $|\psi_6\rangle = (|11\rangle + |pm\rangle) \otimes |m\rangle$ |
| Step 3.3. Application of $CZ(q_1; q_2) \otimes Id$ | $|\psi_7\rangle = (-|11\rangle + (|pm\rangle + |11\rangle)) \otimes |m\rangle$ |
|  | $|\psi_7\rangle = (-|11\rangle + |11\rangle + |pm\rangle) \otimes |m\rangle$ |
|  | $|\psi_7\rangle = |pm\rangle \otimes |m\rangle$ |
| Step 3.4. Application of $X^{\otimes 2} \otimes Id$ | $|\psi_8\rangle = -|pm\rangle \otimes |m\rangle$ |
| Step 3.5. Application of $H^{\otimes 2} \otimes Id$ | $|\psi_9\rangle = -|01\rangle \otimes |m\rangle$ |

### 7.2. Photonic Grover implementation

For the sake of simplicity, and without loss of generality, we focus on the photonic circuit depicted in Figure 41 adapted from (), which constitutes the photonic implementation of Grover's circuit.

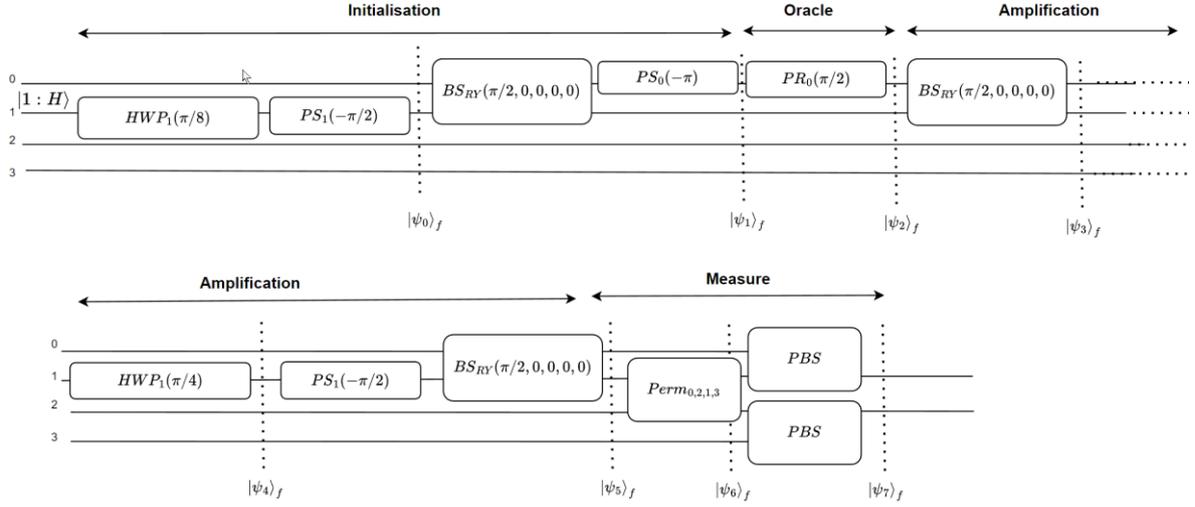

**Figure 41.** Photonic Grover implementation circuit

The circuit leverages photon polarization to encode the four possible states. The equivalence between the conventional qubit states and the encoding using Fock states is presented in Table 5. It is observed that the circuit comprises two modes dedicated to state encoding and two additional modes used exclusively during the measurement process.

**Table 5. Polarization-based encoding**

| Qubit encoding | Fock state encoding |
|---|---|
| $\|00\rangle$ | $\|0, 1:H\rangle_f$ |
| $\|01\rangle$ | $\|0, 1:V\rangle_f$ |
| $\|10\rangle$ | $\|1:H, 0\rangle_f$ |
| $\|11\rangle$ | $\|1:V, 0\rangle_f$ |

**Step 0. Initialization**

Initialization is performed by preparing an equal superposition of the input qubit states, namely:

$$|00\rangle, |01\rangle, |10\rangle \text{ and } |11\rangle$$

These four states are represented as:

$$|0,1:H\rangle_{01}, |0,1:V\rangle_{01}, |1:H,0\rangle_{01} \text{ and } |1:V,0\rangle_{01}$$

It is important to note that only two spatial modes are used in this representation. By incorporating polarization, one obtains a description that is distinct from, yet equivalent to, the conventional qubit-based model: the resulting state space remains four-dimensional. This step consists of four operations that enable the computation of $|\psi_0\rangle_f$ and $|\psi_1\rangle_f$.

$$|\psi_1\rangle = PS_0(-\pi) \cdot BS_{Ry}\left(\frac{\pi}{2}, 0, 0, 0\right) \cdot PS_1\left(-\frac{\pi}{2}\right) \cdot HWP_1\left(\frac{\pi}{8}\right) \cdot |1:H\rangle_1$$

Step 0.1. Application of HWP with $\xi = \frac{\pi}{8}$ on mode 1

Let us note that the operator

$$HWP(\xi) = \begin{pmatrix} i\cos(2\xi) & i\sin(2\xi) \\ i\sin(2\xi) & -i\cos(2\xi) \end{pmatrix}$$

gives

$$HWP\left(\frac{\pi}{8}\right) = \begin{pmatrix} i\cos\left(\frac{\pi}{4}\right) & i\sin\left(\frac{\pi}{4}\right) \\ i\sin\left(\frac{\pi}{4}\right) & -i\cos\left(\frac{\pi}{4}\right) \end{pmatrix} = \begin{pmatrix} i\frac{1}{\sqrt{2}} & i\frac{1}{\sqrt{2}} \\ i\frac{1}{\sqrt{2}} & -i\frac{1}{\sqrt{2}} \end{pmatrix}$$

Thus

$$HWP\left(\frac{\pi}{8}\right).|1:H\rangle = HWP\left(\frac{\pi}{8}\right).\begin{pmatrix}1\\0\end{pmatrix} = \begin{pmatrix} i\frac{1}{\sqrt{2}} & i\frac{1}{\sqrt{2}} \\ i\frac{1}{\sqrt{2}} & -i\frac{1}{\sqrt{2}} \end{pmatrix}.\begin{pmatrix}1\\0\end{pmatrix}$$

$$HWP\left(\frac{\pi}{8}\right).|1:H\rangle = \begin{pmatrix} i\frac{1}{\sqrt{2}} \\ i\frac{1}{\sqrt{2}} \end{pmatrix} = i\frac{1}{\sqrt{2}}|1:H\rangle_1 + i\frac{1}{\sqrt{2}}|1:V\rangle_1$$

So the Fock state considering only the first mode of the circuit is $i\frac{1}{\sqrt{2}}|0,1:H\rangle_{01} + i\frac{1}{\sqrt{2}}|0,1:V\rangle_{01}$):

$$|\psi_0\rangle_f = PS_1\left(-\frac{\pi}{2}\right).HWP_1\left(\frac{\pi}{8}\right).|1:H\rangle$$

$$|\psi_0\rangle_f = PS_1\left(-\frac{\pi}{2}\right)\left(i\frac{1}{\sqrt{2}}|0,1:H\rangle_{01} + i\frac{1}{\sqrt{2}}|0,1:V\rangle_{01}\right)$$

Because

$$PS_1\left(-\frac{\pi}{2}\right)|0\rangle_f = |0\rangle_f \text{ et } PS_1\left(-\frac{\pi}{2}\right)|1\rangle_f = e^{i-\frac{\pi}{2}}.|1\rangle_f = -i.|1\rangle_f$$

We have

$$|\psi_0\rangle_f = PS_1\left(-\frac{\pi}{2}\right).\left(i\frac{1}{\sqrt{2}}|1:H\rangle_1 + i\frac{1}{\sqrt{2}}|1:V\rangle_1\right)$$

$$= -i.\left(i\frac{1}{\sqrt{2}}|1:H\rangle_1 + i\frac{1}{\sqrt{2}}|1:V\rangle_1\right)$$

Considering the two modes simultaneously gives:

$$|\psi_0\rangle_f = \frac{1}{\sqrt{2}}|0,1:H\rangle_{01} + \frac{1}{\sqrt{2}}|0,1:V\rangle_{01}$$

Step 0.2. Application of $BS_{RY}$

Because

$$BS_{Ry}\left(\frac{\pi}{2},0,0,0\right)|0,1:H\rangle_{01} = \frac{1}{\sqrt{2}}(|0,1:H\rangle_{01} - |1:H,0\rangle_{01}))$$

$$BS_{Ry}\left(\frac{\pi}{2},0,0,0\right)|0,1:V\rangle_{01} = \frac{1}{\sqrt{2}}(|0,1:V\rangle_{01} - |1:V,0\rangle_{01}))$$

We have

$$|\psi_1\rangle_f = BS_{Ry}\left(\frac{\pi}{2},0,0,0\right).|\psi_0\rangle_f$$

$$|\psi_1\rangle_f = BS_{Ry}\left(\frac{\pi}{2},0,0,0\right).\frac{1}{\sqrt{2}}.(|0,1:H\rangle_{01} + |0,1:V\rangle_{01})$$

$$|\psi_1\rangle_f = \frac{1}{2}.(|0,1:H\rangle_{01} - |1:H,0\rangle_{01}) + |0,1:V\rangle_{01} - |1:V,0\rangle_{01}))$$

Step 0.3. Application of $PS_0(-\pi)$

$$|\psi_2\rangle_f = PS_0(-\pi).|\psi_1\rangle_f$$

We have

$$PS_p(\phi)|0\rangle_p \mapsto |0\rangle_p \text{ and } PS_p(\phi)|1\rangle_p \mapsto e^{i\phi}.|1\rangle_p$$

and

$$PS_p(-\pi)|0\rangle_p \mapsto |0\rangle_p \text{ and } PS_p(-\pi)|1\rangle_p \mapsto e^{-i\pi}|1\rangle_p = -|1\rangle_p$$

Thus

$$|\psi_1\rangle_f = \frac{1}{2}.PS_0(-\pi).(|0,1:H\rangle_{01} - |1:H,0\rangle_{01}) + |0,1:V\rangle_{01} - |1:V,0\rangle_{01}))$$

$$|\psi_1\rangle_f = \frac{1}{2}.(|1:H,0\rangle_{01} + |0,1:H\rangle_{01} + |1:V,0\rangle_{01} + |0,1:V\rangle_{01})$$

Thus $|\psi_1\rangle_f$ defines a state in which all Fock states occur with equal probability; this constitutes an equiprobable superposition over the entire Fock space.

**Step 1. Oracle for $|1:V,0\rangle_f$**

In this section, we try to mark the state being $|1:V,0\rangle_f$. The Oracle is composed on one quantum operator $PR_0\left(\frac{\pi}{2}\right)$ only leading to:

$$|\psi_2\rangle_f = PR_0\left(\frac{\pi}{2}\right).|\psi_1\rangle_f$$

Since we have

$$PR\left(\frac{\pi}{2}\right) = \begin{pmatrix} 0 & 1 \\ -1 & 0 \end{pmatrix}$$

and

$$PR\left(\frac{\pi}{2}\right).\begin{pmatrix} 1 \\ 0 \end{pmatrix} = \begin{pmatrix} 0 & 1 \\ -1 & 0 \end{pmatrix}.\begin{pmatrix} 1 \\ 0 \end{pmatrix} = -\begin{pmatrix} 0 \\ 1 \end{pmatrix}$$

$$PR\left(\frac{\pi}{2}\right).\begin{pmatrix} 0 \\ 1 \end{pmatrix} = \begin{pmatrix} 0 & 1 \\ -1 & 0 \end{pmatrix}.\begin{pmatrix} 0 \\ 1 \end{pmatrix} = \begin{pmatrix} 1 \\ 0 \end{pmatrix}$$

We have

$$PR_0\left(\frac{\pi}{2}\right)|1:H\rangle_0 = -|1:V\rangle_0$$

$$PR_0\left(\frac{\pi}{2}\right)|1:V\rangle_0 = |1:H\rangle_0$$

$$PR_0\left(\frac{\pi}{2}\right)|1:H\rangle_1 = |1:H\rangle_1$$

$$PR_0\left(\frac{\pi}{2}\right)|1:V\rangle_1 = |1:V\rangle_1$$

Thus

$$|\psi_2\rangle_f = PR_0\left(\frac{\pi}{2}\right) \cdot |\psi_1\rangle_f$$

$$|\psi_2\rangle_f = PR_0\left(\frac{\pi}{2}\right)\left(\frac{1}{2} \cdot (|1:H,0\rangle_f + |0,1:H\rangle_f + |1:V,0\rangle_f + |0,1:V\rangle_f)\right)$$

$$|\psi_2\rangle_f = \left(\frac{1}{2} \cdot \left(PR_0\left(\frac{\pi}{2}\right)|1:H,0\rangle_f + \cancel{PR_0\left(\frac{\pi}{2}\right)}|0,1:H\rangle_f + PR_0\left(\frac{\pi}{2}\right)|1:V,0\rangle_f\right.\right.$$
$$\left.\left. + \cancel{PR_0\left(\frac{\pi}{2}\right)}|0,1:V\rangle_f\right)\right)$$

$$|\psi_2\rangle_f = \frac{1}{2} \cdot (-|1:V,0\rangle_f + |0,1:H\rangle_f + |1:H,0\rangle_f + |0,1:V\rangle_f)$$

The sign of the state $|1:V,0\rangle_f$ has been flipped from negative to positive by the action of the oracle, thereby fulfilling its purpose of marking the state $|1:V,0\rangle_f$

**Step 2. Amplification**

Step 2.1. Application of $BS_{Ry}\left(\frac{\pi}{2},0,0,0\right)$

Considering the $BS_{Ry}\left(\frac{\pi}{2},0,0,0\right)$ definition we have:

$$|\psi_3\rangle_f = BS_{Ry}\left(\frac{\pi}{2},0,0,0\right) \cdot |\psi_2\rangle_f$$

$$|\psi_3\rangle_f = BS_{Ry}\left(\frac{\pi}{2},0,0,0\right) \cdot \frac{1}{2}(-|1:V,0\rangle_f + |0,1:H\rangle_f + |1:H,0\rangle_f + |0,1:V\rangle_f)$$

$$|\psi_3\rangle_f = \frac{1}{2\sqrt{2}}(-|1:V,0\rangle_f - |0,1:V\rangle_f - |1:H,0\rangle_f + |0,1:H\rangle_f + |1:H,0\rangle + |0,1:H\rangle_f - |1:V,0\rangle + |0,1:V\rangle_f)$$

$$|\psi_3\rangle_f = \frac{1}{\sqrt{2}}(-|1:V,0\rangle_f + |0,1:H\rangle_f)$$

Step 2.2. Application of $HWP\left(\frac{\pi}{4}\right)$

We have

$$HWP(\xi) = \begin{pmatrix} i\cos(2\xi) & i\sin(2\xi) \\ i\sin(2\xi) & -i\cos(2\xi) \end{pmatrix}$$

And by consequence

$$HWP\left(\frac{\pi}{4}\right) = \begin{pmatrix} i\cos\left(\frac{\pi}{2}\right) & i\sin\left(\frac{\pi}{2}\right) \\ i\sin\left(\frac{\pi}{2}\right) & -i\cos\left(\frac{\pi}{2}\right) \end{pmatrix} = \begin{pmatrix} 0 & i \\ i & 0 \end{pmatrix}$$

so

$$HWP\left(\frac{\pi}{4}\right).|1:H\rangle_f = \begin{pmatrix} 0 & i \\ i & 0 \end{pmatrix}\begin{pmatrix} 1 \\ 0 \end{pmatrix} = i.\begin{pmatrix} 0 \\ 1 \end{pmatrix} = i.|1:V\rangle_f$$

$$HWP\left(\frac{\pi}{4}\right).|1:V\rangle_f = \begin{pmatrix} 0 & i \\ i & 0 \end{pmatrix}\begin{pmatrix} 0 \\ 1 \end{pmatrix} = i.\begin{pmatrix} 1 \\ 0 \end{pmatrix} = i.|1:H\rangle_f$$

Thus we have

$$|\psi_4\rangle_f = HWP_1\left(\frac{\pi}{4}\right).|\psi_3\rangle_f$$

$$|\psi_4\rangle_f = HWP_1\left(\frac{\pi}{4}\right).\left(\frac{1}{\sqrt{2}}.(-|1:V,0\rangle_f + |0,1:H\rangle_f)\right)$$

$$|\psi_4\rangle_f = \frac{1}{\sqrt{2}}.\left(-HWP_1\left(\frac{\pi}{4}\right).|1:V,0\rangle_f + HWP_1\left(\frac{\pi}{4}\right).|0,1:H\rangle_f\right)$$

$$|\psi_4\rangle_f = \frac{1}{\sqrt{2}}.(-|1:V,0\rangle_f + i|0,1:V\rangle_f)$$

Step 2.3. Application of $PS_1\left(-\frac{\pi}{2}\right)$

$$|\psi_5\rangle_f = PS_1\left(-\frac{\pi}{2}\right).|\psi_4\rangle_f$$

and

$$|\psi_5\rangle_f = PS_1\left(-\frac{\pi}{2}\right)\frac{1}{\sqrt{2}}.(-|1:V,0\rangle_f + i|0,1:V\rangle_f)$$

Because

$$PS\left(-\frac{\pi}{2}\right).|1:H\rangle = -i.|1:H\rangle$$

$$PS\left(-\frac{\pi}{2}\right).|1:V\rangle = -i.|1:V\rangle$$

We have

$$|\psi_5\rangle_f = \frac{1}{\sqrt{2}}.(-|1:V,0\rangle_f + |0,1:V\rangle_f)$$

Step 2.4. Application of $BS_{Ry}\left(\frac{\pi}{2},0,0,0\right)$

$$|\psi_6\rangle_f = BS_{Ry}\left(\frac{\pi}{2},0,0,0\right)|\psi_5\rangle_f$$

Which gives

$$|\psi_6\rangle_f = \frac{1}{\sqrt{2}}.BS_{Ry}\left(\frac{\pi}{2},0,0,0\right).(-|1:V,0\rangle_f + |0,1:V\rangle_f)$$

$$|\psi_6\rangle_f = -\frac{1}{\sqrt{2}}.BS_{Ry}\left(\frac{\pi}{2},0,0,0\right).|1:V,0\rangle_f + \frac{1}{\sqrt{2}}.BS_{Ry}\left(\frac{\pi}{2},0,0,0\right).|0,1:V\rangle_f$$

Using the $BS_{Ry}\left(-\frac{\pi}{2}, 0, 0, 0\right)$ definition we have:

$$|\psi_6\rangle_f = \frac{1}{2}\cdot(-|1:V,0\rangle_f - |0,1:V\rangle_f - |1:V,0\rangle_f + |0,1:V\rangle_f)$$

and

$$|\psi_6\rangle_f = -|1:V,0\rangle_f$$

Such a result meets the result provided in section 2.1 by one classical tensorial computation.

### Step 3. Measure

It is not possible to directly measure the difference between $|1:V,0\rangle_f$ and $|1:H,0\rangle_f$ or between $|0,1:V\rangle_f$ and $|0,1:H\rangle_f$. The solution consists in using a PBS (Polarizing Beam Splitter), which enables the separation of $H$-polarized and $V$-polarized photons in each spatial mode, thereby allowing one, for instance, to distinguish between $|1:V,0\rangle_f$ and $|1:H,0\rangle_f$. The two extra mode in the circuit are used to make a base switch from a two modes encoding into a 4 modes encoding (Table 6).

**Table 6. 4 modes encoding at the end of the circuit**

| Fock state encoding | Fock state encoding on 4 modes |
|---|---|
| $\|0, 1:H\rangle_f$ | $\|0,0,0,1:H\rangle_f$ |
| $\|0, 1:V\rangle_f$ | $\|0,0,1:V,0\rangle_f$ |
| $\|1:H, 0\rangle_f$ | $\|0,1:H,0,0\rangle_f$ |
| $\|1:V, 0\rangle_f$ | $\|1:V,0,0,0\rangle_f$ |

The switch from Fock state encoding on 2 modes and the Fock state encoding on 4 modes is described by the following operator:

$$PBS_{23}.PBS_{01}Perm_{0213}$$

If the Fock state encoded is $|1:H,0,0,0\rangle_f$ the measure gives the following:

$$PBS_{23}.PBS_{01}Perm_{0213}|1:H,0,0,0\rangle_f = PBS_{23}.PBS_{01}.a_{0H}^\dagger|0,0,0,0\rangle_f$$
$$PBS_{23}.PBS_{01}Perm_{0213}|1:H,0,0,0\rangle_f = PBS_{23}.a_{1H}^\dagger|0,0,0,0\rangle_f$$
$$PBS_{23}.PBS_{01}Perm_{0213}|1:H,0,0,0\rangle_f = PBS_{23}.a_{1H}^\dagger|0,0,0,0\rangle_f$$
$$PBS_{23}.PBS_{01}Perm_{0213}.|1:H,0,0,0\rangle_f = |0,1:H,0,0\rangle_f$$

If the Fock state encoded is $|1:V,0,0,0\rangle_f$ the measure gives the following:

$$PBS_{23}.PBS_{01}Perm_{0213}|1:V,0,0,0\rangle_f = PBS_{23}.PBS_{01}.a_{0V}^\dagger|0,0,0,0\rangle_f$$
$$PBS_{23}.PBS_{01}Perm_{0213}|1:V,0,0,0\rangle_f = PBS_{23}.a_{0V}^\dagger|0,0,0,0\rangle_f$$
$$PBS_{23}.PBS_{01}Perm_{0213}|1:V,0,0,0\rangle_f = PBS_{23}.a_{0V}^\dagger|0,0,0,0\rangle_f$$
$$PBS_{23}.PBS_{01}Perm_{0213}.|1:V,0,0,0\rangle_f = |1:V,0,0,0\rangle_f$$

If the Fock state encoded is $|0,1:H,0,0\rangle_f$ the measure gives the following:

$$PBS_{23}.PBS_{01}Perm_{0213}|0,1:H,0,0\rangle_f = PBS_{23}.PBS_{01}.|0,0,1:H,0\rangle_f$$
$$PBS_{23}.PBS_{01}Perm_{0213}|0,1:H,0,0\rangle_f = PBS_{23}.PBS_{01}.a^\dagger_{2H}|0,0,0,0\rangle_f$$
$$PBS_{23}.PBS_{01}Perm_{0213}|0,1:H,0,0\rangle_f = PBS_{23}.a^\dagger_{2H}|0,0,0,0\rangle_f$$
$$PBS_{23}.PBS_{01}Perm_{0213}.|0,1:H,0,0\rangle_f = a^\dagger_{3H}|0,0,0,0\rangle_f$$
$$PBS_{23}.PBS_{01}Perm_{0213}.|0,1:H,0,0\rangle_f = |0,0,0,1:H\rangle_f$$

If the Fock state encoded is $|0,1:V,0,0\rangle_f$ the measure gives the following:

$$PBS_{23}.PBS_{01}Perm_{0213}|0,1:V,0,0\rangle_f = PBS_{23}.PBS_{01}.|0,0,1:V,0\rangle_f$$
$$PBS_{23}.PBS_{01}Perm_{0213}|0,1:V,0,0\rangle_f = PBS_{23}.PBS_{01}.a^\dagger_{2V}|0,0,0,0\rangle_f$$
$$PBS_{23}.PBS_{01}Perm_{0213}|0,1:V,0,0\rangle_f = PBS_{23}.a^\dagger_{2V}|0,0,0,0\rangle_f$$
$$PBS_{23}.PBS_{01}Perm_{0213}.|0,1:V,0,0\rangle_f = a^\dagger_{2V}|0,0,0,0\rangle_f$$
$$PBS_{23}.PBS_{01}Perm_{0213}.|0,1:V,0,0\rangle_f = |0,0,1:V,0\rangle_f$$

The component corresponding to the "readout" step enables the transformation of an encoding based on polarization across two spatial modes into one employing four distinct spatial modes. This readout allows for a binary detection resulting in one of the states $|0,1,0,0\rangle_f$, $|1,0,0,0\rangle_f$, $|0,0,1,0\rangle_f$ or $|0,0,0,1\rangle_f$.

$$|1:H,0,0,0\rangle_f \rightarrow |0,1:H,0,0\rangle_f \rightarrow |0,1,0,0\rangle_f$$
$$|1:V,0,0,0\rangle_f \rightarrow |1:V,0,0,0\rangle_f \rightarrow |1,0,0,0\rangle_f$$
$$|0,1:H,0,0\rangle_f \rightarrow |0,0,0,1:H\rangle_f \rightarrow |0,0,0,1\rangle_f$$
$$|0,1:V,0,0\rangle_f \rightarrow |0,0,1:V,0\rangle_f \rightarrow |0,0,1,0\rangle_f$$

### *7.3. Perceval implementation*

The full Photonic Grover code is inspired of the version provided on the Quandela web page and it is provided in Appendix. The example concern the state $|1:V,0\rangle_f$ which is marked by the oracle using $PR_0\left(\frac{\pi}{2}\right)$. The expected result is $|11\rangle$ according to table 5. The results provided in Figure 42 meets the theoretical requirements with 100% of the probability concentrated on $|11\rangle$.

```
[0.+0.j 0.+0.j 0.+0.j 1.+0.j]
[|0,0,0,1>, |0,0,1,0>, |0,1,0,0>, |1,0,0,0>]
[[0.+0.j 0.+0.j 0.+0.j 1.+0.j]]
[0.+0.j 0.+0.j 0.+0.j 1.+0.j]
['"00"', '"01"', '"10"', '"11"']
```

*Figure 42. Photonic Grover circuit execution*

## 7.4. Remarks

If one considers a circuit that encodes information using 4 photons distributed over 4 spatial modes, where each photon can occupy one of two polarization modes, the corresponding encoding space has a dimension of $2^4 = 16$. This appears equivalent to that of a standard quantum circuit. However, during the measure step, it becomes necessary to introduce 4 additional modes in order to define the binary encoding of the measurement outcome. Consequently, the photonic circuit employs operators acting on 8 modes, whereas a standard quantum circuit (e.g., implemented with Qiskit) would use operators acting on only 4 qubits (in addition to the ancillary qubit required for the oracle).

## 7.5. Remark on the Oracle

It should be noted that the simplest possible versions of the Oracles could be defined in a very different way that Oracles found in (Heurtel et al., 2023) and (Kwiat et al., 2000). The physical component corresponding to $PR_m(\alpha)$ requires manual configuration intervention to transition from $PR_m(\alpha)$ à $PR_n(\alpha')$. The switch from one oracle to a second one in Table 7 requires manual configuration and could be difficult to practice.

**Table 7: Oracle definition after simplification**

| Qubit encoding | Fock state encoding | Oracle definition (original version) |
|---|---|---|
| $\lvert 00\rangle$ | $\lvert 0, 1{:}H\rangle_f$ | $PR_1\left(-\frac{\pi}{2}\right)$ |
| $\lvert 01\rangle$ | $\lvert 0, 1{:}V\rangle_f$ | $PR_1\left(\frac{\pi}{2}\right)$ |
| $\lvert 10\rangle$ | $\lvert 1{:}H, 0\rangle_f$ | $PR_0\left(-\frac{\pi}{2}\right)$ |
| $\lvert 11\rangle$ | $\lvert 1{:}V, 0\rangle_f$ | $PR_0\left(\frac{\pi}{2}\right)$ |

From a practical standpoint, it may be advantageous to employ oracles that all use the same $PR_m(\alpha)$, which leads to the Oracles presented in Table 8: these are longer and more complex than the preceding ones.

**Table 8: Oracle definition based on $PR_0\left(\frac{\pi}{2}\right)$**

| Qubit encoding | Fock state encoding | Oracle definition (original version) |
|---|---|---|
| $\lvert 00\rangle$ | $\lvert 0, 1{:}H\rangle_f$ | $PR_0\left(\frac{\pi}{2}\right).PS_0\left(-\frac{\pi}{2}\right).HWP_0(0)PS_1\left(-\frac{\pi}{2}\right).HWP_1(0)$ |
| $\lvert 01\rangle$ | $\lvert 0, 1{:}V\rangle_f$ | $PS_0\left(-\frac{\pi}{2}\right).HWP_0(0).PR_0\left(\frac{\pi}{2}\right)PS_1\left(-\frac{\pi}{2}\right).HWP_1(0)$ |
| $\lvert 10\rangle$ | $\lvert 1{:}H, 0\rangle_f$ | $PS_0\left(-\frac{\pi}{2}\right).HWP_0(0).PR_0\left(\frac{\pi}{2}\right)PS_0\left(-\frac{\pi}{2}\right).HWP_0(0).$ $PS_1\left(-\frac{\pi}{2}\right).HWP_1(0)PS_1\left(-\frac{\pi}{2}\right).HWP_1(0)$ |
| $\lvert 11\rangle$ | $\lvert 1{:}V, 0\rangle_f$ | $PR_0\left(\frac{\pi}{2}\right)$ |

In the special case where the goal is to mark the state $\lvert 1{:}V, 0\rangle_{01}$, it suffices to apply the gate $PR_0\left(\frac{\pi}{2}\right)$, as shown in the previous calculations.

Let us explicit the oracles to mark the other states. But first we want to introduce a new notation

**Reminder:** On a given spatial mode $m$, we have:
$$PR_m\left(\frac{\pi}{2}\right)|1H\rangle_m = -|1V\rangle_m$$
$$PR_m\left(\frac{\pi}{2}\right)|1V\rangle_m = |1H\rangle_m$$

and

$$HWP_m(0).|1:H\rangle_m = \begin{pmatrix} i & 0 \\ 0 & -i \end{pmatrix}.\begin{pmatrix} 1 \\ 0 \end{pmatrix} = i.|1:H\rangle_m$$
$$HWP_m(0).|1:V\rangle_m = \begin{pmatrix} i & 0 \\ 0 & -i \end{pmatrix}.\begin{pmatrix} 0 \\ 1 \end{pmatrix} = -i.|1:V\rangle_m$$

**Notation:** To simplify the following calculations, let us consider the effect of applying $S_m\left(-\frac{\pi}{2}\right).HWP_m\left(\frac{\pi}{2}\right)$ on a given mode $m$

$$PS_m\left(-\frac{\pi}{2}\right).HWP_m\left(\frac{\pi}{2}\right)|1:H\rangle_m = PS_m\left(-\frac{\pi}{2}\right).i.|1:H\rangle_m = -i^2.|1:H\rangle_m = |1:H\rangle_m$$
$$PS_m\left(-\frac{\pi}{2}\right).HWP_m\left(\frac{\pi}{2}\right)|1:V\rangle_m = PS_m\left(-\frac{\pi}{2}\right).-i.|1:V\rangle_m = (-i)^2.|1:V\rangle_m = -|1:V\rangle_m$$

Now to allow for better reading, we define $\hat{O}_m$ for Operator applied to a mode $m$ :
$$\hat{O}_m = PS_m\left(-\frac{\pi}{2}\right).HWP_m\left(\frac{\pi}{2}\right)$$

We immediately get by definition of $\hat{O}_m$:
$$\hat{O}_m|1:H\rangle_m = |1:H\rangle_m$$
$$\hat{O}_m|1:V\rangle_m = -|1:V\rangle_m$$

Please note that this $\hat{O}_m$ is not a standard definition found elsewhere.

**Property:** The Oracle that allows to mark the $|0, 1:H\rangle_{01}$ state is:
$$\left[PS_1\left(-\frac{\pi}{2}\right).HWP_1(0)\right].\left[PR_0\left(\frac{\pi}{2}\right).PS_0\left(-\frac{\pi}{2}\right).HWP_0(0)\right]$$

**Proof.**

Considering the state $|\psi_1\rangle_{01}$ obtained before the oracle
$$|\psi_1\rangle_{01} = \frac{1}{2}.(|1:H,0\rangle_{01} + |0,1:H\rangle_{01} + |1:V,0\rangle_{01} + |0,1:V\rangle_{01})$$

We apply the oracle

$$|\psi_2\rangle_{01} = PR_0\left(\frac{\pi}{2}\right).PS_1\left(-\frac{\pi}{2}\right).HWP_1(0).PS_0\left(-\frac{\pi}{2}\right).HWP_0(0)\left(-\frac{1}{2}.(|1:H,0\rangle_{01}\right.$$
$$\left. + |0,1:H\rangle_{01} + |1:V,0\rangle_{01} + |0,1:V\rangle_{01})\right)$$

Thus, making use of the notation $\hat{O}_m$, we get

$$|\psi_2\rangle_f = \left[PR_0\left(\frac{\pi}{2}\right)\hat{O}_0\hat{O}_1.\right].\left(\frac{1}{2}.(|1:H,0\rangle_{01} + |0,1:H\rangle_{01} + |1:V,0\rangle_{01} + |0,1:V\rangle_{01})\right)$$

Applying the operators $\hat{O}$ gives

$$|\psi_2\rangle_f = PR_0\left(\frac{\pi}{2}\right)\left(\frac{1}{2}.(|1:H,0\rangle_{01} + |0,1:H\rangle_{01} - |1:V,0\rangle_{01} - |0,1:V\rangle_{01})\right)$$

Applying the polarisation rotator:

$$|\psi_2\rangle_f = \left(\frac{1}{2}.(-|1:V,0\rangle_{01} + |0,1:H\rangle_{01} - |1:H,0\rangle_{01} - |0,1:V\rangle_{01})\right)$$

Finally

$$|\psi_2\rangle_f = \frac{1}{2}(-|1:V,0\rangle_{01} + |0,1:H\rangle_{01} - |1:H,0\rangle_{01} - |0,1:V\rangle_{01})$$

□

**Property:** The Oracle that allows to mark the $|0,1:V\rangle_f$ is:

$$\left[PS_0\left(-\frac{\pi}{2}\right).HWP_0(0).PR_0\left(\frac{\pi}{2}\right)\right].\left[PS_1\left(-\frac{\pi}{2}\right).HWP_1(0)\right]$$

**Proof.**

Considering

$$|\psi_2\rangle_{01} = \left[PS_0\left(-\frac{\pi}{2}\right).HWP_0(0).PR_0\left(\frac{\pi}{2}\right)\right].\left[PS_1\left(-\frac{\pi}{2}\right).HWP_1(0)\right]|\psi_1\rangle_{01}$$

With

$$|\psi_1\rangle_{01} = \frac{1}{2}.(|1:H,0\rangle_{01} + |0,1:H\rangle_{01} + |1:V,0\rangle_{01} + |0,1:V\rangle_{01})$$

Let us rewrite using $\hat{O}_m$ :

$$|\psi_2\rangle_{01} = \left[\hat{O}_0\hat{O}_1.PR_0\left(\frac{\pi}{2}\right)\right]\left(\frac{1}{2}.(|1:H,0\rangle_{01} + |0,1:H\rangle_{01} + |1:V,0\rangle_{01} + |0,1:V\rangle_{01})\right)$$

Because

$$PR_0\left(\frac{\pi}{2}\right)|1:H\rangle_0 = -|1:V\rangle_0$$

$$PR_0\left(\frac{\pi}{2}\right)|1:V\rangle_0 = |1:H\rangle_0$$

We have

$$|\psi_2\rangle_{01} = [\hat{O}_0\hat{O}_1].\left(\frac{1}{2}.(-|1:V,0\rangle_{01} + |0,1:H\rangle_{01} + |1:H,0\rangle_{01} + |0,1:V\rangle_{01})\right)$$

Applying the two $\hat{O}_m$, we get:

$$|\psi_2\rangle_{01} = [\hat{O}_0\hat{O}_1].\left(\frac{1}{2}.(-|1:V,0\rangle_{01} + |0,1:H\rangle_{01} + |1:H,0\rangle_{01} + |0,1:V\rangle_{01})\right)$$

and finally

$$|\psi_2\rangle_{01} = \frac{1}{2}(|1:V,0\rangle_{01} + |0,1:H\rangle_{01} + |1:H,0\rangle_{01} - |0,1:V\rangle_{01})$$

□

**Property**

The Oracle that allows to mark the $|1:H,0\rangle_{01}$ is:

$$\left[PS_0\left(-\frac{\pi}{2}\right).HWP_0(0).PR_0\left(\frac{\pi}{2}\right)PS_0\left(-\frac{\pi}{2}\right).HWP_0(0)\right].$$
$$\left[PS_1\left(-\frac{\pi}{2}\right).HWP_1(0).PS_1\left(-\frac{\pi}{2}\right).HWP_1(0)\right]$$

**Proof.**

Applied to:

$$|\psi_1\rangle_{01} = \frac{1}{2}.(|1:H,0\rangle_{01} + |0,1:H\rangle_{01} + |1:V,0\rangle_{01} + |0,1:V\rangle_{01})$$

The oracle gives

$$|\psi_2\rangle_{01} = \left[PS_0\left(-\frac{\pi}{2}\right).HWP_0(0).PR_0\left(\frac{\pi}{2}\right)PS_0\left(-\frac{\pi}{2}\right).HWP_0(0)\right]$$
$$.\left[PS_1\left(-\frac{\pi}{2}\right).HWP_1(0)PS_1\left(-\frac{\pi}{2}\right).HWP_1(0)\right]$$
$$.\frac{1}{2}(|1:H,0\rangle_{01} + |0,1:H\rangle_{01} + |1:V,0\rangle_{01} + |0,1:V\rangle_{01})$$

Also written

$$|\psi_2\rangle_{01} = \left[\hat{O}_0.PR_0\left(\frac{\pi}{2}\right).\hat{O}_0\right].\frac{1}{2}(|1:H,0\rangle_{01} + |0,1:H\rangle_{01} + |1:V,0\rangle_{01} + |0,1:V\rangle_{01})$$

Applying $\hat{O}_0$:

$$|\psi_2\rangle_{01} = \left[\hat{O}_0.PR_0\left(\frac{\pi}{2}\right)\right].\frac{1}{2}(|1:H,0\rangle_{01} + |0,1:H\rangle_{01} - |1:V,0\rangle_{01} + |0,1:V\rangle_{01})$$

After the polarization rotator:

$$|\psi_2\rangle_{01} = [\hat{O}_0].\frac{1}{2}(-|1:V,0\rangle_{01} + |0,1:H\rangle_{01} - |1:H,0\rangle_{01} + |0,1:V\rangle_{01})$$

Finally the last $\hat{O}_0$:

$$|\psi_2\rangle_{01} = \frac{1}{2}(|1:V,0\rangle_{01} + |0,1:H\rangle_{01} - |1:H,0\rangle_{01} + |0,1:V\rangle_{01})$$

□

*7.6. Polarization Summary*

Photon polarization provides another approach to designing and defining circuits. The design of polarization-based photonic circuits relies on components specific to polarization, which have been presented in this work solely through computational modelling. As illustrated by the example of Grover's circuit, the use of polarization does not confer a clear advantage to photonic circuits, as the readout process requires specific operations that depend on new spatial modes essential for retrieving the results. This article provides a comprehensive introduction to help understand the concepts behind polarization and the associated components.

Let us note that in the description of Grover's oracle, the use of polarization allows for the avoidance of CNOT gates, whose success rates remain problematic, particularly for circuits with many CNOT gates.

**Table 9. Oracle definition of (Kiwat et al., 2000)**

| Fock state encoding | Oracle definition | Oracle definition (original version) |
|---|---|---|
| $\|0, 1:H\rangle_f$ | $PR_1\left(-\frac{\pi}{2}\right)$ | $PR_0\left(\frac{\pi}{2}\right).PS_0\left(-\frac{\pi}{2}\right).HWP_0(0)PS_1\left(-\frac{\pi}{2}\right).HWP_1(0)$ |
| $\|0, 1:V\rangle_f$ | $PR_1\left(+\frac{\pi}{2}\right)$ | $PS_0\left(-\frac{\pi}{2}\right).HWP_0(0).PR_0\left(\frac{\pi}{2}\right)PS_1\left(-\frac{\pi}{2}\right).HWP_1(0)$ |
| $\|1:H, 0\rangle_f$ | $PR_0\left(-\frac{\pi}{2}\right)$ | $PS_0\left(-\frac{\pi}{2}\right).HWP_0(0).PR_0\left(\frac{\pi}{2}\right)PS_0\left(-\frac{\pi}{2}\right).HWP_0(0).$ $PS_1\left(-\frac{\pi}{2}\right).HWP_1(0).PS_1\left(-\frac{\pi}{2}\right).HWP_1(0)$ |
| $\|1:V, 0\rangle_f$ | $PR_0\left(+\frac{\pi}{2}\right)$ | $PR_0\left(\frac{\pi}{2}\right)$ |

In this document, we chose to describe the oracle from (Kiwat et al., 2000) which is specifically tailored to respect physical constraints. It is possible to control the $PS_0\left(-\frac{\pi}{2}\right).HWP_0(0)$ apparition or not using Pockels cell (PC) and Crystal phase retarder (LC) using an electrical current. But it is not possible to have a controllable polarization rotator in the scheme proposed by (Kiwat et al., 2000) this is why we have a fixed polarization rotator in every oracle. Table 9 gives a comparison between the initial Kiwat et al' oracle proposition and the basic version of the oracle. The Proof that the basic version of the oracle. meets the same final results that oracle introduced by (Kiwat et al., 2000) can be achieved easily considering the quantum gates.

# 8. Conclusion

This article presents the foundational concepts for photonic computing with an operator- and computation-oriented perspective. Each gate is described in a matrix form, and we demonstrate how Pauli gates can be implemented using photonic operators. Basic elements for programming with Quandela's Perceval library are provided in the article and its appendices, to facilitate the reuse of these concepts. Photon polarization provides a novel approach to designing and defining circuits. The design of polarization-based photonic circuits relies on specific gates, which have been presented in this work solely through computational modeling. As illustrated by the example of Grover's circuit, the use of polarization does not confer a clear advantage to photonic circuits, as the readout process requires specific operations that depend on new modes essential for retrieving the results. Let us note that, the use of polarization allows for the avoidance of CNOT gates, whose success rates remain problematic, particularly for circuits with a large number of CNOT gates.

**Remark**

Authors have used Chatgpt to improve English skill

# Appendix: *X* gate implementation

```python
from perceval.rendering import DisplayConfig, SymbSkin
from perceval.components import PS, BS, PERM

# 1. initialisation

# qubit 1 -> |1> et qubit 2 -> |0>
# definition des deux états de fock
qubit_istate = [1,0]
istate = toFockState(qubit_istate)

print("Input qubit state:", strState(qubit_istate))
print("Corresponding input Fock state:", strState(istate))

# 2.ajout d'une porte X
circuit = pcvl.Circuit(4)
circuit.add(0,PERM([1,0, 2,3]))

# 3. affichage du circuit
pcvl.pdisplay(circuit.compute_unitary())
print("")
pcvl.pdisplay(circuit)

# 4. Simulation

backend = pcvl.BackendFactory().get_backend("Naive")
backend.set_circuit(circuit)
backend.set_input_state(pcvl.BasicState(istate))

output_qubit_states = [
    [x1,x2]
    for x1 in [0,1] for x2 in [0,1]
]

# 5. Affichage des résultats

print("Output state amplitudes: (post-selected on qubit states, not renormalized)")
print("|x1,x2>")
for oqstate in output_qubit_states:
    ostate = toFockState(oqstate)
    a = backend.prob_amplitude(pcvl.BasicState(ostate))

    print(strState(ostate)," -> ",strState(oqstate), a)
```

# Appendix: Grover implementation

```python
from perceval.components import catalog

# define the gates used
t_gate = catalog['t'].build_processor()
h_gate = catalog['h'].build_processor()
t_dag_gate = catalog['tdag'].build_processor()
cnot = catalog['heralded cnot'].build_processor()
X = catalog['x'].build_processor()

# Dual-rail encoding: 3 qubits = 6 modes
# q0 = [0,1], q1 = [2,3], q2 = [4,5]
q0, q1, q2 = [0,1], [2,3], [4,5]

# Initialisation
processor_toffoli = pcvl.Processor("SLOS",6)
processor_toffoli.add(q2, X)
processor_toffoli.add(q0, h_gate)
processor_toffoli.add(q1, h_gate)
processor_toffoli.add(q2, h_gate)

# BEGIN Naive oracle
processor_toffoli.add(q0, X)

# BEGIN TOFFOLI IMPLEMENTATION FROM CNOT
processor_toffoli.add(q2, h_gate)                   # (q2, H)
processor_toffoli.add(q1 + q2, cnot)                # (q1+q2, CNOT)
processor_toffoli.add(q2, t_dag_gate)               # (q2, T dag)
processor_toffoli.add(q0 + q2, cnot)                # (q0+q2, CNOT)
processor_toffoli.add(q2, t_gate)                   # (q2, T)
processor_toffoli.add(q1 + q2, cnot)                # (q1+q2, CNOT)
processor_toffoli.add(q2, t_dag_gate)               # (q2, T†)
processor_toffoli.add(q0 + q2, cnot)                # (q0+q2, CNOT)
processor_toffoli.add(q1, t_gate)                   # (q1, T)
processor_toffoli.add(q2, t_gate)                   # (q2, T)
processor_toffoli.add(q0 + q1, cnot)                # (q0+q1, CNOT)
processor_toffoli.add(q2, h_gate)                   # (q2, H)
processor_toffoli.add(q0, t_gate)                   # (q0, T)
processor_toffoli.add(q1, t_dag_gate)               # (q1, T†)
processor_toffoli.add(q0 + q1, cnot)                # (q0+q1, CNOT)

# END TOFFOLI IMPLEMENTATION FROM CNOT
processor_toffoli.add(q0, X)

# END Naive oracle

# Amplification
processor_toffoli.add(q0, h_gate)
processor_toffoli.add(q1, h_gate)
processor_toffoli.add(q0, X)
processor_toffoli.add(q1, X)
processor_toffoli.add(q1, h_gate)

processor_toffoli.add(q0+q1, cnot)
```

```
processor_toffoli.add(q1, h_gate)
processor_toffoli.add(q0, X)
processor_toffoli.add(q1, X)
processor_toffoli.add(q0, h_gate)
processor_toffoli.add(q1, h_gate)

# printing the circuit
pcvl.pdisplay(processor_toffoli,recursive=False)

# Simulation and execution of the circuit
from perceval.algorithm import Sampler # import the Sampler class

# Defining the input state.
istate=[1,0,1,0,1,0] # qubit input state : |000⟩
processor_toffoli.min_detected_photons_filter(0)   # Do not filter out any output state
processor_toffoli.with_input(pcvl.BasicState(istate))

# The sampler holds 'probs', 'sample_count' and 'samples' calls. You can use the one that fits your needs!
sampler = Sampler(processor_toffoli)

# Get the results.
sample_count = sampler.sample_count(1000)
print(sample_count['results'])
```

# Appendix: Grover with polarization

```python
from tabulate import tabulate
import numpy as np
import sympy as sp
import matplotlib.pyplot as plt

import perceval as pcvl

states = [pcvl.BasicState("|0,{P:H}>"),
          pcvl.BasicState("|0,{P:V}>"),
          pcvl.BasicState("|{P:H},0>"),
          pcvl.BasicState("|{P:V},0>"),
         ]

states_modes = [
    pcvl.BasicState([0, 0, 0, 1]),
    pcvl.BasicState([0, 0, 1, 0]),
    pcvl.BasicState([0, 1, 0, 0]),
    pcvl.BasicState([1, 0, 0, 0])
]

BS = pcvl.BS.Ry()

def HWPandPS(xsi):
    hwp = pcvl.Circuit(m=1)
    hwp.add(0, pcvl.HWP(xsi)).add(0, pcvl.PS(-sp.pi/2))
    return hwp

# construction du circuit
# ----------------------

init_circuit = (pcvl.Circuit(m=2, name="Initialization")
                // (1,pcvl.HWP(sp.pi / 8))
                // (1, pcvl.PS(-sp.pi/2))
                // BS
                // pcvl.PS(-sp.pi) )

pcvl.pdisplay(init_circuit)

def oracle():
    oracle_circuit = pcvl.Circuit(m=2, name='Oracle')

    # pour marquer |1V;0> i.e. 100% sur 11
    # oracle_circuit.add(0, pcvl.PR(sp.pi/2))

    # pour marquer |0;1H> i.e. 100% sur 00
    #oracle_circuit.add(1, pcvl.PR(-sp.pi/2))

    # pour marquer |0;1V> i.e. 100% sur 01

    #oracle_circuit.add(1, pcvl.HWP(0))
    #oracle_circuit.add(1, pcvl.PS(-sp.pi/2))
    #oracle_circuit.add(0, pcvl.PR(sp.pi/2))
    #oracle_circuit.add(0, pcvl.HWP(0))
```

```python
        #oracle_circuit.add(0, pcvl.PS(-sp.pi/2))

        # pour marquer |1H,0> i.e. 100% sur 10
        oracle_circuit.add(0, pcvl.HWP(0))
        oracle_circuit.add(0, pcvl.PS(-sp.pi/2))
        oracle_circuit.add(0, pcvl.PR(sp.pi/2))
        oracle_circuit.add(0, pcvl.HWP(0))
        oracle_circuit.add(0, pcvl.PS(-sp.pi/2))

    return oracle_circuit

pcvl.pdisplay(oracle())

inversion_circuit = (pcvl.Circuit(m=2, name='Inversion')
                    // BS
                    // (1,pcvl.HWP(sp.pi / 4))
                    // (1,pcvl.PS(-sp.pi / 2))
                    // BS)

pcvl.pdisplay(inversion_circuit)

detection_circuit = pcvl.Circuit(m=4, name='Detection')
detection_circuit.add(1, pcvl.PERM([1, 0]))
detection_circuit.add((0, 1), pcvl.PBS())
detection_circuit.add((2, 3), pcvl.PBS())

pcvl.pdisplay(detection_circuit)

def grover_circuit():
    grover_circuit = pcvl.Circuit(m=4, name='Grover')
    grover_circuit.add(0,     init_circuit).add(0,     oracle()).add(0,
inversion_circuit)
    grover_circuit.add(0, detection_circuit)
    return grover_circuit

print('Grover optical circuit for searching database element "00":')
pcvl.pdisplay(grover_circuit(), recursive=True)

# Circuit simulation
input_state = pcvl.BasicState("|0,{P:H}, 0, 0>")
results_list = []  # probability amplitudes storage

sim = pcvl.Processor("SLOS", grover_circuit())
ca = pcvl.algorithm.Analyzer(sim,
                             input_states=[input_state],
                             output_states=states_modes,
                            )
print(ca.distribution[0])
print(ca.output_states_list)
print(ca._distribution)

results_list.append(ca.distribution[0])
labels = ['"00"', '"01"', '"10"', '"11"']
state_0_prob_list = results_list[0]
print(state_0_prob_list)
print(labels)
```

# Appendix: *Ralph Cnot* implementation

```python
import perceval as pcvl
import numpy as np

def toFockState(qubitState):
            pe = {0:[1,0],  1:[0,1],  2:[0,1]}
            res =  pe[qubitState[0]] + pe[qubitState[1]]+ pe[qubitState[2]]
            return res
def strState(state):
         return str(pcvl.BasicState(state))

from perceval.components import PS, BS, PERM

theta_13 = BS.r_to_theta(r=1/3)
cnot = (pcvl.Circuit(6, name = "Ralph CNOT")
       .add((0, 1), BS.H(theta_13, phi_bl = np.pi,
             phi_tr = np.pi/2, phi_tl = -np.pi/2))
       .add((3, 4), BS.H())
       .add((2, 3), BS.H(theta_13, phi_bl = np.pi,
              phi_tr = np.pi/2 , phi_tl = -np.pi/2))
       .add((4, 5), BS.H(theta_13))
       .add((3, 4), BS.H()))

# 1. initialisation
istate=[0,1,0,1,0,0]

# 2. test of the cnot
circuit = pcvl.Circuit(6)

circuit.add(0,PERM([0,1,2,4,3,5]))
circuit.add(0,PERM([0,1,3,2,4,5]))
circuit.add(0,PERM([0,2,1,3,4,5]))
circuit.add(0,PERM([1,0,2,3,4,5]))
circuit.add(0,cnot)
circuit.add(0,PERM([1,0,2,3,4,5]))
circuit.add(0,PERM([0,2,1,3,4,5]))
circuit.add(0,PERM([0,1,3,2,4,5]))
circuit.add(0,PERM([0,1,2,4,3,5]))

# 3. Displaying the circuit
pcvl.pdisplay(circuit.compute_unitary())
print("")
pcvl.pdisplay(circuit)

# 4. Simulation
p = pcvl.Processor("SLOS", circuit)
p.set_postselection(pcvl.PostSelect("[0,1]==1 & [2,3]==1 & [4]==0 & [5]==0"))
nsample = 50000
from perceval.algorithm import Sampler , Analyzer
sampler = Sampler ( p )
p.with_input ( pcvl.BasicState ([0, 1, 0, 1, 0, 0]))
output = sampler.sample_count ( 1000 )
pcvl.pdisplay ( output['results'], output_format = pcvl.Format.TEXT )
```

# Appendix: *Heralded* Cnot

```python
import perceval as pcvl
import numpy as np
from perceval.components import PS, BS, PERM

from perceval.components import catalog

cnot = catalog['heralded cnot'].build_circuit()

# 1. initialization
istate=[0,1,0,1,1,1]

# 2. test of the cnot
circuit = pcvl.Circuit(6)
circuit.add(0,cnot)

# 3. Displaying the circuit
pcvl.pdisplay(circuit.compute_unitary())
print("")
pcvl.pdisplay(circuit)

# 4 calculation
p = pcvl.Processor("SLOS", circuit)
p.set_postselection(pcvl.PostSelect("[4]==1 & [5]==1"))
nsample = 50000
from perceval.algorithm import Sampler , Analyzer

sampler = Sampler ( p )
p.with_input ( pcvl.BasicState (istate)) #Corresponds to logical qubit state

output = sampler.sample_count ( 1000 )

pcvl.pdisplay ( output['results'], output_format = pcvl.Format.TEXT )
print('end')
```

# Appendix: *Ralph Cnot* with the library Perceval

```python
import perceval as pcvl
import numpy as np
from perceval.components import PS, BS, PERM

from perceval.components import catalog

cnot = catalog['postprocessed cnot'].build_circuit()

# 1. initialization
istate=[0,1,0,1,0,0]
# 2. test of cnot
circuit = pcvl.Circuit(6)
circuit.add(0,cnot)
# 3. Print of the circuit
pcvl.pdisplay(circuit.compute_unitary())
print("")
pcvl.pdisplay(circuit)
# 4 Computation
p = pcvl.Processor("SLOS", circuit)
p.set_postselection(pcvl.PostSelect("[0,1]==1 & [2,3]==1 & [4]==0 & [5]==0"))
nsample = 50000
from perceval.algorithm import Sampler , Analyzer

sampler = Sampler ( p )
p.with_input ( pcvl.BasicState (istate)) #Corresponds to logical qubit state

output = sampler.sample_count ( 1000 )
pcvl.pdisplay ( output['results'], output_format = pcvl.Format.TEXT )
print('end')
```

# Appendix: Implementation of *CCNOT*

```python
import perceval as pcvl
import numpy as np
from perceval.components import PS, BS, PERM
from perceval.components import catalog
ccnot = catalog['toffoli'].build_circuit()

# 1. initialisation
istate=[0, 1, 0, 1, 1, 0, 0, 0, 0, 0, 0, 0]
# 2. Cnot
circuit = pcvl.Circuit(12)
circuit.add(0,ccnot)
# 3. print of the circuit
pcvl.pdisplay(circuit.compute_unitary())
print("")
pcvl.pdisplay(circuit)
# 4 Computation
p = pcvl.Processor("SLOS", circuit)
p.set_postselection(pcvl.PostSelect("[0,1]==1 & [2,3]==1 & [6]==0 & [7]==0 &
[8]==0 & [9]==0 & [10]==0 & [11]==0"))

nsample = 1000
from perceval.algorithm import Sampler , Analyzer
sampler = Sampler ( p )
p.with_input ( pcvl.BasicState (istate)) #Corresponds to logical qubit state
output = sampler.sample_count ( nsample )
print(output['results'])
pcvl.pdisplay ( output['results'], output_format = pcvl.Format.TEXT )
print('end')
```

# Appendix: Perceval implementation of $PERM_{1023}$

```python
import perceval as pcvl
import numpy as np
from perceval.components import PS, BS, PERM

def toFockState(qubitState):
              pe = {0:[1,0],  1:[0,1]}
              res =  pe[qubitState[0]] + pe[qubitState[1]]
              return res
def strState(state):
           return str(pcvl.BasicState(state))

qubit_istate = [1,0]
istate = toFockState(qubit_istate)

circuit = pcvl.Circuit(4)

circuit.add(0,PERM([1,0, 2,3]))
#or circuit.add(0,PERM([1,0]))

pcvl.pdisplay(circuit.compute_unitary())
print("")
pcvl.pdisplay(circuit)

backend = pcvl.BackendFactory().get_backend("Naive")
backend.set_circuit(circuit)
backend.set_input_state(pcvl.BasicState(istate))

output_qubit_states = [
    [x1,x2]
    for x1 in [0,1] for x2 in [0,1]
]

for oqstate in output_qubit_states:
    ostate = toFockState(oqstate)
    a = backend.prob_amplitude(pcvl.BasicState(ostate))
    print(strState(ostate)," -> ",strState(oqstate), a)

print('end')
```